# Variance of the Hellings-Downs correlation

Bruce Allen[*]
*Max Planck Institute for Gravitational Physics (Albert Einstein Institute), Leibniz Universität Hannover, Callinstrasse 38, D-30167 Hannover, Germany*

Gravitational waves (GWs) create correlations in the arrival times of pulses from different pulsars. The expected correlation $\mu(\gamma)$ as a function of the angle $\gamma$ between the directions to two pulsars was calculated by Hellings and Downs for an isotropic and unpolarized GW background, and several pulsar timing array (PTA) collaborations are working to observe these. We ask: given a set of noise-free observations, are they consistent with that expectation? To answer this, we calculate the expected variance $\sigma^2(\gamma)$ in the correlation for a single GW point source, as pulsar pairs with fixed separation angle $\gamma$ are swept around the sky. We then use this to derive simple analytic expressions for the variance produced by a set of discrete point sources uniformly scattered in space for two cases of interest: (1) point sources radiating GWs at the same frequency, generating confusion noise, and (2) point sources radiating GWs at distinct nonoverlapping frequencies. By averaging over all pulsar sky positions at fixed separation angle $\gamma$, we show how this variance may be cleanly split into cosmic variance and pulsar variance, also demonstrating that measurements of the variance can provide information about the nature of GW sources. In a series of technical appendices, we calculate the mean and variance of the Hellings-Downs correlation for an arbitrary (polarized) point source, quantify the impact of neglecting pulsar terms, and calculate the pulsar and cosmic variance for a Gaussian ensemble. The mean and variance of the Gaussian ensemble may be obtained from the previous discrete-source confusion-noise model in the limit of a high density of weak sources.

## I. INTRODUCTION

The idea of observing low-frequency gravitational waves (GWs) via their influence on radio pulsar arrival times first appears in work of Sazhin [1]. Subsequently Detweiler [2] described the improvements obtained by cross-correlating data from multiple pulsars. Such detectors, called "pulsar timing arrays" (PTAs), have been studied in detail by many authors [3–5]. Currently, a number of groups are working to detect gravitational waves in the nano-Hz regime with these techniques [6–8].

The basic idea is as follows. Select a pulsar which is a very stable clock, emitting pulses at uniform intervals in its rest frame. Determine the arrival times $t$ of the pulses at the solar system barycenter (SSB) (calculated based on the arrival times at Earth [9,10]). In the absence of GWs, those pulses would have uniform time offsets from each other, whereas in the presence of a GW, the offsets $\Delta\tau$ from regular arrival times are nonzero. These form a time series of *timing residuals* $\Delta\tau(t)$ that vary at the frequency of the GW, with an amplitude proportional to the strain amplitude of the GW.

In 1983, Hellings and Downs [11] predicted an important property of pulsar timing residuals, which is central to their detection with PTAs. The timing residuals from different pulsars are correlated, in a way that depends upon the angle $\gamma$ between the lines of sight to the pulsars as viewed from Earth. This is "on the average" because the correlation depends upon the actual positions of the pulsars on the sky, even if the angle between them is fixed. A clear explanation of these effects may be found in [12]. To observe this, one averages the product of the timing residual time-series over years or decades to form a correlation

$$\rho = \overline{\Delta\tau_1(t)\Delta\tau_2(t)}, \quad (1.1)$$

where the overbar denotes a time average, and 1 and 2 label the pulsars. The expected value of $\rho$ vanishes if the timing residuals are uncorrelated, and is nonzero if they are correlated.

The function which describes the behavior of the average or expected correlation $\langle\rho\rangle$ as a function of the angle $\gamma$ between the directions to the two pulsars is called the Hellings-Downs curve $\mu_u(\gamma)$, where the subscript means

---

[*]bruce.allen@aei.mpg.de

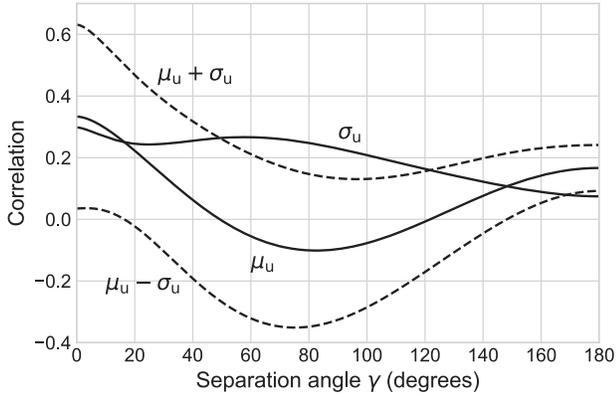

FIG. 1. The Hellings-Downs correlation mean $\langle\rho\rangle = \mu_u(\gamma)$ and standard deviation $\Delta\rho = \sigma_u(\gamma)$ for a single unpolarized point source with the pulsar term neglected, from Eqs. (2.4) and (2.6). The dashed lines show a $\pm 1\sigma_u$ range about the mean.

"unpolarized." This function is given in Eq. (2.4) and illustrated in Fig. 1. On the average, the expected correlation is largest when the two pulsars are along the same line of sight, and (for example) vanishes when the two pulsars are separated by about 50°. A convincing PTA detection of GWs is expected to show this pattern of correlation among the different possible pulsar pairs [13].

Here, we ask: how close to this expectation would the observed correlations lie, in the absence of experimental noise or errors? Simulations have demonstrated that fluctuations away from the mean can be large [14,15]. These fluctuations away from the mean have two contributions: a "pulsar variance" and a "cosmic variance." The first, which can in principle be measured and removed, reflects the variations in correlation between pairs of pulsars at different sky locations, but separated by the same angle $\gamma$. The second cannot be directly observed but can be inferred. It arises because GW sources radiating at overlapping frequencies generate interference patterns, whose average over pulsar pairs differs from the Hellings-Downs curve.

Previous work on estimating the variance relies on "Gaussian ensemble" techniques which were developed for the detection of audio-frequency GW stochastic backgrounds with interferometric detectors [16,17]. These methods were extended to PTAs in [18,19], and used to estimate the variances in [20–22]. In contrast, our calculations use a finite collection of discrete sources and smoothly transition from one source to many. Similar models have been employed in the context of audio-frequency GW stochastic backgrounds [23]. If we let the density of sources approach infinity while keeping the mean-squared GW amplitude at Earth fixed, then our discrete source models reproduce the results of the Gaussian ensemble (see Appendix C 3).

One of the most important contributions of this paper is a straightforward method to separately identify the pulsar and the cosmic variance contributions to the total variance. The distinction is relevant, because observations with access to enough low-noise pulsars, uniformly distributed around the sky, could reduce the impact of pulsar variance as much as desired, by averaging the Hellings-Downs correlation over many pairs of pulsars separated by the same angle. However, the effects of cosmic variance cannot be removed. Our calculational technique to distinguish these effects corresponds closely to experimental practice. To compute the cosmic variance, we average the correlation over pulsar pairs separated by the same angle, *before* the first and second moments of the correlation are calculated with ensemble averages. The same technique can also be applied to the traditional Gaussian ensemble (see Appendix C 5).

Simulations modeling sources in the nearest few hundred Mpc to Earth suggest that these can cause significant deviations from isotropy [15,24]. It has been argued that these will not lead to substantial variations from the Hellings-Downs curve [13,25]. Our findings tend to support this conclusion for the mean of the correlation, but not for the variance, which is also an observable quantity. In this sense, the predicted variance of the Hellings-Downs curve contains additional information about the nature of the sources.

A brief outline of the paper is as follows. In Sec. II, we present and discuss pulsar variance for a single unpolarized point source, neglecting pulsar terms. In Sec. III, we show how those single-point results can be used to estimate the variance for a large collection of point sources, based on a simple model for the ensemble of sources, also including pulsar terms. As a simple example of the calculation technique (unrelated to PTAs) we calculate the expected value and variance of the GW energy density. In Sec. III A, we consider point sources which are all radiating at the same GW frequency, resulting in confusion noise, and in Sec. III B, we consider point sources radiating at distinct frequencies. Finally, in Sec. IV, we show how these variances can be cleanly split into pulsar variance and cosmic variance. This is followed by a short Conclusion.

The paper relies on detailed calculations given in a series of technical appendices, which are referenced throughout. In Appendix A, we calculate the mean and variance of the Hellings-Downs correlation for a general GW point source and explicitly demonstrate that the "pulsar average" over three variables is equivalent to the "GW sky direction" average over two variables. As an example, including Earth terms only, we calculate the mean and variance for a binary inspiral GW source model (which is polarized if the orbit is not face-on or face-off) and its population average.

In Appendix B, we examine the effects of the pulsar terms on the mean and variance of the Hellings-Downs correlation. These do not affect the mean, but for sources with slow evolution they increase the variance by a factor of four. As an example, we calculate the mean and variance for a nonevolving (single frequency) binary inspiral GW source model, and its population average.

In Appendix C, we employ the standard Gaussian ensemble from [17]. We first estimate the importance of the pulsar terms in the Hellings-Downs correlation. We then compute the total variance (one pulsar pair) and the cosmic variance (many pulsar pairs) of the Hellings and Downs correlation for the Gaussian ensemble. We also compute the variance of the GW energy density. These Gaussian ensemble results are limits of our discrete-source confusion-noise ensemble results from Secs. III A and IV. Finally, we derive the cosmic variance as a harmonic sum.

Appendix D is the first of three in which we carry out the integrals needed to obtain the second moments of the Hellings-Downs correlation. Appendix D does the unpolarized case corresponding to the "$I$" term in the Stokes parameters. Appendix E does the polarized case corresponding to the "$Q$" Stokes parameter, and Appendix F does the cross term.

Appendix G calculates the Hellings-Downs two-point function $\mu(\gamma, \beta)$ and its second moment $\tilde{\mu}^2(\gamma)$ with respect to $\beta$.

Notes on terminology and notation:
 (i) The term "Hellings-Downs correlation" refers to correlations among pulsar timing residuals, which might or might not follow the "Hellings-Downs curve" $\mu_u(\gamma)$.
 (ii) Correlations between pulsar timing residuals or redshifts are denoted by $\rho$, and we use $\mu$ and $\sigma^2$ for the mean and variance of $\rho$. These symbols are used for general purposes, for specific models or systems, or for the Universe as a whole.
 (iii) In contrast, the expressions $\mu_u(\gamma)$, $\sigma_u^2(\gamma)$, $\sigma_p^2(\gamma)$, $\sigma_c^2(\gamma)$, $\mu(\gamma, \beta)$ and $\tilde{\mu}^2(\gamma)$ are used for specific mean and variance functions which are calculated in the Appendix and given in Eqs. (D29), (D37), (E8), (F2), (G5), and (G11).
 (iv) Angle brackets $\langle Q \rangle$ indicate the average or expectation value of a function $Q$, with respect to certain variables. This is always normalized so that $\langle 1 \rangle = 1$. If unclear from context, we add a subscript after the right-hand bracket to indicate the type of averaging.
 (v) Three-vectors are indicated in boldface $\boldsymbol{x}$; spatial indices $a$, $b$, $c$ and $d$ denote vector or tensor components. For these indices, we adopt the Einstein summation convention.
 (vi) Indices $A, A', \ldots$ denote one of two orthogonal GW polarizations "+" and "×". For these, repeated indices are *not* summed.
 (vii) We use fixed Cartesian coordinates centered at the SSB with $\hat{\boldsymbol{x}}, \hat{\boldsymbol{y}}, \hat{\boldsymbol{z}}$ constant unit-length vectors along three orthogonal directions. Time $t$ is that of a clock at rest at the SSB. Since the distance from Earth to the SSB (which is located near the Sun, about 500 seconds from Earth) is much less than the GW wavelengths or other distance of interest here (years to millenia), quantities evaluated at the SSB are called "Earth" terms.
 (viii) Unit vectors on the two-sphere are written $\boldsymbol{\Omega}$.
 (ix) The angles $\theta \in [0, \pi]$ and $\phi \in [0, 2\pi)$ are conventional spherical-polar coordinates on the two-sphere, where $\theta$ is the angle down from the z-axis and $\phi$ is the angle counterclockwise from the x-axis after projection into the x–y plane.
 (x) Indices $j$, $k$, $\ell$, and $m$ in the range $1, 2, \ldots, N$ label GW sources. For these, repeated indices are *not* summed.
 (xi) The two pulsars with correlated pulse arrival times or redshifts are labeled "1" and "2" and are at distances $L_1$ and $L_2$ from Earth. Unit vectors pointing to them are $\boldsymbol{p}_1$ and $\boldsymbol{p}_2$, and we use $\gamma \in [0, \pi]$ for the angle between them on the sky: $\cos \gamma = \boldsymbol{p}_1 \cdot \boldsymbol{p}_2$.
 (xii) Some calculations incorporate a variable $\chi$ which is set to unity to properly incorporate pulsar terms, or set to zero to neglect pulsar terms.
 (xiii) Unless indicated, we use units in which the speed of light $c = 1$. We assume that GWs also propagate at this speed, since observations demonstrate that the fractional difference is at most a few parts in $10^{15}$ [26–28].

## II. MEAN AND VARIANCE OF THE HELLINGS-DOWNS CORRELATION FOR ONE UNPOLARIZED GW POINT SOURCE

Here, we briefly describe the mean and variance of the Hellings-Downs correlation for a single unpolarized point source, which are computed in Appendix D and used throughout the paper. In this section, we neglect pulsar terms.

Hellings and Downs calculate the effect (on pulse redshift or timing residual) from a plane GW with unit amplitude traveling in direction $\boldsymbol{\Omega}$. For a wave with polarization $A = +$ or $A = \times$, this effect is proportional to the antenna pattern function

$$F^A(\boldsymbol{\Omega}) = \frac{1}{2} \frac{p^a p^b}{1 + \boldsymbol{\Omega} \cdot \boldsymbol{p}} e^A_{ab}(\boldsymbol{\Omega}), \qquad (2.1)$$

where $\boldsymbol{p}$ is a unit-length spatial vector pointing from Earth to pulsar, and the $e^A_{ab}(\boldsymbol{\Omega})$ are a pair of normalized orthogonal spin-two polarization tensors, defined in Eq. (D6). Clear derivations of these equations starting from first principles may be found in Appendixes A and B of Ref. [18], noting the redshift sign correction in [29].

The correlation between two pulsars (subscripts 1 and 2) summed uniformly over both GW polarizations is

$$\rho = F_1^+(\boldsymbol{\Omega}) F_2^+(\boldsymbol{\Omega}) + F_1^\times(\boldsymbol{\Omega}) F_2^\times(\boldsymbol{\Omega}). \qquad (2.2)$$

The Hellings-Downs curve [[11], Eq. 5] is obtained by averaging this correlation over the directions of the GWs. We use angle brackets to denote this uniform average over directions: if $Q(\boldsymbol{\Omega})$ is any function of direction, define

$$\langle Q \rangle = \frac{1}{4\pi} \int d\boldsymbol{\Omega} \, Q(\boldsymbol{\Omega})$$
$$= \frac{1}{4\pi} \int_0^\pi \sin\theta \, d\theta \int_0^{2\pi} d\phi \, Q(\theta, \phi). \qquad (2.3)$$

If we average the correlation $\rho$ of Eq. (2.2) in this way, we obtain the Hellings-Downs curve

$$\mu_{\rm u}(\gamma) = \langle \rho \rangle$$
$$= \frac{1}{4} + \frac{1}{12}\cos\gamma + \frac{1}{2}(1-\cos\gamma)\log\left(\frac{1-\cos\gamma}{2}\right), \quad (2.4)$$

where $\gamma$ is the angle between the lines of sight to the two pulsars $\cos\gamma = \boldsymbol{p}_1 \cdot \boldsymbol{p}_2$. A detailed calculation is given in Appendix D; this result is obtained in Eq. (D29) and is plotted in Fig. 1.

Note that we follow the Hellings-Downs normalization convention $\mu_{\rm u}(0) = 1/3$. Some authors normalize this correlation function to $1/2$ at $\gamma = 0$, and some incorporate an additional delta-function term which only contributes if pulsar 1 is identical to pulsar 2 (same distance and direction) as shown in Eq. (C15). This additional delta-function comes from the "pulsar terms," and doubles the correlation for time-stationary GW sources. It arises from Eq. (3.41) for individual sources and from Eq. (C24) for a Gaussian ensemble of sources.

Our interpretation of the function $\mu_{\rm u}(\gamma)$ is based on an important argument from [25]. Place an unpolarized point source of GWs at a fixed point on the sky. Then consider the correlation Eq. (2.2) as a function of the three variables that define the positions of the two pulsars on the sky, with fixed angle $\gamma$ between them. As the pulsar pair is shifted around the sky, keeping the angle between them fixed, the correlation will vary. Map that function $\rho$ of three variables (a mountainous landscape) and add a flat level surface at the mean value $\mu_{\rm u}(\gamma) = \langle \rho \rangle$, where the angle brackets now refer to that average over three variables. The deviations of $\rho$ from that level surface are the amount $\Delta\rho$ by which the correlation for a particular configuration of pulsars at separation angle $\gamma$ differs from the mean. The scale of those variations away from the mean (in the absence of any other sources of noise) is characterized by the (in this case "pulsar") variance

$$\sigma_{\rm u}^2(\gamma) = \langle \Delta\rho^2 \rangle = \langle(\rho - \langle\rho\rangle)^2\rangle = \langle\rho^2\rangle - \langle\rho\rangle^2, \quad (2.5)$$

where the subscript "u" means "unpolarized." This variance cannot be determined from the Hellings-Downs curve itself, which is the first moment $\langle\rho\rangle$. A new integral needs to be evaluated,to obtain the second moment of the product of the antenna patterns: $\langle\rho^2\rangle$.

There are two equivalent ways to calculate that second moment, which give exactly the same result. One can (a) fix the source position and average over the three variables that define the directions to the pulsars at fixed relative angle $\gamma$ (we do this in Appendix A) or one can (b) fix the pulsar locations, with angle $\gamma$ between them, and average over the two variables that define the direction to the source (we do this in Appendix D). Both approaches give the second moment of Eq. (D36). From that second moment, one finds the variance from Eqs. (2.4) and (2.5), obtaining

$$\sigma_{\rm u}^2(\gamma) = \frac{97}{80} + \frac{1}{24}\cos\gamma - \frac{839}{720}\cos^2\gamma$$
$$+ \frac{1}{12}(1-\cos\gamma)\log\left(\frac{1-\cos\gamma}{2}\right) \quad (2.6)$$
$$\times \left(18 - 10\cos\gamma - 3(1-\cos\gamma)\log\left(\frac{1-\cos\gamma}{2}\right)\right).$$

A detailed calculation is given in Appendix D; this result is obtained in Eq. (D37).

The square root of the variance is the standard deviation $\Delta\rho = \sigma_{\rm u}$. This is also plotted in Fig. 1. The standard deviation indicates the approximate range in which, with the Hellings and Downs assumptions, the correlation for any particular pulsar pair should fall, if the Universe contains a single distant pointlike unpolarized GW source, and the pulsar terms can be neglected. To indicate this graphically, the dashed lines in Fig. 1 show the $\pm 1\Delta\rho$ range about $\langle\rho\rangle$. (Figure 3 shows a similar plot for a constant amplitude GW source, to which the pulsar terms necessarily contribute.)

In Appendix A we show that the mean correlation produced by a single (possibly polarized) point source is always described by the Hellings-Downs curve $\mu_{\rm u}(\gamma)$. However, the variance is a sum of two functions, if pulsar terms are neglected, or three functions, if pulsar terms are included.

## III. VARIANCE OF THE HELLINGS-DOWNS CORRELATION WITH MANY SOURCES

The mean and variance of the correlation for a single point source are not directly applicable. We expect that our Universe contains large numbers of PTA point sources, probably in the form of supermassive black-hole binaries [30,31] with the closest at 50–100 Mpc from Earth, and the most distant at the Hubble radius $\approx 4$ Gpc. Here, we use the single point source results to predict the mean and variance for a large collection of such point sources.

We begin by constructing a model Universe containing $N$ point sources at sky locations $-\boldsymbol{\Omega}_j$, where $j = 1, 2, \ldots, N$. The minus sign ensures that the GWs of relevance to us propagate in direction $\boldsymbol{\Omega}_j$, in agreement with our conventions elsewhere in this paper. (Nevertheless, we often call $\boldsymbol{\Omega}_j$ the "source direction.") Each source has its own GW waveforms $h_j^+(t)$ and $h_j^\times(t)$ for the two GW polarizations, as measured at Earth. We adopt simple but realistic statistical models for the locations and waveforms of these sources, and use them to compute the mean and variance of the Hellings-Downs correlation. For convenience, we describe the correlation in terms of pulse redshifts rather than timing residuals. At fixed GW frequency, these are proportional, see [18] for further discussion.

For this fixed set of GW sources, the pulse redshift of pulsar $n = 1, 2$ at time $t$ is a sum over the individual sources

$$Z_n(t) = \sum_j \Delta h_j^+(t, L_n\mathbf{p}_n) F_n^+(\mathbf{\Omega}_j) + \Delta h_j^\times(t, L_n\mathbf{p}_n) F_n^\times(\mathbf{\Omega}_j), \quad (3.1)$$

where $L_n\mathbf{p}_n$ is the vector from Earth to pulsar $n$, and $\mathbf{p}_n$ has unit length. (This is derived starting from the fundamentals of general relativity in Appendixes A and B of [18].) The antenna pattern functions $F^{+,\times}$ are defined in Eq. (2.1), or more explicitly as functions of the pulsar positions in Eq. (A16), and the strain differences are

$$\Delta h_j^A(t, L\mathbf{p}) = h_j^A(t) - \chi h_j^A\big(t - L(1 + \mathbf{p} \cdot \mathbf{\Omega}_j)\big), \quad (3.2)$$

where the first term is called the "Earth term" and the second is called the "pulsar term." Here, $A = +, \times$ denotes polarization, and we have assumed that the sources are distant enough that their GWs are described by plane waves in the Earth/pulsar neighborhood. Note that (consistent with the notation) the right-hand side (rhs) of Eq. (3.2) is a function of time $t$ and of the pulsar position $L\mathbf{p}$, since $L = (L\mathbf{p} \cdot L\mathbf{p})^{1/2}$.

The constant $\chi$ in Eq. (3.2) allows us to control the pulsar term. Setting $\chi = 1$ is physically correct: the redshift then properly incorporates the pulsar term. However, by setting $\chi = 0$, we can "turn off" the pulsar term, and thus isolate and identify its effects. We will see that for the models of most interest, the mean value of the Hellings-Downs correlation is independent of $\chi$, but the variance depends on $\chi$ [32].

Note that the ensembles of GW sources which we employ in Secs. III A and III B have frequencies and amplitudes that do not change with time. For such ensembles, or for any time-stationary ensemble such as the Gaussian ensemble of Appendix C, the pulsar terms cannot be neglected when computing the variance, since they make contributions comparable to the Earth terms. Hence, for such ensembles, one must set $\chi = 1$ to make physically meaningful predictions for the variance. Setting $\chi = 0$ should only be done as a computational check and to gain insight.

To calculate the correlations between the pulse redshifts, it is sometimes convenient to work in a circular polarization basis rather than in a linear polarization basis. For this, define the complex waveforms

$$h_j(t) = h_j^+(t) + i h_j^\times(t) \quad (3.3)$$

and the corresponding complex antenna pattern functions

$$F_n(\mathbf{\Omega}) = F_n^+(\mathbf{\Omega}) - i F_n^\times(\mathbf{\Omega}), \quad (3.4)$$

where $n = 1, 2$ labels the pulsar. In this circular polarization basis the redshift Eq. (3.1) of pulsar $n$ is the real part

$$Z_n(t) = \Re \sum_j \Delta h_j(t, L_n\mathbf{p}_n) F_n(\mathbf{\Omega}_j), \quad (3.5)$$

where the complex strain differences are

$$\Delta h_j(t, L\mathbf{p}) = \Delta h_j^+(t, L\mathbf{p}) + i\Delta h_j^\times(t, L\mathbf{p})$$
$$= h_j(t) - \chi h_j(t - L(1 + \mathbf{p} \cdot \mathbf{\Omega})). \quad (3.6)$$

This can be verified by using Eqs. (3.3), (3.4), and (3.6) to obtain Eq. (3.1).

We now compute the correlation between two pulsars. For each, take the real part of Eq. (3.5) by summing half of the rhs and its complex conjugate, then multiply and time average. The resulting correlation is the sum of $N^2$ terms

$$\overline{Z_1 Z_2} = \frac{1}{4} \sum_{j=1}^{N} \sum_{k=1}^{N} \Big[ \overline{\Delta h_j(t, L_1\mathbf{p}_1) \Delta h_k(t, L_2\mathbf{p}_2)} F_1(\mathbf{\Omega}_j) F_2(\mathbf{\Omega}_k)$$
$$+ \overline{\Delta h_j^*(t, L_1\mathbf{p}_1) \Delta h_k(t, L_2\mathbf{p}_2)} F_1^*(\mathbf{\Omega}_j) F_2(\mathbf{\Omega}_k)$$
$$+ \overline{\Delta h_j(t, L_1\mathbf{p}_1) \Delta h_k^*(t, L_2\mathbf{p}_2)} F_1(\mathbf{\Omega}_j) F_2^*(\mathbf{\Omega}_k)$$
$$+ \overline{\Delta h_j^*(t, L_1\mathbf{p}_1) \Delta h_k^*(t, L_2\mathbf{p}_2)} F_1^*(\mathbf{\Omega}_j) F_2^*(\mathbf{\Omega}_k) \Big], \quad (3.7)$$

where the overbars denote the time averages of the quantities beneath them. Here, we explicitly indicate that there are $N^2$ terms in the double sum, but henceforth we use a single summation symbol over $j, k$ for this. Note that for distant sources, this equation is exact. If we were given the GW source locations and their waveforms, then we could use Eq. (3.7) to compute the redshift correlation for any pair of pulsars.

Since we do not know the locations or waveforms of the relevant PTA sources, to estimate the correlation and its variance, we need to adopt some statistical model. This creates an ensemble of universes, which we can then employ for averaging. We will pick the minimum set of assumptions that will permit us to derive an estimate.

### A. Confusion-noise case

PTAs measure timing residuals with a cadence of about a week. Since the observations span several decades, this means that the fluctuations are spread among $O(10^3)$ distinct frequency bins. Since it is expected that there will be at least $O(10^6)$ sources, on average each frequency bin will have at least $O(1000)$ sources, though it may be more at low frequencies and fewer at high frequencies. In the bins containing many sources, the amplitudes will add like a random walk (i.e., with random phases) producing a central-limit theorem Gaussian process. Here, we consider this "confusion-noise" case. Section III B examines the opposite situation, where the distinct sources do not "interfere" because they radiate at distinct frequencies.

For our model, we give all sources an identical fixed GW angular frequency $\omega$ so that they all lie in the same frequency bin, picking $\omega$ to be an integer multiple of the Rayleigh frequency $2\pi/T$, where $T$ is the total observation time (typically decades) [33]. Assume that the sources are unpolarized, with the same intrinsic GW amplitude (at some fiducial distance) in both polarizations. Thus,

$$h_j^+(t) = \mathcal{A}_j \cos(\omega t + \phi_j), \quad \text{and}$$
$$h_j^\times(t) = \mathcal{A}_j \sin(\omega t + \phi_j). \quad (3.8)$$

(Note that, consistent with Eq. (3.1), these equations define plus and cross polarization amplitudes with respect to a polarization basis that depends upon the direction $\mathbf{\Omega}$ to the source according to Eq. (D6): see the first line of Eq. (3.11) below.) Here, the $\mathcal{A}_j$ are real amplitudes, and the $\phi_j$ are phases associated with each source. We assume that these phases are uniformly distributed on $[0, 2\pi]$, and are statistically independent for different sources. These independent random phases create the "confusion noise."

We locate the sources uniformly in 3-dimensional space, labeling them with increasing distance from Earth, so $j=1$ is the closest source, $j=2$ is the next closest source, and so on. Each source has the same intrinsic GW amplitude, so the amplitude at Earth is proportional to the inverse distance. Thus, a uniform density of sources corresponds to

$$\mathcal{A}_1 = \mathcal{A}, \quad \mathcal{A}_2 = 2^{-1/3}\mathcal{A},$$
$$\mathcal{A}_3 = 3^{-1/3}\mathcal{A}, \ldots, \quad \mathcal{A}_N = N^{-1/3}\mathcal{A}, \quad (3.9)$$

where the dimensionless real positive quantity $\mathcal{A} = \mathcal{A}_1$ denotes the GW amplitude of the closest source to Earth. (Rather than fixed amplitudes, we could also have drawn the amplitudes from some probability distribution.) To avoid a GW Olbers' paradox [34], the number of sources $N$ is finite.

Finally, to finish specifying our statistical ensemble of model universes, we need to specify the directions $\mathbf{\Omega}_j$ to the sources. We assume that these are drawn from a uniform distribution on the sphere, independent of the phases $\phi_j$ and independent of the directions to the other sources. These two assumptions (independent random phases, independent random directions) are sufficient.

We begin by computing the time average $s = \overline{h_{ab}h^{ab}}$ of the squared GW amplitude at Earth for any representative of the ensemble, and the ensemble average and variance of $s$. This is partly for illustration: the variance in $s$ arises from the same mechanism which is responsible for the variance in the Hellings and Downs correlation, but is easier to calculate. In additional, $s$ is directly related to physically observable quantities, since the energy density in GWs is given by $c^2\omega^2 s/32\pi G$ [[17], Eq. (2.13)], where we explicitly include the gravitational constant $G$ and speed of light $c$.

The value of $s$ in any representative universe depends upon the specific values of $\phi_j$ and $\mathbf{\Omega}_j$ in that universe. We find $s$ starting from the complex GW strain for each GW source, obtained from Eqs. (3.3) and (3.8) as

$$h_j(t) = h_j^+(t) + ih_j^\times(t) = \mathcal{A}_j e^{i(\omega t + \phi_j)}. \quad (3.10)$$

The total GW strain at Earth is obtained by summing this over all sources

$$h_{ab}(t) = \sum_j h_j^+(t) e_{ab}^+(\mathbf{\Omega}_j) + h_j^\times(t) e_{ab}^\times(\mathbf{\Omega}_j)$$
$$= \Re \sum_j h_j(t) e_{ab}(\mathbf{\Omega}_j) \quad (3.11)$$
$$= \frac{1}{2}\sum_j \mathcal{A}_j [e^{i(\omega t+\phi_j)} e_{ab}(\mathbf{\Omega}_j) + e^{-i(\omega t+\phi_j)} e_{ab}^*(\mathbf{\Omega}_j)],$$

where $e_{ab} = e_{ab}^+ - ie_{ab}^\times$ and its complex conjugate are a circular-polarization basis. For any representative of the ensemble, the time average of the squared GW amplitude at Earth is

$$s = \overline{h_{ab}h^{ab}} = \frac{1}{4}\sum_{j,k}\mathcal{A}_j\mathcal{A}_k\left[e^{i(\phi_j-\phi_k)}e^{ab}(\mathbf{\Omega}_j)e_{ab}^*(\mathbf{\Omega}_k) + e^{-i(\phi_j-\phi_k)}e^{ab*}(\mathbf{\Omega}_j)e_{ab}(\mathbf{\Omega}_k)\right]$$
$$= 2\sum_j \mathcal{A}_j^2 + \frac{1}{4}\sum_{j\neq k}\mathcal{A}_j\mathcal{A}_k\left[e^{i(\phi_j-\phi_k)}e^{ab}(\mathbf{\Omega}_j)e_{ab}^*(\mathbf{\Omega}_k) + e^{-i(\phi_j-\phi_k)}e^{ab*}(\mathbf{\Omega}_j)e_{ab}(\mathbf{\Omega}_k)\right] \quad (3.12)$$

To obtain the second equality of Eq. (3.12), square the final line of Eq. (3.11), contract polarization tensor indices and average over time: two terms vanish because the frequency is an integer multiple of the inverse observation time. For the final equality, break the sum into diagonal terms ($j=k$) and off-diagonal terms ($j \neq k$) and use $e^{ab}(\mathbf{\Omega})e_{ab}^*(\mathbf{\Omega}) = 4$.

While $s$ is time-independent, its value in any particular representative universe depends upon the values of the phases $\phi_j$ and source directions $\mathbf{\Omega}_j$ in that particular realization. What is the expected value of $s$ and of its square, averaged over the entire ensemble? To evaluate such ensemble averages, we will need to compute averages over the random phases $\phi_j$. Those averages are defined as follows. If $Q(\phi_1, \phi_2, \ldots, \phi_N)$ is any function of those phases, its expected value is

$$\langle Q \rangle_\phi = \int_0^{2\pi}\frac{d\phi_1}{2\pi}\cdots\int_0^{2\pi}\frac{d\phi_N}{2\pi}Q(\phi_1,\ldots,\phi_N), \quad (3.13)$$

where the subscript on the angle brackets indicates an average over random phases. For example, if we pick $Q = e^{i(\phi_j-\phi_k)}$, then Eq. (3.13) gives an ensemble average

$$\langle e^{i(\phi_j-\phi_k)} \rangle_\phi = \delta_{jk}, \quad (3.14)$$

where $\delta_{jk}$ is the Kronecker delta.

The expected value of $s$ is the ensemble average of Eq. (3.12). Evaluating this with Eq. (3.14) eliminates the off-diagonal terms $j \neq k$, leaving

$$\langle s \rangle_\phi = 2\sum_j \mathcal{A}_j^2 = 2\mathcal{H}_2. \quad (3.15)$$

Since $\langle s \rangle_\phi$ is independent of the sky positions $\mathbf{\Omega}_j$ of the GW sources, it is the final ensemble expectation value: $\langle s \rangle = \langle s \rangle_\phi$. Here, the mean squared strain at Earth is

$$\mathcal{H}_2 = \sum_{j=1}^N \mathcal{A}_j^2 = \mathcal{A}^2 \sum_{j=1}^N j^{-2/3} = \mathcal{A}^2 N_s. \quad (3.16)$$

For any fixed $+, \times$ linear polarization basis at Earth, $h_{ab}h^{ab} = 2(h^+)^2 + 2(h^\times)^2$, so Eq. (3.15) implies that the average squared strain in either linear polarization is $\mathcal{H}_2/2$. The energy density $c^2\omega^2 \mathcal{H}_2/16\pi G$ in GW follows directly from Eq. (3.15).

In Eq. (3.16) we define the number of shells $N_s$ by

$$N_s = \sum_{n=1}^N n^{-2/3} \approx \int_0^N n^{-2/3} dn \approx 3N^{1/3}, \quad (3.17)$$

where the approximation is valid for larger numbers of sources. $N_s$ may be thought of as the number of radial "shells" containing the sources, if the radial thickness of each shell is the same, and if the radius of the first shell is just sufficient to enclose the nearest source to Earth.

Thus, $N_s$ is the ratio of the distance to the farthest typical source and the distance to the nearest typical source. If all PTA sources were in a single frequency bin, then $N_s \approx 4000 \text{ Mpc}/50 \text{ Mpc} \approx 80$, where the numerator is the Hubble radius and the denominator is the distance to the closest supermassive black-hole binary in the PTA frequency band. (An exact formula for $N_s$ as a function of $N$ may be given in terms of the generalized harmonic numbers, or the Hurwitz zeta function.)

For later use, it is also helpful to define another measure of strain

$$\mathcal{H}_4 = \sum_{j=1}^N \mathcal{A}_j^4 = \mathcal{A}^4 \sum_{j=1}^N j^{-4/3}, \quad (3.18)$$

noting that if there is a single source, then $\mathcal{H}_4 = \mathcal{H}_2^2 = \mathcal{A}^4$. On the other hand, if the number of sources is large, then

$$\mathcal{H}_4 \approx \mathcal{A}^4 \zeta(4/3), \quad (3.19)$$

where $\zeta(4/3) \approx 3.601$ is the Riemann zeta function. So as the number of sources grows, $\mathcal{H}_2$ increases without bound, but $\mathcal{H}_4$ converges to a maximum value.

Having found the first moment of $s = \overline{h_{ab}h^{ab}}$ in Eq. (3.15), we now compute the second moment. For this, we square $s$ as given in Eq. (3.12) and then take its ensemble average over the source phases $\phi_j$. The square of Eq. (3.12) consists of three terms: the square of the first term, twice the cross-term, and the square of the second term. The cross-term has $j \neq k$, so its ensemble average vanishes because of Eq. (3.14). The remaining two terms are

$$\langle s^2 \rangle_\phi = \left(2\sum_j \mathcal{A}_j^2\right)^2 + \frac{1}{16}\sum_{j\neq k}\sum_{\ell\neq m} \mathcal{A}_j\mathcal{A}_k\mathcal{A}_\ell\mathcal{A}_m \Big[ e^{ab}(\mathbf{\Omega}_j)e^*_{ab}(\mathbf{\Omega}_k)e^{cd}(\mathbf{\Omega}_\ell)e^*_{cd}(\mathbf{\Omega}_m)\langle e^{i(\phi_j-\phi_k+\phi_\ell-\phi_m)}\rangle_\phi$$
$$+ e^*_{ab}(\mathbf{\Omega}_j)e^{ab}(\mathbf{\Omega}_k)e^*_{cd}(\mathbf{\Omega}_\ell)e^{cd}(\mathbf{\Omega}_m)\langle e^{-i(\phi_j-\phi_k+\phi_\ell-\phi_m)}\rangle_\phi + e^{ab}(\mathbf{\Omega}_j)e^*_{ab}(\mathbf{\Omega}_k)e^*_{cd}(\mathbf{\Omega}_\ell)e^{cd}(\mathbf{\Omega}_m)\langle e^{i(\phi_j-\phi_k-\phi_\ell+\phi_m)}\rangle_\phi$$
$$+ e^*_{ab}(\mathbf{\Omega}_j)e^{ab}(\mathbf{\Omega}_k)e^{cd}(\mathbf{\Omega}_\ell)e^*_{cd}(\mathbf{\Omega}_m)\langle e^{-i(\phi_j-\phi_k-\phi_\ell+\phi_m)}\rangle_\phi \Big]. \quad (3.20)$$

The four ensemble averages which appear in Eq. (3.20) are evaluated using Eq. (3.13). Since $j \neq k$ and $\ell \neq m$, the first two give $\langle e^{i(\phi_j-\phi_k+\phi_l-\phi_m)}\rangle_\phi = \delta_{jm}\delta_{k\ell}$, and the second two give $\delta_{j\ell}\delta_{km}$. Thus, Eq. (3.20) simplifies to a double sum. The four products of polarization tensors are equal (just relabel the indices $a, b, c, d$) so one obtains

$$\langle s^2 \rangle_\phi = (2\mathcal{H}_2)^2$$
$$+ \frac{1}{4}\sum_{j\neq k} \mathcal{A}_j^2\mathcal{A}_k^2 e^{ab}(\mathbf{\Omega}_j)e^*_{cd}(\mathbf{\Omega}_j)e^{cd}(\mathbf{\Omega}_k)e^*_{ab}(\mathbf{\Omega}_k). \quad (3.21)$$

To complete our calculation of the second moment of $s$, we now average $\langle s^2 \rangle_\phi$ over the directions $\mathbf{\Omega}_j$ to the GW sources. Because the sum in Eq. (3.21) only contains terms with $j \neq k$, these spherical averages may be easily computed using Eq. (C33) from the Appendix. Each average yields a term $\eta_{abcd}$, as given by Eq. (C34) with $\alpha = -4/15$ and $\beta = 2/5$. Thus, we obtain the second moment

$$\langle s^2 \rangle = \langle s^2 \rangle_{\phi,\Omega}$$
$$= (2\mathcal{H}_2)^2 + \frac{1}{4}\sum_{j\neq k}\mathcal{A}_j^2\mathcal{A}_k^2 \eta^{ab}{}_{cd}\eta^{cd}{}_{ab}$$
$$= 4\mathcal{H}_2^2 + \frac{4}{5}\sum_{j\neq k}\mathcal{A}_j^2\mathcal{A}_k^2, \quad (3.22)$$

where we have used $\eta^{ab}{}_{cd}\eta^{cd}{}_{ab} = \eta_{abcd}\eta^{abcd} = 16/5$ as found after Eq. (C38).

The variance of the average squared strain $s = \overline{h_{ab}h^{ab}}$ follows immediately from Eqs. (3.15) and (3.22), and is

$$\sigma_s^2 = \langle s^2 \rangle - \langle s \rangle^2 = \frac{4}{5}\sum_{j \neq k} \mathcal{A}_j^2 \mathcal{A}_k^2 = \frac{4}{5}(\mathcal{H}_2^2 - \mathcal{H}_4), \quad (3.23)$$

where we used Eq. (3.47) to evaluate the double sum. Thus, the fractional (cosmic) variance in the squared strain is

$$\frac{\sigma_s^2}{\langle s \rangle^2} = \frac{1}{5}\left(1 - \frac{\mathcal{H}_4}{\mathcal{H}_2^2}\right). \quad (3.24)$$

The fractional fluctuation $\sigma_s/\langle s \rangle$ vanishes for a single source, is $\approx 21\%$ for $N = 2$ sources, $\approx 32\%$ for $N = 4$ sources, $\approx 40\%$ for $N = 14$ sources and approaches $\approx 45\%$ for large numbers of sources. The corresponding result for the Gaussian ensemble is computed in Appendix C 4 and given by Eq. (C40). The confusion-noise model of this section corresponds to $\hbar^4/h^4 = 1/2$, so in the large $N$ limit, the model presented here gives the same variance as the Gaussian ensemble.

We now compute the variance in the Hellings-Downs correlation for this confusion noise model. As a first step, we need to compute the correlation $\overline{Z_1 Z_2}$ given by Eq. (3.7) and its square for any one of our model universes. Inserting Eq. (3.10) into Eq. (3.6), the complex strain differences are

$$\Delta h_j(t, L\mathbf{p}) = \mathcal{A}_j e^{i(\omega t + \phi_j)}\left[1 - \chi e^{-i\Delta_j(L\mathbf{p})}\right], \quad (3.25)$$

where the Earth-pulsar phase offset for a given GW source and pulsar is

$$\Delta_j(L\mathbf{p}) = \omega L(1 + \mathbf{p} \cdot \mathbf{\Omega}_j). \quad (3.26)$$

(As discussed earlier, in Eq. (3.25) the physically correct value is $\chi = 1$, whereas setting $\chi = 0$ corresponds to dropping the pulsar term.) The redshift of the pulse from pulsar $n$ is given by Eqs. (3.5) and (3.25) as

$$Z_n(t) = \Re \sum_j \mathcal{A}_j e^{i(\omega t + \phi_j)}\left[1 - \chi e^{-i\Delta_j(L_n\mathbf{p}_n)}\right] F_n(\mathbf{\Omega}_j)$$

$$= \frac{1}{2}\sum_j \mathcal{A}_j \left[e^{i(\omega t + \phi_j)} R_n(\mathbf{\Omega}_j) + e^{-i(\omega t + \phi_j)} R_n^*(\mathbf{\Omega}_j)\right], \quad (3.27)$$

where the modified antenna pattern function is defined from Eq. (3.27) by

$$R_n(\mathbf{\Omega}_j) = \left[1 - \chi e^{-i\Delta_j(L_n\mathbf{p}_n)}\right] F_n(\mathbf{\Omega}_j). \quad (3.28)$$

Thus, setting $n = 1$ and $n = 2$ in Eq. (3.27), the redshifts of pulsars 1 and 2 are given by

$$Z_1(t) = \sum_j \left[c_j e^{i(\omega t + \phi_j)} + c_j^* e^{-i(\omega t + \phi_j)}\right], \quad \text{and}$$

$$Z_2(t) = \sum_k \left[d_k e^{i(\omega t + \phi_k)} + d_k^* e^{-i(\omega t + \phi_k)}\right], \quad (3.29)$$

where the coefficients that appear in the sums are

$$c_j = \frac{1}{2}\mathcal{A}_j\left[1 - \chi e^{-i\omega L_1(1 + \mathbf{p}_1 \cdot \mathbf{\Omega}_j)}\right]\left(F_1^+(\mathbf{\Omega}_j) - iF_1^\times(\mathbf{\Omega}_j)\right), \quad \text{and}$$

$$d_k = \frac{1}{2}\mathcal{A}_k\left[1 - \chi e^{-i\omega L_2(1 + \mathbf{p}_2 \cdot \mathbf{\Omega}_k)}\right]\left(F_2^+(\mathbf{\Omega}_k) - iF_2^\times(\mathbf{\Omega}_k)\right). \quad (3.30)$$

Note that the $c_j$ coefficients are for pulsar 1, and the $d_k$ coefficients are for pulsar 2. The correlation is given by the time-averaged product of the redshifts. Taking these from Eq. (3.29), the time-averaged product is

$$\rho = \overline{Z_1(t)Z_2(t)} = \sum_{j,k}\left[c_j d_k^* e^{i(\phi_j - \phi_k)} + c_j^* d_k e^{-i(\phi_j - \phi_k)}\right]$$

$$= \sum_j \left(c_j d_j^* + c_j^* d_j\right) + \sum_{j \neq k}\left[c_j d_k^* e^{i(\phi_j - \phi_k)} + c_j^* d_k e^{-i(\phi_j - \phi_k)}\right], \quad (3.31)$$

where two terms have been eliminated on the first line of Eq. (3.31) by assuming that the angular frequency $\omega$ is an integer multiple of $2\pi$ divided by the observation time. In the final line of Eq. (3.31) we have broken up the double sum over sources into "diagonal" terms for which $j = k$ and "off-diagonal" terms $j \neq k$. Equation (3.31) gives the pulsar-pair correlation for any representative member of our ensemble of model universes, which are defined by specific values of the phases $\phi_j$ and source positions $\mathbf{\Omega}_j$.

To obtain the mean (first moment) of the Hellings-Downs correlation, we calculate the ensemble average of $\rho$ in Eq. (3.31) by making use of Eq. (3.14). Since the Kronecker delta in Eq. (3.14) vanishes for $j \neq k$, the average of Eq. (3.31) over the random phases eliminates the off-diagonal terms, leaving

$$\langle \rho \rangle_\phi = \sum_j \left(c_j d_j^* + c_j^* d_j\right). \quad (3.32)$$

To complete the ensemble average, we average this over random source directions $\mathbf{\Omega}_j$, obtaining

$$\langle \rho \rangle = \langle \rho \rangle_{\phi, \Omega} = \sum_j \left(\langle c_j d_j^* \rangle_\Omega + \langle c_j^* d_j \rangle_\Omega\right). \quad (3.33)$$

Here, the angle bracket with subscript $\Omega$ indicates an average over statistically independent source directions $\mathbf{\Omega}_j$ uniformly distributed on a sphere. This corresponds to integrating $\mathbf{\Omega}_j$ over a unit two-sphere and dividing by $4\pi$.

To evaluate the source-direction average of the product $c_j d_j^*$ that appears in Eq. (3.33), first examine the definitions of $c$ and $d$ in Eq. (3.30). If the interpulsar separation $L_1 \mathbf{p}_1 - L_2 \mathbf{p}_2$ is much larger than the typical GW wavelength $2\pi/\omega$, then the product of the terms in square brackets from Eq. (3.30) can be replaced by unity

$$\left[1 - \chi e^{-i\omega L_1(1+\mathbf{p}_1\cdot\mathbf{\Omega}_j)}\right]\left[1 - \chi e^{i\omega L_2(1+\mathbf{p}_2\cdot\mathbf{\Omega}_j)}\right] \approx 1. \quad (3.34)$$

This is because the product of the terms in square brackets yields four terms. Three are rapidly oscillating (complex exponential) functions of the source direction $\mathbf{\Omega}_j$. When multiplied by the slowly varying antenna pattern functions, these three terms average to zero, as illustrated in Fig. 16 of Ref. [35]. Only the unity term remains, so the average over source directions gives

$$\begin{aligned}\langle c_j d_j^* \rangle_{\Omega} &= \frac{1}{4}\mathcal{A}_j^2 \langle (F_1^+(\mathbf{\Omega}) - iF_1^\times(\mathbf{\Omega}))(F_2^+(\mathbf{\Omega}) + iF_2^\times(\mathbf{\Omega}))\rangle_\Omega \\ &= \frac{1}{4}\mathcal{A}_j^2 \Big[\langle F_1^+(\mathbf{\Omega})F_2^+(\mathbf{\Omega}) + F_1^\times(\mathbf{\Omega})F_2^\times(\mathbf{\Omega})\rangle_\Omega \\ &\quad + i\langle F_1^+(\mathbf{\Omega})F_2^\times(\mathbf{\Omega}) - F_1^\times(\mathbf{\Omega})F_2^+(\mathbf{\Omega})\rangle_\Omega\Big] \\ &= \frac{1}{4}\mathcal{A}_j^2 \mu_u(\gamma), \quad (3.35)\end{aligned}$$

where $\gamma$ is the angle between the lines of sight to pulsars 1 and 2. The second and third line of Eq. (3.35) has two source-direction averages, which are evaluated in Appendixes D and E respectively. The first average is the mean value of the standard Hellings-Downs "unpolarized" correlation. This is given by Eq. (D3), and evaluates to $\mu_u$, given in Eq. (D29). The second average is the mean value of the "polarized" integrand Eq. (E1). That mean value evaluates to zero, as explained after Eq. (E3), which means that the imaginary part of $\langle c_j d_j^*\rangle_\Omega$ vanishes.

The mean or first moment of the Hellings-Downs correlation is found by substituting Eq. (3.35) and its complex conjugate into Eq. (3.33) and summing over the sources:

$$\mu(\gamma) = \langle \rho \rangle = \frac{1}{2}\mathcal{H}_2 \mu_u(\gamma). \quad (3.36)$$

The result is expressed in terms of the squared strain amplitude at Earth, defined by Eq. (3.16).

Next, we compute the second moment of $\rho$. For this, we first square $\rho$ from expression Eq. (3.31) and average it over the random phases. The diagonal term does not depend upon the random phases, so its square appears "as is." The product of the diagonal and off-diagonal terms contains $\langle e^{i(\phi_j - \phi_k)}\rangle_\phi$, whose phase average vanishes for $j \neq k$ as shown by Eq. (3.14). Thus, only the squares of the diagonal and off-diagonal terms remain, giving

$$\langle \rho^2 \rangle_\phi = \left(\sum_j (c_j d_j^* + c_j^* d_j)\right)^2 + \sum_{j \neq k}\sum_{\ell \neq m}\Big[c_j d_k^* c_\ell d_m^* \langle e^{i(\phi_j - \phi_k + \phi_\ell - \phi_m)}\rangle_\phi + c_j d_k^* c_\ell^* d_m \langle e^{i(\phi_j - \phi_k - \phi_\ell + \phi_m)}\rangle_\phi$$
$$+ c_j^* d_k c_\ell^* d_m \langle e^{-i(\phi_j - \phi_k + \phi_\ell - \phi_m)}\rangle_\phi + c_j^* d_k c_\ell d_m^* \langle e^{-i(\phi_j - \phi_k - \phi_\ell + \phi_m)}\rangle_\phi\Big]. \quad (3.37)$$

The four ensemble averages which appear in Eq. (3.37) may be easily evaluated using Eq. (3.13). Since $j \neq k$ and $\ell \neq m$, the first gives $\langle e^{i(\phi_j - \phi_k + \phi_\ell - \phi_m)}\rangle_\phi = \delta_{jm}\delta_{k\ell}$. The remaining three ensemble averages give $\delta_{j\ell}\delta_{km}$, $\delta_{jm}\delta_{k\ell}$, and $\delta_{j\ell}\delta_{km}$ respectively. Thus, Eq. (3.37) simplifies to the double sum

$$\langle \rho^2 \rangle_\phi = \left(\sum_j (c_j d_j^* + c_j^* d_j)\right)^2 + \sum_{j \neq k}\Big[c_j d_k^* c_k d_j^* + c_j d_k^* c_j^* d_k + c_j^* d_k c_k^* d_j + c_j^* d_k c_j d_k^*\Big]. \quad (3.38)$$

To evaluate the ensemble average over source directions, it is helpful to rewrite the first term (a perfect square) as a double sum, also breaking it into "diagonal" and "off-diagonal" terms. Noting also that the second and fourth term of the double sum in Eq. (3.38) are each equal to $|c_j|^2|d_k|^2$, we obtain

$$\begin{aligned}\langle \rho^2\rangle_\phi &= \sum_j (c_j d_j^* + c_j^* d_j)^2 + \sum_{j\neq k}\Big[(c_j d_j^* + c_j^* d_j)(c_k d_k^* + c_k^* d_k) + c_j d_k^* c_k d_j^* + c_j^* d_k c_k^* d_j + 2|c_j|^2|d_k|^2\Big] \\ &= \sum_j (c_j d_j^* + c_j^* d_j)^2 + 2\sum_{j\neq k}\Big[c_j d_j^* c_k d_k^* + c_j^* d_j c_k d_k^* + c_j^* d_j c_k^* d_k + |c_j|^2|d_k|^2\Big], \quad (3.39)\end{aligned}$$

where we have used symmetry in the sum over $j$, $k$ to combine terms in the final line. To obtain the second moment, we now average this expression over source directions. The off-diagonal terms factor into products of two first moments, because $j \neq k$. In contrast, the diagonal terms give second moments of the types evaluated in Appendixes D–F.

The source-direction averages that appear in the double-sum terms of Eq. (3.39) are evaluated by first factoring them (since $j \neq k$) then using the definitions of Eq. (3.30) and proceeding exactly as for Eq. (3.35). This gives

$$\langle |c_j|^2 |d_k|^2 \rangle_\Omega = \langle |c_j|^2 \rangle_\Omega \langle |d_k|^2 \rangle_\Omega = \frac{1}{16} \mathcal{A}_j^2 \mathcal{A}_k^2 (1+\chi^2)^2 \mu_u^2(0),$$

$$\langle c_j d_j^* c_k d_k^* \rangle_\Omega = \langle c_j d_j^* \rangle_\Omega \langle c_k d_k^* \rangle_\Omega = \frac{1}{16} \mathcal{A}_j^2 \mathcal{A}_k^2 \mu_u^2(\gamma),$$

$$\langle c_j^* d_j c_k d_k^* \rangle_\Omega = \langle c_j^* d_j \rangle_\Omega \langle c_k d_k^* \rangle_\Omega = \frac{1}{16} \mathcal{A}_j^2 \mathcal{A}_k^2 \mu_u^2(\gamma), \text{ and}$$

$$\langle c_j^* d_j c_k^* d_k \rangle_\Omega = \langle c_j^* d_j \rangle_\Omega \langle c_k^* d_k \rangle_\Omega = \frac{1}{16} \mathcal{A}_j^2 \mathcal{A}_k^2 \mu_u^2(\gamma). \quad (3.40)$$

In the first line above, the factors of $1 + \chi^2$ arise because in the expression for $|c_j|^2$, the products of the terms in square brackets from Eq. (3.30) give

$$\left| 1 - \chi e^{-i\omega L_1(1+p_1 \cdot \Omega_j)} \right|^2 = 1 + \chi^2 - 2\chi \cos(\omega L_1(1 + p_1 \cdot \Omega_j))$$
$$\approx 1 + \chi^2, \quad (3.41)$$

and a similar factor arises within $|d_k|^2$. The reasoning is similar to that given after Eqs. (3.34) and (3.35), but here two of the four terms contribute, rather than just one.

The source-direction average of the single-sum term in Eq. (3.39) may be evaluated by first expanding it as

$$\left\langle (c_j d_j^* + c_j^* d_j)^2 \right\rangle_\Omega = \left\langle (c_j d_j^*)^2 + (c_j^* d_j)^2 + 2|c_j|^2 |d_j|^2 \right\rangle_\Omega. \quad (3.42)$$

From the definitions of Eq. (3.30), the first two terms on the rhs of Eq. (3.42) give

$$\left\langle (c_j d_j^*)^2 + (c_j^* d_j)^2 \right\rangle_\Omega = \frac{1}{16} \mathcal{A}_j^4 \left\langle 2 \left( F_1^+(\Omega) F_2^+(\Omega) + F_1^\times(\Omega) F_2^\times(\Omega) \right)^2 - 2 \left( F_1^+(\Omega) F_2^\times(\Omega) - F_1^\times(\Omega) F_2^+(\Omega) \right)^2 \right\rangle_\Omega$$
$$= \frac{1}{16} \mathcal{A}_j^4 \left( 2 \left( \mu_u^2(\gamma) + \sigma_u^2(\gamma) \right) - 2\sigma_p^2(\gamma) \right)$$
$$= \frac{1}{8} \mathcal{A}_j^4 \left( 2\mu_u^2(\gamma) + 2\sigma_u^2(\gamma) - \sigma_c^2(\gamma) \right). \quad (3.43)$$

The first term on the rhs of the first line of Eq. (3.43) is the second moment of the standard unpolarized Hellings-Downs correlation Eq. (D5). This is given in Eq. (D36) of Appendix D, and we write it as $\langle \rho^2 \rangle = \mu_u^2 + \sigma_u^2$. The second term on the rhs of the first line of Eq. (3.43) is the second moment of the "polarized" Hellings-Downs correlation Eq. (E1), evaluated in Appendix E. This is given by the variance $\sigma_p^2$ of Eq. (E8), since the first moment vanishes. For the final line of Eq. (3.43) we have used the identity $\sigma_c^2 = \sigma_u^2 + \sigma_p^2 + \mu_u^2$ from Eq. (F4) to eliminate $\sigma_p^2$.

The final term of Eq. (3.42) is evaluated using the definitions of Eqs. (3.30) and (3.41) to obtain

$$\left\langle 2|c_j|^2 |d_j|^2 \right\rangle_\Omega = \frac{2}{16} \mathcal{A}_j^4 (1+\chi^2)^2 \left\langle \left( F_1^+(\Omega) F_1^+(\Omega) + F_1^\times(\Omega) F_1^\times(\Omega) \right) \left( F_2^+(\Omega) F_2^+(\Omega) + F_2^\times(\Omega) F_2^\times(\Omega) \right) \right\rangle_\Omega$$
$$= \frac{1}{8} \mathcal{A}_j^4 (1+\chi^2)^2 \sigma_c^2(\gamma). \quad (3.44)$$

This is evaluated by noting that the product of antenna pattern functions on the first line is the same as Eq. (F1), whose source-direction average is evaluated in Appendix F, giving Eq. (F2).

Combining the results of Eqs. (3.42)–(3.44), we obtain the source-direction average for the single-sum term of Eq. (3.39) as

$$\left\langle (c_j d_j^* + c_j^* d_j)^2 \right\rangle_\Omega = \frac{1}{8} \mathcal{A}_j^4 \left( \mu_u^2(\gamma) + \sigma_u^2(\gamma) - \sigma_p^2(\gamma) + (1+\chi^2)^2 \sigma_c^2(\gamma) \right)$$
$$= \frac{1}{8} \mathcal{A}_j^4 \left( 2\mu_u^2(\gamma) + 2\sigma_u^2(\gamma) + \left( (1+\chi^2)^2 - 1 \right) \sigma_c^2(\gamma) \right). \quad (3.45)$$

This completes the evaluation of the source-direction averages for the different terms appearing in the Eq. (3.39) expression of $\langle \rho^2 \rangle_\phi$.

We now compute the total variance for this simple confusion-noise model. The ensemble-averaged second moment $\langle \rho^2 \rangle = \langle \rho^2 \rangle_{\phi,\Omega}$ is obtained by starting with $\langle \rho^2 \rangle_\phi$ as given by Eq. (3.39), using Eqs. (3.40) and (3.45) to average over source directions, and summing over the sources. This gives

$$\langle \rho^2 \rangle = \frac{1}{8}\mathcal{H}_4\left(2\mu_u^2(\gamma) + 2\sigma_u^2(\gamma) + \left((1+\chi^2)^2 - 1\right)\sigma_c^2(\gamma)\right) + \frac{1}{8}(\mathcal{H}_2^2 - \mathcal{H}_4)\left(3\mu_u^2(\gamma) + \left(1+\chi^2\right)^2 \mu_u^2(0)\right), \quad (3.46)$$

where we have used Eqs. (3.16) and (3.18) to carry out the sums over sources, noting that

$$\sum_{j \neq k} \mathcal{A}_j^2 \mathcal{A}_k^2 = \sum_{j,k} \mathcal{A}_j^2 \mathcal{A}_k^2 - \sum_j \mathcal{A}_j^4 = \left(\sum_j \mathcal{A}_j^2\right)^2 - \sum_j \mathcal{A}_j^4 = \mathcal{H}_2^2 - \mathcal{H}_4. \quad (3.47)$$

To obtain the variance $\sigma^2 = \langle \rho^2 \rangle - \langle \rho \rangle^2$, we subtract the square of the mean given in Eq. (3.36) from Eq. (3.46). This eliminates the first $\mathcal{H}_4$ term and 2/3 of the first $\mathcal{H}_2^2 - \mathcal{H}_4$ term from Eq. (3.46), giving

$$\sigma^2 = \frac{1}{8}\mathcal{H}_4\left(2\sigma_u^2(\gamma) + \left((1+\chi^2)^2 - 1\right)\sigma_c^2(\gamma)\right) + \frac{1}{8}(\mathcal{H}_2^2 - \mathcal{H}_4)\left(\mu_u^2(\gamma) + \left(1+\chi^2\right)^2 \mu_u^2(0)\right). \quad (3.48)$$

This is one of the main results of the paper: the expected variance in the Hellings-Downs correlation for a confusion-noise model containing $N$ sources (set $\chi = 1$ to include the pulsar terms, or $\chi = 0$ to exclude them). Note that this variance includes both pulsar variance and cosmic variance. In Sec. IV, we calculate the cosmic variance, which can be subtracted from Eq. (3.48) to obtain the pulsar variance.

As a consistency check, consider the case where there is only a single source, so that $\mathcal{H}_4 = \mathcal{H}_2^2 = \mathcal{A}^4$. In this case Eq. (3.48) reduces to

$$\sigma^2 = \frac{\mathcal{A}^4}{8}\left(2\sigma_u^2(\gamma) + \left((1+\chi^2)^2 - 1\right)\sigma_c^2(\gamma)\right). \quad (3.49)$$

If the pulsar term is neglected (set $\chi = 0$) we obtain the correct $\sigma^2 = \mathcal{A}^4 \sigma_u^2(\gamma)/4$ for a single unpolarized point source. (This agrees with the general result Eq. (A30): set $c_1 = c_2 = \mathcal{A}^2/2$ and $c_3 = c_4 = 0$.) This also demonstrates that Eq. (3.48) includes both the pulsar variance and cosmic variance, because in the case $N = 1$ there is no cosmic variance: the expected correlation for pulsars at angle $\gamma$ is independent of the source location. If the pulsar term is included by setting $\chi = 1$, we obtain the correct result $\sigma^2 = \mathcal{A}^4(2\sigma_u^2(\gamma) + 3\sigma_c^2(\gamma))/8$. This agrees with Eq. (B10) after setting $A_c = A_s = \mathcal{A}$.

Another important limit is that of many sources. For current PTAs, with at least $N = O(10^3)$ sources per frequency bin, one can see immediately from Eqs. (3.16) and (3.18) that $\mathcal{H}_2^2 \gg \mathcal{H}_4$, because $\mathcal{H}_2$ grows proportional to $N_s \propto N^{1/3}$, whereas $\mathcal{H}_4 \to \mathcal{A}^4 \zeta(4/3) \approx 3.6\mathcal{A}^4$ converges to a constant for large $N$. Hence, the second term in Eq. (3.48) dominates, and (setting $\chi = 1$ to properly include the pulsar term)

$$\sigma^2 \approx \frac{1}{8}\mathcal{H}_2^2\left(\mu_u^2(\gamma) + 4\mu_u^2(0)\right). \quad (3.50)$$

This should be compared with the mean taken from Eq. (3.36):

$$\mu(\gamma) = \frac{1}{2}\mathcal{H}_2\mu_u(\gamma) \quad (3.51)$$

Notice that for large numbers of sources the standard deviation $\sigma$ does not become small in comparison with the mean $\mu$. The fractional variations in the Hellings-Downs correlation are determined by the ratio of the square root of the variance from Eq. (3.50) to the mean value from Eq. (3.51), which are shown in Fig. 2. When there are many confusion-noise sources,

$$\frac{\sigma}{\mu} \approx \sqrt{\frac{1}{2}\left[1 + 4\frac{\mu_u^2(0)}{\mu_u^2(\gamma)}\right]^{1/2}} \quad (3.52)$$

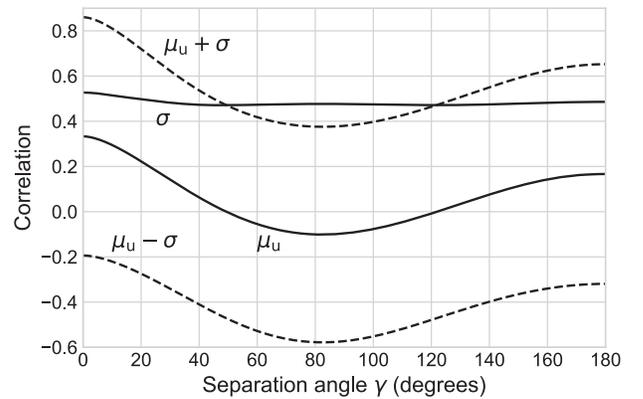

FIG. 2. The Hellings-Downs correlation mean $\mu$ [from Eq. (3.51)] and standard deviation $\sigma$ [from Eq. (3.50)] for an unpolarized confusion-limited stochastic background from many point sources, where all sources have the same frequency. The dashed lines show a $\pm 1\sigma$ range about the mean. In Sec. IV, we show that this variance includes both pulsar and cosmic variance. In the plot, $\mu$ and $\sigma$ are divided by a factor $\mathcal{H}_2/2 = \langle (h^+)^2 \rangle = \langle (h^\times)^2 \rangle$, which is the mean squared GW strain (per polarization) at Earth.

for the sources in one frequency bin. (Note: if the source frequency is not commensurate with the observation time, then the factor $\sqrt{1/2}$ in Eq. (3.52) can increase to be as large as unity; see comments and footnote before Eq. (3.8) and the discussion following Eq. (C30).)

The effective number of frequency bins may be quite small ($\lesssim 4$). This is because the spectrum of characteristic strain from black-hole binaries falls steeply with increasing frequency. One might be concerned that the low-frequency growth in the spectrum could invalidate our results, which assume that the GW frequency is a multiple of the inverse observation time $T$. However, in practice, frequencies below $1/T$ are effectively cut off, because pulsar timing residuals at lower frequencies are typically well-matched by a change in the intrinsic rotation parameters of the pulsar, which are fit to the data to reduce the timing residuals as much as possible [36]. For further details, see the "transmission functions" illustrated in Fig. 2 of Ref. [36].

If there are $M$ independent confusion-noise limited frequency bins labeled by $i = 1, \ldots, M$, then the means will add linearly and the variances will add in quadrature, giving

$$\mu = \mu_u(\gamma) \sum_{i=1}^{M} h_i^2, \quad \text{and}$$

$$\sigma^2 = \frac{1}{2}\left(\mu_u^2(\gamma) + 4\mu_u^2(0)\right) \sum_{i=1}^{M} h_i^4, \quad (3.53)$$

where $h_i^2 = (\mathcal{H}_2/2)_i$ denotes the mean-squared GW strain at Earth from the $i$th frequency bin. (The justification may be found in Appendix C 5: compare Eqs. (C30) and (C32).) If the squared strain is distributed uniformly over frequency, then the fractional fluctuations $\sigma/\mu$ will be proportional to $M^{-1/2}$. Note: for the fractional fluctuation in the correlation of timing residuals (rather than redshifts) a factor of GW frequency $f_i^{-2}$ should be added on the rhs of the definition of $h_i^2$.

In Appendix C 3 we calculate the variance for a Gaussian ensemble of sources as employed in [17,20,21]. The Gaussian ensemble variance has the term proportional to $\mu_u^2(\gamma) + 4\mu_u^2(0)$, but not the other terms found in Eq. (3.48). This is because the Gaussian ensemble corresponds to the limit as the number of sources $N$ goes to infinity with $\mathcal{A}^2 N^{1/3}$ held constant, which keeps the mean squared strain at Earth constant. The terms proportional to $\mathcal{H}_4$ in our calculation arise from the handful of closest sources. In the Gaussian ensemble limit, those become arbitrarily weak and hence do not contribute.

Why does the standard deviation $\sigma$ not decrease in comparison with the mean $\mu$ as the number of shells increases? This can be traced back to the behavior of the sum of the diagonal and off-diagonal terms in Eq. (3.31). In our simple model, the diagonal terms have nonzero mean but zero variance. The off-diagonal terms have zero mean but nonzero variance. Moreover, (if we turn off the pulsar terms) the variance of the second quantity is the squared mean of the first quantity. So, on average, the square of the second term contributes as much as the square of the first term. The situation is identical to that which describes the errors in periodogram estimation of power spectra [[37] see discussion on the final page of Sec. 13.4].

### B. Noninterfering source case

The previous subsection considers a set of sources radiating at the same GW frequency, which generate confusion noise from interference between the sources. Here, we consider the opposite extreme, in which each source is radiating at a different GW frequency, or following a different track in time-frequency space, so that their waveforms are uncorrelated in time.

The calculation is very similar to that of the previous Sec. III A for the confusion-noise case. The fundamental difference is that in the place of the GW waveforms of Eq. (3.8), we use waveforms

$$h_j^+(t) = \mathcal{A}_j \cos(\omega_j t + \phi_j), \quad \text{and}$$
$$h_j^\times(t) = \mathcal{A}_j \sin(\omega_j t + \phi_j), \quad (3.54)$$

where *we assume that the GW angular frequencies $\omega_j$ are different for all of the $j = 1, \ldots, N$ sources.* As before, we also assume that the $\omega_j$ are integer multiples of $2\pi/T$, where $T$ is the total observation time. We use the same set of GW amplitudes $\mathcal{A}_j$ as in Eq. (3.9), corresponding to a uniform density of sources, as well as the same set of random phases $\phi_j$, and the same ensemble of source directions $\mathbf{\Omega}_j$.

In this model, it is easy to see that the variance of the time-averaged squared strain $s = \overline{h_{ab}h^{ab}}$ is zero. This is because, in the equivalent of Eq. (3.12), all of the $j \neq k$ terms are absent. The reason is simple: since the waveforms from different sources oscillate at different frequencies, their time-averaged products vanish. Thus $s = \langle s \rangle = 2\mathcal{H}_2$, $\langle s^2 \rangle = 4\mathcal{H}_2^2$, and $\sigma_s^2 = 0$. Because the time-averaged GW energy density is proportional to $s$, its variance also vanishes.

The effect on the Hellings and Downs correlations is more subtle: while the one-pulsar-pair (total) variance does not vanish, it is reduced. Again, the critical difference between the independent source model of this subsection and the confusion-noise model of the previous subsection lies in the form taken by the time-averaged correlation $\rho$, which in the confusion-noise case is given by Eq. (3.31). With the "independent source" assumptions we have made here, because every source is emitting GW at a different frequency, $\rho$ has the simple "diagonal" form

$$\rho = \overline{Z_1(t)Z_2(t)} = \sum_j \left(c_j d_j^* + c_j^* d_j\right). \quad (3.55)$$

The off-diagonal $j \neq k$ terms in Eq. (3.31) are absent. *In contrast to the confusion-noise case of Eq. (3.31), $\rho$ has no*

dependence on the random phases $\phi_j$. In fact, the random phases play no role for noninterfering sources, and are not needed in our source-model ensemble. (Note: in the definitions of $c_j$ and $d_k$ given in Eq. (3.30), the quantity $\omega$ should be replaced by $\omega_j$ and $\omega_k$ respectively.)

For the independent sources which we consider here, the ensemble average of the correlation $\rho$ over random phases is also given by Eq. (3.55), since $\rho$ has no dependence on the random phases. This is the same expression as the ensemble average of $\rho$ over random phases in the confusion-noise case Eq. (3.32). Thus, after completing the average over source directions, the independent-source case has the same first moment as the confusion-noise case

$$\mu(\gamma) = \langle \rho \rangle = \frac{1}{2}\mathcal{H}_2\mu_u(\gamma), \quad (3.56)$$

which is identical to Eq. (3.36).

Because $\rho$ does not depend upon random phases, its second moment is smaller than in the confusion-noise case. (In the confusion-noise case, both the random phases and the random sky positions contribute to the second moment. Here, only the random sky positions contribute.) In comparison with the confusion-noise expression given in Eq. (3.37), for the case of independent sources, all of the double-sum terms are absent. So, in contrast with Eq. (3.38), one has the simpler expression

$$\langle \rho^2 \rangle_\phi = \left( \sum_j \left( c_j d_j^* + c_j^* d_j \right) \right)^2. \quad (3.57)$$

The notation here correctly indicates an average over the random phases, but note that $\rho$ (and hence $\rho^2$) are *independent* of the random phases. Thus, the averaging over random phases is trivial, and Eq. (3.57) follows directly from Eq. (3.55). Hence, the equivalent of Eq. (3.39) is

$$\langle \rho^2 \rangle_\phi = \sum_j \left( c_j d_j^* + c_j^* d_j \right)^2$$
$$+ \sum_{j \neq k} \left[ c_j d_j^* c_k d_k^* + 2 c_j^* d_j c_k d_k^* + c_j^* d_j c_k^* d_k \right]. \quad (3.58)$$

The average of Eq. (3.58) over source directions is obtained from Eqs. (3.40) and (3.45). After summing over $j$ and $k$ we obtain the final ensemble-averaged second moment

$$\langle \rho^2 \rangle = \frac{1}{8}\mathcal{H}_4 \left( 2\mu_u^2(\gamma) + 2\sigma_u^2(\gamma) + \left( (1+\chi^2)^2 - 1 \right)\sigma_c^2(\gamma) \right)$$
$$+ \frac{1}{8}(\mathcal{H}_2^2 - \mathcal{H}_4)\left( 2\mu_u^2(\gamma) \right). \quad (3.59)$$

This may be compared with the confusion-noise second moment given in Eq. (3.46).

The variance in the Hellings-Downs correlation is obtained by forming $\sigma^2 = \langle \rho^2 \rangle - \langle \rho \rangle^2$ from Eqs. (3.56)

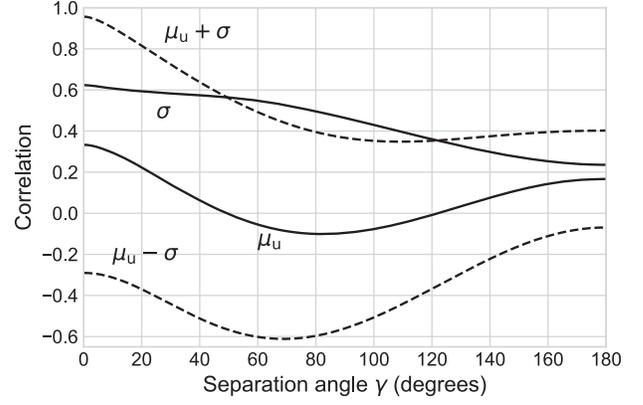

FIG. 3. The Hellings-Downs correlation mean and standard deviation for a single unpolarized point source, including the pulsar terms. These are given by Eqs. (3.56) and (3.60) with $\chi = 1$, $\mathcal{H}_2 = 2$ and $\mathcal{H}_4 = 4$, so that $\mu = \mu_u(\gamma)$ and $\sigma^2 = \sigma_u^2(\gamma) + 3\sigma_c^2(\gamma)/2$. The dashed lines show a $\pm 1\sigma$ range about the mean. If there are many independent point sources, then $\sigma$ has the same shape but smaller relative amplitude. Comparison with Fig. 1 shows that including the pulsar term increases the variance by about a factor of four.

and (3.59). This cancels the $\mathcal{H}_2^2$ terms, giving the variance for noninterfering sources

$$\sigma^2(\gamma) = \frac{1}{8}\mathcal{H}_4\left( 2\sigma_u^2(\gamma) + \left((1+\chi^2)^2 - 1\right)\sigma_c^2(\gamma) \right). \quad (3.60)$$

For any number of sources, this is exactly proportional to the single-point-source variance (without the pulsar term, if $\chi = 0$, or with the pulsar term, if $\chi = 1$) as discussed after Eq. (3.49). This is shown in Fig. 3, including the pulsar term.

The variance Eq. (3.60) lacks the term proportional to $\mathcal{H}_2^2$, which dominates the variance in the confusion-noise case when there are many sources. Here, for independent sources, the mean of the correlation is the sum of the correlations arising from each source term individually, and the variance of the correlation is the sum of the variances arising from each source term individually. This has an important consequence in the limit of large numbers of sources. In that limit, using Eqs. (3.16) and (3.19) in Eq. (3.60), and setting $\chi = 1$ to properly include the pulsar term, the fractional fluctuations of the correlation are

$$\frac{\sigma}{\mu} \approx \frac{\sqrt{\zeta(4/3)}}{N_s}\sqrt{\frac{\sigma_u^2(\gamma) + 3\sigma_c^2(\gamma)/2}{\mu_u^2(\gamma)}}, \quad (3.61)$$

where $\zeta(4/3) \approx 3.601$ is the Riemann zeta function, and the number of shells $N_s$ is the ratio of the distance to the farthest source divided by the distance to the closest source. In contrast to the confusion-noise case, the fractional fluctuations in the correlation vanish for large numbers of sources.

For independent noninterfering sources with constant amplitude and frequency, in contrast with the confusion-noise case given in Eq. (3.52), the variance is dominated by the closest handful of sources in the numerator, whereas the mean in the denominator is proportional to the total number of radial shells of sources. So, for large numbers of sources, the standard deviation $\sigma$ drops in comparison with the mean $\mu$. The shape of the variance is the same as that shown in Fig. 3 for a single unpolarized point source, but the fractional fluctuations are reduced in proportion to the number of source shells $N_s$.

## IV. COSMIC VARIANCE

The variance quantifies the uncertainty in the correlation between pairs of pulsars. Part of this uncertainty arises because, for a given source, different pairs of pulsars, separated by the same angle on the sky, have different correlations. This is called "pulsar variance." The remainder of the uncertainty arises because of interference between GW sources radiating at the same frequency, which creates interference patterns on the sky. These do not average to give the Hellings-Downs curve $\mu_u(\gamma)$ and are called "cosmic variance." Here, we show how to distinguish these, and how to remove the pulsar variance from our calculations, thus obtaining the cosmic variance alone.

How do we eliminate the pulsar variance? The variance calculated in Secs. III A and III B is the sum of both cosmic and pulsar variance. We can *remove* the pulsar variance from these calculations, by replacing the correlation $\rho$ with a pulsar-averaged correlation $\Gamma(\gamma)$. $\Gamma(\gamma)$ is obtained by averaging $\rho$ over all pairs of pulsars separated by angle $\gamma$, *before* determining the ensemble mean and the ensemble second moment. This approach lets us calculate the pure cosmic variance, because it completely removes the pulsar variance.

This also reflects observational reality. In experiments with many pulsars, the correlation is typically "binned" by separation angle and averaged within each bin. In fact, an experiment with access to large numbers of pulsars can produce a correlation curve $\Gamma(\gamma)$, *which is only a function of angle $\gamma$*, by averaging the measured correlation over all pulsars separated by angle $\gamma$, uniformly distributed around the sky. This is what we mean by a "pulsar-averaged correlation curve." (Note that the optimal way to perform such averaging takes into account the expected correlations between different pairs of pulsars. Careful consideration of this point provides an alternative way to derive the cosmic variance [38]. It also shows that PTAs are sensitive to GW energy density in a way which differs from "local" measurements of $s = \overline{h_{ab}h^{ab}}$ or $\overline{\dot{h}_{ab}\dot{h}^{ab}}$.)

For our Universe, where we believe that the GW sources interfere with one other, this pulsar-averaged correlation curve will probably *not* agree exactly in shape with the Hellings-Downs curve $\mu_u(\gamma)$. In our approach, this deviation arises from cosmic variance. How large do we expect it to be? The reply to this statistical question depends upon the ensemble of universes that we select to answer it.

For example, consider the ensemble of independent noninterfering GW point sources used in Sec. III B, and look closely at Eq. (3.55), which is the correlation $\rho$ for any member of the ensemble, making reference to the definitions of $c_j$ and $d_k$ given in Eq. (3.30). If we average $\rho$ over all pulsar positions $\mathbf{p}_1$ and $\mathbf{p}_2$ separated by angle $\gamma$ (we denote such averages by appending a subscript "p" after the angle brackets) only the Earth terms survive [39] and we obtain

$$\begin{aligned}\Gamma(\gamma) &= \langle \rho \rangle_p \\ &= \sum_j \langle c_j d_j^* + c_j^* d_j \rangle_p \\ &= \frac{1}{2}\sum_j \mathcal{A}_j^2 \langle F_1^+(\mathbf{\Omega}_j)F_2^+(\mathbf{\Omega}_j) + F_1^\times(\mathbf{\Omega}_j)F_2^\times(\mathbf{\Omega}_j)\rangle_p \\ &= \frac{1}{2}\mathcal{H}_2\mu_u(\gamma).\end{aligned} \quad (4.1)$$

Thus, after averaging over pulsar positions, we obtain a correlation curve $\Gamma(\gamma)$ which is *exactly* proportional to the Hellings-Downs curve $\mu_u(\gamma)$ for *any member* of the ensemble, regardless of the source locations $\mathbf{\Omega}_j$. (Note: both [25] and Appendix A prove that the pulsar average of the antenna pattern functions in Eq. (4.1) yields $\mu_u$.) Moreover, in this ensemble, the amplitude is exactly the same for every realization. So, in that ensemble of universes, there is no cosmic variance. Stated another way: for any realization in this ensemble, the curve $\Gamma(\gamma)$ obtained by an experimenter with access to large numbers of low-noise pulsars would always agree exactly with $\mu_u(\gamma)$ in shape, and would always have the same overall normalization.

In contrast, consider the confusion-noise-limited ensemble of Sec. III A, where Eq. (3.31) gives the correlation $\rho$ for any member of the ensemble. If we pick a single representative universe from the ensemble, and average $\rho$ over all pulsar positions $\mathbf{p}_1$ and $\mathbf{p}_2$ separated by angle $\gamma$, we will get a function $\Gamma(\gamma)$ that is *not* proportional to the Hellings-Downs curve $\mu_u(\gamma)$ [40]. However, if we examine many representatives from the ensemble, we'll find that *on the average* the curve $\Gamma(\gamma)$ is proportional to $\mu_u(\gamma)$. How much variation should be expected between the $\Gamma(\gamma)$ found in any particular realization of this ensemble, and the ensemble average? Our answer: the cosmic variance.

To calculate the cosmic variance, we must remove the pulsar variance. To do this, begin with the general expression for the correlation between pulsar redshifts given in Eq. (3.7), and average it over pulsar positions using the method from [25], which also eliminates the pulsar terms. In Appendix A, we show how to compute such averages over the three variables $(\theta, \phi, \lambda)$ that define the pulsar positions at fixed separation angle $\gamma$. In Appendix G, we evaluate these averages for the four products of real antenna pattern functions which appear in Eq. (3.7).

The corresponding complex averages are

$$\langle F_1(\mathbf{\Omega}_j)F_2^*(\mathbf{\Omega}_k)\rangle_p = \mu_{++}(\gamma,\beta_{jk}) + \mu_{\times\times}(\gamma,\beta_{jk}) = \mu(\gamma,\beta_{jk}), \quad \text{and}$$
$$\langle F_1(\mathbf{\Omega}_j)F_2(\mathbf{\Omega}_k)\rangle_p = \mu_{++}(\gamma,\beta_{jk}) - \mu_{\times\times}(\gamma,\beta_{jk}) = \mu(\pi-\gamma,\pi-\beta_{jk}). \quad (4.2)$$

The two-point function $\mu(\gamma,\beta)$ which appears here is calculated and illustrated in Appendix G, and given by Eq. (G5). The individual polarization functions $\mu_{++}$ and $\mu_{\times\times}$ are defined by Eq. (G6) and given in Eq. (G9). The function $\mu(\gamma,\beta)$ is a function of the angle $\gamma$ between pulsar directions and of the angle $\beta_{jk}$ between the $j$th and $k$th sources: $\cos\beta_{jk} = \mathbf{\Omega}_j \cdot \mathbf{\Omega}_k$. This two-point function is a generalization of the normal Hellings-Downs curve and reduces to it in the limit $\beta = 0$, where $\mu(\gamma,0) = 2\mu_{++}(\gamma,0) = 2\mu_{\times\times}(\gamma,0) = \mu_u(\gamma)$. Using Eq. (4.2) to compute the pulsar average of Eq. (3.7) gives

$$\Gamma(\gamma) = \langle \overline{Z_1 Z_2}\rangle_p = \frac{1}{4}\sum_{j,k}\left[\left(\overline{h_j(t)h_k^*(t)} + \overline{h_j^*(t)h_k(t)}\right)\mu(\gamma,\beta_{jk}) + \left(\overline{h_j(t)h_k(t)} + \overline{h_j^*(t)h_k^*(t)}\right)\mu(\pi-\gamma,\pi-\beta_{jk})\right]$$
$$= \sum_{j,k}\left(\overline{h_j^+(t)h_k^+(t)}\mu_{++}(\gamma,\beta_{jk}) + \overline{h_j^\times(t)h_k^\times(t)}\mu_{\times\times}(\gamma,\beta_{jk})\right). \quad (4.3)$$

Here, we have given equivalent expressions for complex and real waveforms. As previously mentioned, because the pulsar averaging eliminates the pulsar terms, the time averages in Eq. (4.3) only include the Earth terms.

We now evaluate this pulsar-averaged correlation for the confusion-noise case of Sec. III A. Substitute the complex GW waveform Eq. (3.10) into the first line of Eq. (4.3). The time averaging eliminates the second term, giving

$$\Gamma(\gamma) = \frac{1}{4}\sum_{j,k}\mathcal{A}_j\mathcal{A}_k\left(e^{i(\phi_j-\phi_k)} + e^{-i(\phi_j-\phi_k)}\right)\mu(\gamma,\beta_{jk})$$
$$= \frac{1}{2}\mathcal{H}_2\mu_u(\gamma)$$
$$+ \frac{1}{4}\sum_{j\neq k}\mathcal{A}_j\mathcal{A}_k\left(e^{i(\phi_j-\phi_k)} + e^{-i(\phi_j-\phi_k)}\right)\mu(\gamma,\beta_{jk}). \quad (4.4)$$

On the second line we have partitioned the sum into diagonal and off-diagonal terms; for the diagonal terms with $j = k$, the two-point function reduces to the normal Hellings-Downs curve since $\beta_{jj} = 0$. One can see immediately from Eq. (4.4) that the pulsar-averaged correlation curve $\Gamma(\gamma)$ in a typical representative universe will not have the same shape as the Hellings-Downs curve $\mu_u(\gamma)$. This is because the terms contributed by the double sum, which arise from the interference between different GW sources radiating in the same frequency bin, are cross sections (at various fixed $\beta$ values) of the function plotted in Fig. 12. Those cross-sections have a shape that differs from $\mu_u(\gamma)$. It is these deviations which give rise to the cosmic variance.

We stress that in obtaining Eq. (4.4) from Eq. (3.7), we have not done *anything* to the sources, which have the exact waveforms and sky positions appropriate to that representative of the ensemble. All that we have done is to average over the pulsar locations, as would be done if the average correlation at separation angle $\gamma$ were measured with many pulsar pairs in that representative universe.

To compute the ensemble average mean and variance of $\Gamma$, we first average Eq. (4.4) over the random phases. From Eq. (3.14) we know that $\langle e^{i(\phi_j-\phi_k)}\rangle_\phi$ vanishes for $j \neq k$, so the ensemble average of $\Gamma$ is

$$\langle \Gamma(\gamma)\rangle = \langle \Gamma(\gamma)\rangle_\phi = \frac{1}{2}\mathcal{H}_2\mu_u(\gamma). \quad (4.5)$$

Thus, *on the average*, an observer in this ensemble of model universes would obtain a pulsar-averaged correlation curve $\Gamma(\gamma)$ that follows the shape of the Hellings-Downs curve $\mu_u(\gamma)$ perfectly.

To compute the second moment of $\Gamma$, we first square Eq. (4.4) and evaluate the average over random phases. The square of the first term is independent of random phases and averages to itself. The cross term is zero because $\langle e^{i(\phi_j-\phi_k)}\rangle_\phi$ vanishes for $j \neq k$. The square of the double sum simplifies in the same way as for Eq. (3.37), giving

$$\langle \Gamma^2(\gamma)\rangle_\phi = \frac{1}{4}\mathcal{H}_2^2\mu_u^2(\gamma) + \frac{1}{8}\sum_{j\neq k}\sum_{\ell\neq m}\mathcal{A}_j\mathcal{A}_k\mathcal{A}_\ell\mathcal{A}_m(\delta_{jm}\delta_{k\ell} + \delta_{j\ell}\delta_{km})\mu(\gamma,\beta_{jk})\mu(\gamma,\beta_{\ell m})$$
$$= \frac{1}{4}\mathcal{H}_2^2\mu_u^2(\gamma) + \frac{1}{8}\sum_{j\neq k}\mathcal{A}_j^2\mathcal{A}_k^2\left(\mu(\gamma,\beta_{jk})\mu(\gamma,\beta_{jk}) + \mu(\gamma,\beta_{jk})\mu(\gamma,\beta_{kj})\right)$$
$$= \frac{1}{4}\mathcal{H}_2^2\mu_u^2(\gamma) + \frac{1}{4}\sum_{j\neq k}\mathcal{A}_j^2\mathcal{A}_k^2\mu^2(\gamma,\beta_{jk}), \quad (4.6)$$

where to obtain the final line we have used $\beta_{jk} = \beta_{kj}$, since the angle between two sources is independent of the ordering.

To complete the ensemble average, we now need to average Eq. (4.6) over the random directions $\mathbf{\Omega}_j$ and $\mathbf{\Omega}_k$ to the sources, which enter via the angle $\beta_{jk} = \cos^{-1} \mathbf{\Omega}_j \cdot \mathbf{\Omega}_k$. This gives

$$\langle \Gamma^2(\gamma) \rangle = \frac{1}{4}\mathcal{H}_2^2 \mu_u^2(\gamma) + \frac{1}{4}\sum_{j\neq k}\mathcal{A}_j^2 \mathcal{A}_k^2 \tilde{\mu}^2(\gamma), \quad (4.7)$$

where $\tilde{\mu}^2(\gamma)$ is the average of $\mu^2(\gamma,\beta)$ with respect to $\beta$, with measure $\sin\beta d\beta$. This is computed in Appendix G and given in Eq. (G11). Subtracting the square of the first moment Eq. (4.5) from Eq. (4.7) gives the cosmic variance

$$\begin{aligned}\sigma_{\text{cosmic}}^2(\gamma) &= \langle \Gamma^2(\gamma) \rangle - \langle \Gamma(\gamma) \rangle^2 \\ &= \frac{1}{4}\sum_{j\neq k}\mathcal{A}_j^2 \mathcal{A}_k^2 \tilde{\mu}^2(\gamma) \\ &= \frac{1}{4}\left(\mathcal{H}_2^2 - \mathcal{H}_4\right)\tilde{\mu}^2(\gamma)\end{aligned} \quad (4.8)$$

for our confusion-noise-limited ensemble. This is another one of the main results of this paper: the cosmic variance for the discrete-source confusion-noise model. It may be compared with the total variance given earlier in Eq. (3.48).

For large numbers of sources $\mathcal{H}_2^2 \gg \mathcal{H}_4$, so we can use Eqs. (3.16), (3.50), (4.5), and (4.8) to write the mean, cosmic variance, and total variance as

$$\begin{aligned}\langle \Gamma(\gamma) \rangle &= \mu(\gamma) = \frac{1}{2}\mathcal{H}_2\, \mu_u(\gamma) = \frac{1}{2}N_s\mathcal{A}^2 \mu_u(\gamma),\\ \sigma_{\text{cosmic}}^2(\gamma) &\approx \frac{1}{4}\mathcal{H}_2^2\, \tilde{\mu}^2(\gamma) = \frac{1}{4}N_s^2\mathcal{A}^4 \tilde{\mu}^2(\gamma), \quad \text{and}\\ \sigma^2(\gamma) &\approx \frac{1}{8}\mathcal{H}_2^2\left(\mu_u^2(\gamma) + 4\mu_u^2(0)\right)\\ &= \frac{1}{8}N_s^2\mathcal{A}^4\left(\mu_u^2(\gamma) + 4\mu_u^2(0)\right).\end{aligned} \quad (4.9)$$

Thus, the predicted scale of fluctuations in the pulsar-averaged correlation for the confusion-noise ensemble is

$$\frac{\sigma_{\text{cosmic}}}{\langle \Gamma \rangle} = \frac{\sigma_{\text{cosmic}}}{\mu(\gamma)} = \sqrt{\frac{\tilde{\mu}^2(\gamma)}{\mu_u^2(\gamma)}}. \quad (4.10)$$

This is illustrated in Fig. 4, and can be generalized to multiple frequency bins as shown in Eq. (3.53). In Appendix C 5 we use the Gaussian ensemble to derive the same results in the limit of large numbers of weak sources.

If the cosmic variance were zero, then an experiment could (at least in principle) average the correlation over large numbers of low-noise pulsar pairs, and approach the Hellings-Downs curve with arbitrary precision. If the cosmic variance is nonzero, then this is not possible. In that case, once enough pulsar pairs have been employed, the pulsar variance will be reduced to well below the cosmic variance and a fundamental limit is reached. The shape mismatch between the correlation measured in our Universe and the Hellings-Downs curve can no longer be reduced by incorporating additional low-noise pulsar pairs. This topic is studied in more detail in [38], which determines the number of pulsar pairs required to reach the limits imposed by the cosmic variance.

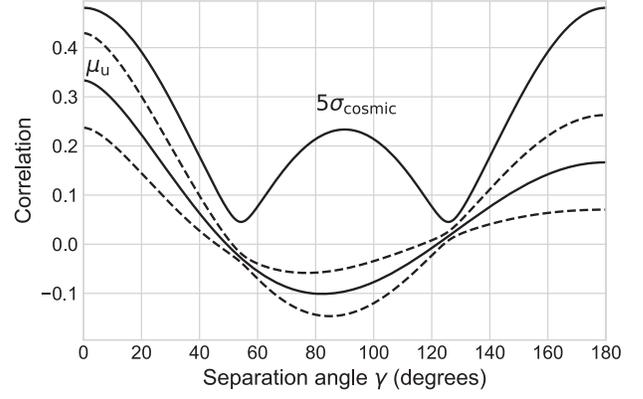

FIG. 4. The cosmic variance for an ensemble of confusion-noise universes in the limit of large numbers of sources (with the mean GW amplitude at Earth $N_s\mathcal{A}^2 = \mathcal{H}_2 = 2$). The mean pulsar-averaged correlation is $\langle \Gamma(\gamma) \rangle = \mu_u(\gamma)$. The dashed curves show $\mu_u(\gamma) \pm \sigma_{\text{cosmic}}$. The upper solid curve shows $5\sigma_{\text{cosmic}}$ from Eqs. (4.10) and (G11) (scaled ×5 so that it does not overlap the other plots).

## V. CONCLUSION

The goal of this paper is to assess how closely the Hellings-Downs correlation curve would be matched in our Universe, in the absence of any experimental or pulsar noise.

We show that the variance in the Hellings-Downs correlation has two parts. Pulsar variance arises from the fluctuations that occur when the correlation is measured with pulsar pairs at different sky locations, but separated by the same angle. In principle, this can be eliminated by using many pulsar pairs and averaging. Cosmic variance arises because the waveforms of the different GW sources interfere with one another. This gives rise to a deviation between the angle-average correlation $\Gamma(\gamma)$ and the Hellings-Downs curve $\mu_u(\gamma)$. We constructed two ensembles of point sources to study these effects.

The first ensemble consists of sources radiating GWs at the same frequency and with constant (in time) amplitude. When there are many such sources, uniformly distributed in space, the sum of the pulsar variance and the cosmic variance is described by Eq. (3.48) with $\chi = 1$. It has a term proportional to the variance expected for a single unpolarized point source, which arises from the closest sources. But for large numbers of sources, this term becomes insignificant in comparison with the sum of variances arising from the interference between the more distant sources, which creates a central-limit-theorem Gaussian process. For this ensemble, the cosmic variance is given by Eq. (4.8).

We also constructed a similar ensemble of point sources, but with each radiating GWs at a different frequency. For these, the variance is given by Eq. (3.60). It is proportional to the variance expected for a single unpolarized point source, because it arises from the closest sources only. There is no interference term, and in this ensemble, the cosmic variance vanishes.

Comparison of these two models leads to an important conclusion. For PTAs, information is carried not just in the mean $\mu(\gamma)$ of the pulsar timing residuals, but also in how the variance $\sigma^2(\gamma)$ behaves with angular separation, and perhaps also with GW frequency. This variance carries information about the nature of the sources.

Recent work [38] provides an alternative way to understand the cosmic variance for a Gaussian ensemble. Suppose that our goal was to determine the pulsar-averaged correlation $\Gamma$ at angle $\gamma$, and that we had a collection of pulsar pairs separated by that angle, with measured correlation values for those pairs. The optimal estimator for $\Gamma(\gamma)$ is not obtained by uniformly averaging the sets of measured values, because there are correlations between different pairs. For a Gaussian ensemble, we can form an optimal estimator for $\Gamma(\gamma)$. In [38], we demonstrate that the variance of that optimal estimator, when there are many pulsars, is exactly $\sigma^2_{\text{cosmic}}$.

The work in this paper could be usefully extended. Our expectation is that the variance in the Hellings-Downs correlation is not described by either of the simple models presented here, but rather by some combination of them [20]. Our formalism can still be employed, but more realistic statistical ensembles of sources are needed. Specifically, we need better models for the ensemble of amplitudes Eq. (3.8) which create the correlations between the GW waveforms of different sources. One could construct a statistical model containing a realistic distribution of supermassive black-hole binary systems with different masses and other properties [31]. This could then be used to define the correlation $\rho$ [41], which in turn can be used to predict the mean of the Hellings-Downs correlation, and the pulsar and cosmic variance. A promising approach for creating such ensembles is to create synthetic catalogs based on cosmological simulations, which can be used to predict and model the mean and variance of the Hellings-Downs correlation [15].

It may be some years before there is a definitive detection of the gravitational-wave background with pulsar timing arrays. Even when that takes place, the correlations between individual pulsar pairs may be dominated by pulsar timing noise and experimental noise. So it is difficult to predict if the variance can be observed in the near future. But we are hopeful that as telescopes improve and pulsar populations increase, this will eventually become possible.

## ACKNOWLEDGMENTS

I gratefully acknowledge the members of the International Pulsar Timing Array Detection Committee (DC), and the speakers who made presentations to the DC, for teaching me something about PTAs and for stimulating my interest. I thank Xavier Siemens, Alberto Sesana, Curt Cutler, and Serena Valtolina for useful conversations and Joe Romano and Neil Cornish for pointing me to earlier work on this topic and for identifying a significant mistake in the first version of the manuscript. I also acknowledge the support of the anonymous referee, who identified a number of misstatements and omissions. Very special thanks to Joe Romano for a close and critical reading of many versions of the manuscript, and extended discussions which resulted in many corrections, clarifications and improvements.

## APPENDIX A: GENERAL FORM OF THE HELLINGS-DOWNS CORRELATION MEAN AND VARIANCE FOR A SINGLE POINT SOURCE

Here, we derive the general form of the mean and variance of the Hellings-Downs correlation for a single distant GW point source, without making the assumption that the source is unpolarized or that the pulsar terms can be neglected. Such single point source results can be combined for a collection of sources following the argument given in Sec. III.

In electromagnetism, the polarization of a light or radio source may be described in terms of four Stokes parameters. The same is possible for gravitational waves. For example, see Eq. (31) of [42] for the GW Stokes parameters of a binary system in a circular orbit emitting GWs.

In the literature an unpolarized source is often assumed to be one for which the time correlation between the cross and plus polarizations vanishes. For example, in [25], the authors remark following their Eq. (10), "it has been assumed that $\langle R_+ R_\times \rangle = 0$, which holds for cosmological stochastic backgrounds and binary systems." This assumption is enough to obtain the Hellings-Downs correlation mean $\mu_u(\gamma)$, but unless the squared amplitudes of the two polarizations have the same amplitude, the variance in the correlation contains both polarized and unpolarized components. For example, we will see that for a binary system in a circular orbit, the "unpolarized" variance is only obtained for inclination angles of $\iota = 0°$ or $\iota = 180°$, for which the cross and plus components have the same amplitude.

We now derive the general formula for the mean and variance of the Hellings-Downs correlation. Our starting point is the effect on pulse arrival times or redshifts induced by a gravitational wave propagating in direction $\mathbf{\Omega}$, where $\mathbf{\Omega}$ is a unit vector on the two-sphere (celestial sphere).

This is derived starting from the fundamentals of general relativity in [[18], Appendixes A and B].

A plane gravitational wave propagating in direction $\mathbf{\Omega}$ is completely described by two arbitrary functions of a single variable, one for each polarization, which we call $h^+$ and $h^\times$. In terms of these functions, the spatial components of the metric perturbation (away from the flat-space Minkowski metric) are

$$h_{ab}(t, \mathbf{x}) = h^+(t - \mathbf{x} \cdot \mathbf{\Omega})e_{ab}^+ + h^\times(t - \mathbf{x} \cdot \mathbf{\Omega})e_{ab}^\times, \quad \text{(A1)}$$

where $t$ is the time coordinate and $\mathbf{x}$ are spatial coordinates, and the transverse traceless symmetric polarization tensors are defined in Eq. (D6).

The effect of the gravitational wave on the pulse arrival times is most simply described by the redshift $Z$ in those arrival times. (One may integrate $Z$ with respect to time to obtain timing delays. For a fixed-frequency GW source, this amounts to dividing by the angular frequency of the GWs.) Consider a pulsar located at position $\mathbf{p}L$ in the sky, where $\mathbf{p}$ is a unit vector and $L$ is the pulsar distance. The redshift of the pulse arriving at time $t$ is

$$Z(t) = \frac{1}{2}\frac{p^a p^b}{1 + \mathbf{\Omega} \cdot \mathbf{p}} \Delta h_{ab}(t), \quad \text{(A2)}$$

where the gravitational wave enters through two terms

$$\Delta h_{ab}(t) = h_{ab}(t, 0) - h_{ab}(t - L, \mathbf{p}L), \quad \text{(A3)}$$

which are called the "Earth term" and the "pulsar term" respectively, since they are the amplitude of the gravitational wave at the time and spatial location where the pulse is received on Earth, and at the time and spatial location where it departed the pulsar. These terms may be simplified and written in terms of the two free functions that define the wave, by

$$\begin{aligned}\Delta h_{ab}(t) &= R^+(t)e_{ab}^+ + R^\times(t)e_{ab}^\times \\ &= \left[h^+(t) - h^+(t - L(1 + \mathbf{p} \cdot \mathbf{\Omega}))\right]e_{ab}^+ + \\ &\quad \left[h^\times(t) - h^\times(t - L(1 + \mathbf{p} \cdot \mathbf{\Omega}))\right]e_{ab}^\times.\end{aligned} \quad \text{(A4)}$$

As shown in [18], for a combination of waves propagating in different directions, these effects may be summed. But for our purposes, it is sufficient to consider the waves coming from individual sources, under the assumption that they are uncorrelated over the (decades-long) pulsar observations.

Here, without loss of generality, we will place our source on the $-z$ axis, so that $\mathbf{\Omega} = \hat{z}$, and assume that the source is much farther from Earth than the pulsars being timed, so that we can treat the GW as a plane wave over the extent of the Earth-pulsar systems. The two polarization tensors are then

$$\begin{aligned}e_{ab}^+ &= \hat{x}_a\hat{x}_b - \hat{y}_a\hat{y}_b, \quad \text{and} \\ e_{ab}^\times &= \hat{x}_a\hat{y}_b + \hat{y}_a\hat{x}_b.\end{aligned} \quad \text{(A5)}$$

Note that GW waveforms are often calculated with respect to one set of polarization axes, which must be rotated (often by an angle denoted $\psi$) to the set of axes appropriate to the application. In this case, because we will be averaging the correlation over all pulsar sky locations, we may assume that our $x$ and $y$ axes are aligned with the ones in which the waveform was obtained. This simplifies matters, with no loss of generality.

To compute the correlations between pulsars 1 and 2, it is convenient to define "antenna pattern" functions

$$\begin{aligned}F_1^+ &= \frac{1}{2}\frac{p_1^a p_1^b}{1 + \mathbf{\Omega} \cdot \mathbf{p}_1}e_{ab}^+, \quad \text{and} \\ F_1^\times &= \frac{1}{2}\frac{p_1^a p_1^b}{1 + \mathbf{\Omega} \cdot \mathbf{p}_1}e_{ab}^\times,\end{aligned} \quad \text{(A6)}$$

where $\mathbf{p}_1$ is a unit vector pointing to the first pulsar, and similar functions for the second pulsar, containing $\mathbf{p}_2$.

The correlation $\rho = \overline{Z_1 Z_2}$ between the pulse redshifts, averaged over time, then takes the form

$$\rho = c_1 F_1^+ F_2^+ + c_2 F_1^\times F_2^\times + c_3 F_1^+ F_2^\times + c_4 F_1^\times F_2^+, \quad \text{(A7)}$$

where the coefficients are the time averages:

$$\begin{aligned}c_1 &= \overline{R_1^+(t)R_2^+(t)}, \\ c_2 &= \overline{R_1^\times(t)R_2^\times(t)}, \\ c_3 &= \overline{R_1^+(t)R_2^\times(t)}, \quad \text{and} \\ c_4 &= \overline{R_1^\times(t)R_2^+(t)}.\end{aligned} \quad \text{(A8)}$$

Here, the subscripts on $R$ are needed because of the pulsar term in Eq. (A4). But note that if the pulsar terms can be neglected, then the subscripts on $R$ are not needed, and $c_3 = c_4$.

Although we are treating the general case, it is helpful to have a specific example in mind. For this, take a pair of orbiting point masses $m_1$ and $m_2$ in a slow circular orbit at distance $r$ from Earth, with orbital angular frequency $\omega$. The angle between the orbital angular momentum and the line of sight (the orbital inclination angle) is $\iota$. To fully define the orbit (for orbital inclination angles other than 0 or 180 degrees) we must specify the orientation of the ellipse that is formed when the orbit is projected onto the plane of the sky. To do this, examine the line formed by the major axis of the ellipse, which lies in the $xy$ plane. For convenience, but without loss of generality, take this line to be parallel to the $x$-axis. Finally, let $\phi_c$ denote the orbital phase at the time of coalescence, which is equivalent to picking an origin of time.

For this system, if we neglect the pulsar terms, we have

$$R^+(t) = h^+(t) = \frac{1}{2}\mathcal{A}(1 + \cos^2\iota)\cos 2(\omega t + \phi_c), \quad \text{and}$$

$$R^\times(t) = h^\times(t) = \mathcal{A}\cos\iota\sin 2(\omega t + \phi_c), \quad \text{(A9)}$$

where the dimensionless GW strain amplitude is

$$\mathcal{A} = \frac{4}{r}\left(\frac{G}{c^2}\right)^{5/3}\left(\frac{\omega}{c}\right)^{2/3}\frac{m_1 m_2}{(m_1+m_2)^{1/3}}, \quad \text{(A10)}$$

and we assume that the orbital evolution is slow enough to treat $\omega$ as a constant but fast enough to decorrelate the pulsar terms. (Notes: in the Appendix B, we will treat the other case, where the pulsar term is significant. Also, to accompany the gravitational constant $G$, we include the speed of light $c$ in the equation for the amplitude $\mathcal{A}$; it is set to unity everywhere else.) So for this system, radiating GWs of angular frequency $2\omega$, the four coefficients appearing in the correlation Eq. (A7) are:

$$c_1 = \frac{1}{8}\mathcal{A}^2(1+\cos^2\iota)^2,$$

$$c_2 = \frac{1}{2}\mathcal{A}^2\cos^2\iota,$$

$$c_3 = 0, \quad \text{and}$$

$$c_4 = 0. \quad \text{(A11)}$$

For the most general source, all four coefficients can be nonzero and different from one another.

We want to compute the average correlation and its variance as the two pulsars sweep around the sky with a fixed angle between them. The first pulsar at distance $L_1$ is located on the sky at angular position $(\theta, \lambda)$ with familiar Cartesian components

$$\boldsymbol{p}_1 = \hat{\boldsymbol{x}}\cos\lambda\sin\theta + \hat{\boldsymbol{y}}\sin\lambda\sin\theta + \hat{\boldsymbol{z}}\cos\theta. \quad \text{(A12)}$$

(Note: it is deliberate that we are using the variable $\lambda$ rather than the more conventional $\phi$ for the polar angle.) We could parametrize the location $\boldsymbol{p}_2$ of the second pulsar using the identical form with different angles $\theta', \phi'$, but that complicates the calculations.

To parametrize the location of the second pulsar, we adopt a nice trick from [25], and use variables $\gamma$ and $\phi$, via

$$\boldsymbol{p}_2 = \hat{\boldsymbol{x}}\Big[\cos\lambda(\sin\theta\cos\gamma - \cos\theta\sin\gamma\cos\phi) + \sin\lambda\sin\gamma\sin\phi\Big] +$$
$$\hat{\boldsymbol{y}}\Big[\sin\lambda(\sin\theta\cos\gamma - \cos\theta\sin\gamma\cos\phi) - \cos\lambda\sin\gamma\sin\phi\Big] +$$
$$\hat{\boldsymbol{z}}\Big[\cos\theta\cos\gamma + \sin\theta\sin\gamma\cos\phi\Big]. \quad \text{(A13)}$$

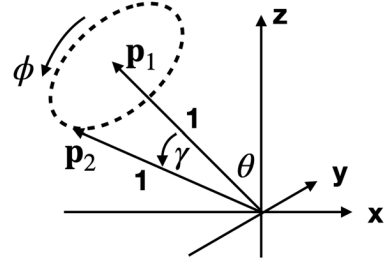

FIG. 5. The direction to the first pulsar $\boldsymbol{p}_1$ is specified with conventional spherical polar angles $(\theta, \lambda)$ in Eq. (A12). The direction to the second pulsar $\boldsymbol{p}_2$ is specified in Eq. (A13) via variables $\gamma$ and $\phi$. Here $\gamma$ is the angle between $\boldsymbol{p}_1$ and $\boldsymbol{p}_2$, and $\phi \in [0, 2\pi)$ is the location along the dashed circle, which lies in the plane perpendicular to $\boldsymbol{p}_1$ and has radius $\sin\gamma$.

The variable $\gamma$ is the angular separation between $\boldsymbol{p}_1$ and $\boldsymbol{p}_2$, as we have been using it throughout this paper. The variable $\phi \in [0, 2\pi)$ is illustrated in Fig. 5, and uniformly parametrizes the position of $\boldsymbol{p}_2$ in the cone around $\boldsymbol{p}_1$. The reader can confirm this picture, by showing that the "velocity vector" $\partial \boldsymbol{p}_2/\partial \phi$ has constant length and is orthogonal to both $\boldsymbol{p}_1$ and $\boldsymbol{p}_2$, and that $\boldsymbol{p}_2$ has unit length and satisfies $\boldsymbol{p}_1 \cdot \boldsymbol{p}_2 = \cos\gamma$.

This parametrization of $\boldsymbol{p}_2$ and $\boldsymbol{p}_1$ makes it easy to average over all pairs of pulsars at a given angular separation $\gamma$. To do this, we have to average over the three variables that define the positions of the pulsars. Generally speaking, we will do that averaging in two steps, which we denote with square brackets and angle brackets. Given a function $Q(\theta, \phi, \lambda)$, the first average (over $\lambda$) is

$$[Q](\theta, \phi) = \frac{1}{2\pi}\int_0^{2\pi} d\lambda\, Q(\theta, \phi, \lambda), \quad \text{(A14)}$$

and the second average (over $\theta$ and $\phi$) is

$$\langle [Q] \rangle = \frac{1}{4\pi}\int_0^\pi d\theta\,\sin\theta\int_0^{2\pi} d\phi\,[Q](\theta, \phi). \quad \text{(A15)}$$

These averages are properly normalized, $[1] = \langle 1 \rangle = 1$, and they treat all pulsars with angular separation $\gamma$ on equal footing (see Fig. 5).

The geometric antenna pattern functions for the first pulsar can now be evaluated by substituting the polarization tensors Eq. (A5) and pulsar-position vector Eq. (A12) into Eq. (A6). One finds

$$F_1^+ = \frac{1}{2}\frac{p_1^a p_1^b}{1 + \boldsymbol{\Omega}\cdot\boldsymbol{p}_1}e_{ab}^+ = U\cos 2\lambda, \quad \text{and}$$

$$F_1^\times = \frac{1}{2}\frac{p_1^a p_1^b}{1 + \boldsymbol{\Omega}\cdot\boldsymbol{p}_1}e_{ab}^\times = U\sin 2\lambda, \quad \text{(A16)}$$

where the function $U = (1 - \cos\theta)/2$. For the second pulsar we use the other pulsar-position vector Eq. (A13), and obtain

$$F_2^+ = V\cos 2\lambda + W\sin 2\lambda, \quad \text{and}$$
$$F_2^\times = V\sin 2\lambda - W\cos 2\lambda, \quad (A17)$$

where to simplify notation we have introduced three functions that are independent of $\lambda$:

$$U = \frac{1-\cos\theta}{2},$$
$$V = \frac{1}{2}\frac{A^2-B^2}{1+q},$$
$$W = \frac{1}{2}\frac{2AB}{1+q}. \quad (A18)$$

Here, we have introduced the three functions

$$A = \sin\theta\cos\gamma - \cos\theta\sin\gamma\cos\phi,$$
$$B = \sin\gamma\sin\phi, \quad \text{and}$$
$$q = \cos\theta\cos\gamma + \sin\theta\sin\gamma\cos\phi. \quad (A19)$$

One can use Eq. (A19) to verify that

$$A^2 - B^2 = (1-q)(1+q) - 2\sin^2\gamma\sin^2\phi. \quad (A20)$$

We are now ready to average over pulsar positions.

We first compute the average correlation $\langle[\rho]\rangle$. Starting with the definition of $\rho$ in Eq. (A7), we insert the antenna pattern functions from Eqs. (A16) and (A17) and integrate over $\lambda$, obtaining

$$[\rho] = \frac{1}{2}(c_1+c_2)UV - \frac{1}{2}(c_3-c_4)UW. \quad (A21)$$

If we now do the remaining average, we obtain

$$\langle[\rho]\rangle = \frac{1}{2}(c_1+c_2)\langle UV\rangle, \quad (A22)$$

because $\langle UW\rangle = 0$. To see this, consider the $\phi$ dependence of $UW$, which is of the form

$$\frac{q_1\sin\phi(1+q_2\cos\phi)}{1+q_3\cos\phi}, \quad (A23)$$

where the $q_i$ are independent of $\phi$. One can see that $UW$ changes sign under reflection about $\pi$, i.e., the transformation $\phi \to 2\pi - \phi$, because under that reflection $\cos\phi$ is even and $\sin\phi$ is odd. Hence, the integral over $\phi \in [0, 2\pi]$ vanishes.

The careful reader will have noticed that the coefficients $c_1, \ldots, c_4$ have an implicit dependence on the pulsar distances and locations, whereas we are treating them as constants. If we ignore the pulsar terms, then the $c_i$ are independent of the pulsar positions, and this is justified. In Appendix B, we will show how to include these pulsar terms if desired: one simply replaces these constants by their mean values obtained by averaging over the pulsar distances, which eliminates their dependence on the pulsar directions.

It is easy to see that $\langle UV\rangle$ is exactly the Hellings-Downs mean correlation $\mu_u(\gamma)$. Use Eq. (A15) to define the average, replace $U$ and $V$ by their definitions in Eq. (A18), and use Eq. (A20) to simplify the integrand. One immediately obtains the form Eq. (D14) computed in Appendix D, where the pulsars are fixed and the average is over wave directions. Thus,

$$\langle[\rho]\rangle = \frac{1}{2}(c_1+c_2)\mu_u(\gamma). \quad (A24)$$

We now compute the second moment of the correlation $\rho$. Starting with Eq. (A7), replace the antenna pattern functions with Eqs. (A16) and (A17), and combine the trig functions to obtain

$$\rho = \frac{1}{2}\Big[(c_1+c_2)UV + (c_4-c_3)UW\Big] +$$
$$\frac{1}{2}\Big[(c_1-c_2)UV - (c_3+c_4)UW\Big]\cos 4\lambda +$$
$$\frac{1}{2}\Big[(c_1-c_2)UW + (c_3+c_4)UV\Big]\sin 4\lambda. \quad (A25)$$

(Note that the average of this over $\lambda$ immediately gives Eq. (A21).) Square Eq. (A25) and average over $\lambda$. One immediately obtains

$$[\rho^2] = \frac{U^2}{4}\Big[(c_1+c_2)V + (c_4-c_3)W\Big]^2 +$$
$$\frac{U^2}{8}\Big[(c_1-c_2)V - (c_3+c_4)W\Big]^2 +$$
$$\frac{U^2}{8}\Big[(c_1-c_2)W + (c_3+c_4)V\Big]^2. \quad (A26)$$

Now average this over $\theta$ and $\phi$ using Eq. (A15).

The average $\langle U^2VW\rangle$ vanishes by the same symmetry argument as before, giving a second moment (we now drop the square brackets in the averaging)

$$\langle\rho^2\rangle = \frac{1}{8}\Big[2(c_1+c_2)^2 + (c_1-c_2)^2 + (c_3+c_4)^2\Big]\langle U^2V^2\rangle +$$
$$\frac{1}{8}\Big[2(c_3-c_4)^2 + (c_1-c_2)^2 + (c_3+c_4)^2\Big]\langle U^2W^2\rangle. \quad (A27)$$

In the same way we demonstrated that $\langle UV\rangle$ was equal to Eq. (D14), one can show that $\langle U^2V^2\rangle$ is equal to Eqs. (D31) and (D36), that $\langle U^2W^2\rangle$ is equal to Eq. (E8), and that $\langle U^2(V^2+W^2)\rangle$ is equal to Eq. (F2). This means that the averages above can be replaced by quantities that we calculate in the Appendix:

$$\langle UV \rangle = \mu_u(\gamma),$$
$$\langle U^2 V^2 \rangle = \mu_u^2(\gamma) + \sigma_u^2(\gamma),$$
$$\langle U^2 W^2 \rangle = \sigma_p^2(\gamma), \quad \text{and}$$
$$\langle U^2(V^2 + W^2) \rangle = \sigma_c^2(\gamma) = \mu_u^2(\gamma) + \sigma_u^2(\gamma) + \sigma_p^2(\gamma). \quad (A28)$$

Because the final line above is the sum of the two previous lines, there are many equivalent ways to write the expressions for the second moment and variance of the correlation $\rho$.

Using Eqs. (A27) and (A28), the second moment may be written

$$\langle \rho^2 \rangle = \frac{1}{4}(c_1 + c_2)^2(\mu_u^2 + \sigma_u^2) + \frac{1}{4}(c_3 - c_4)^2 \sigma_p^2 + \frac{1}{8}\left[(c_1 - c_2)^2 + (c_3 + c_4)^2\right]\sigma_c^2. \quad (A29)$$

The variance is calculated by subtracting from the second moment Eq. (A29) the square of the mean given in Eq. (A24). We find

$$\mu = \frac{1}{2}(c_1 + c_2)\mu_u(\gamma), \quad \text{and}$$
$$\sigma^2 = \frac{1}{4}(c_1 + c_2)^2 \sigma_u^2(\gamma) + \frac{1}{4}(c_3 - c_4)^2 \sigma_p^2(\gamma) + \frac{1}{8}\left[(c_1 - c_2)^2 + (c_3 + c_4)^2\right]\sigma_c^2(\gamma), \quad (A30)$$

which are our final expressions for the mean and variance of the Hellings-Downs correlation in the general case. The mean $\mu_u(\gamma)$ is given in Eq. (D29), the unpolarized variance $\sigma_u^2$ is given in Eq. (D37), the polarized variance $\sigma_p^2$ is given in Eq. (E8), and the cross variance $\sigma_c^2$ is given in Eq. (F2).

The sums and differences of the $c_i$ coefficients, which are defined by Eq. (A8), play the role of Stokes parameters in this formula. If the pulsar terms can be neglected, then $c_3 = c_4$ and the polarized $\sigma_p^2$ term is absent. If, in addition, $c_1 = c_2$, then only the unpolarized $\sigma_u^2$ term remains.

Returning now to the binary inspiral example from Eq. (A9), where pulsar terms are neglected, and inserting the Stokes coefficients from Eq. (A11) into Eq. (A30), the mean and variance of the Hellings-Downs correlation for a single source are

$$\mu = \langle \rho \rangle = \frac{\mathcal{A}^2}{16}\left[1 + 6\cos^2\iota + \cos^4\iota\right]\mu_u(\gamma), \quad \text{and}$$
$$\sigma^2 = \langle \Delta \rho^2 \rangle = \frac{\mathcal{A}^4}{256}\left[1 + 6\cos^2\iota + \cos^4\iota\right]^2 \sigma_u^2(\gamma)$$
$$+ \frac{\mathcal{A}^4}{512}\left[\sin^8\iota\right]\sigma_c^2(\gamma). \quad (A31)$$

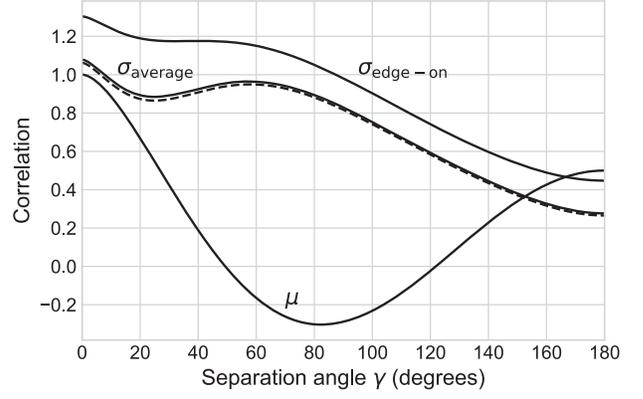

FIG. 6. Mean $\mu_{\text{average}} = \langle \rho \rangle$ and standard deviation $\sigma_{\text{average}} = \sqrt{\langle \Delta \rho^2 \rangle}$ of the Hellings-Downs correlation $\rho$ for a single circular binary GW source drawn from a uniform population, from Eq. (A32) with the amplitude set to $\mathcal{A}^2 = 15$. The dashed curve shows the standard deviation $\sigma_{\text{average}}$ if the term proportional to $\sigma_c^2$ in Eq. (A32) is set to zero; the term has no significant effect on the average. The case where the polarization term has the largest effect (an edge-on binary, $\iota = 90°$) is also shown, taken from Eq. (A31). In this case the amplitude is set to $\mathcal{A}^2 = 48$, which gives the same mean $\mu$ shown in the plot.

If $\iota$ is 0° or 180°, corresponding to a face-on or face-off orbit with $c_1 = c_2$, then only the unpolarized term is present in the variance. If not, then the cross term also appears.

To estimate these for a typical circular binary, we average the mean and variance from Eq. (A31) over orbital inclinations, assuming that $\cos \iota$ is uniformly distributed on $[-1, 1]$. We obtain

$$\mu_{\text{average}} = \frac{1}{5}\mathcal{A}^2 \mu_u(\gamma), \quad \text{and}$$
$$\sigma_{\text{average}}^2 = \frac{71}{1260}\mathcal{A}^4 \sigma_u^2(\gamma) + \frac{1}{1260}\mathcal{A}^4 \sigma_c^2(\gamma). \quad (A32)$$

These are plotted in Fig. 6. In this population average, the $\sin^8 \iota$ term averages to a very small value, and so the $\sigma_c^2$ polarization term has very little effect on the variance: the effect of removing this term for the average system is shown by the dashed curve, and is small. The $\sigma_c^2$ polarization term has the largest effect on the variance for an edge-on binary ($\iota = 90°$) also shown in the figure ($\sigma_{\text{average}}^2$ as defined here does not include all sources of variance [43]).

In the next Appendix, we will consider an example where the pulsar terms have the same magnitude as the Earth terms, so that the polarization term proportional to $(c_3 - c_4)^2$ is also present.

## APPENDIX B: THE HELLINGS-DOWNS CORRELATION MEAN AND VARIANCE INCLUDING PULSAR TERMS

Here, we calculate the mean and variance of the Hellings-Downs correlation for GWs emitted by a slowly evolving binary system, *including the pulsar terms*. We will see that the mean is unchanged from the case where only the Earth terms are included, but that the variance gets larger by about a factor of four and has a slightly different functional form.

The GW source is the same binary system as in the previous Appendix [defined around Eq. (A9)] but now we will assume that it is a very long time before the system will merge. This means that the timescale on which the waveform is changing in frequency and amplitude is much longer than the light travel time between the pulsars and Earth. This system violates the original Hellings and Downs assumptions, because (depending upon the exact separations) there can be perfect correlation or anticorrelation (or anything in between) between the Earth and pulsar terms. In this model, the GW waveforms at the Earth and pulsars are identical in frequency and amplitude, and only differ in phase.

The time averages Eq. (A8) that define the coefficients $c_1, \ldots, c_4$ give:

$$
\begin{aligned}
c_1 &= 2\sin\Delta_1 \sin\Delta_2 \cos(\Delta_2 - \Delta_1) A_c^2, \\
c_2 &= 2\sin\Delta_1 \sin\Delta_2 \cos(\Delta_2 - \Delta_1) A_s^2, \\
c_3 &= 2\sin\Delta_1 \sin\Delta_2 \sin(\Delta_2 - \Delta_1) A_c A_s, \quad \text{and} \\
c_4 &= -c_3.
\end{aligned}
\quad \text{(B1)}
$$

Here, the phase offsets are determined by the pulsar positions relative to the GW source. If $L_1$ and $L_2$ are the distances from Earth to the two pulsars, then the phase offsets are

$$
\begin{aligned}
\Delta_1 &= \omega L_1 (1 + \hat{z} \cdot \boldsymbol{p}_1), \quad \text{and} \\
\Delta_2 &= \omega L_2 (1 + \hat{z} \cdot \boldsymbol{p}_2).
\end{aligned}
\quad \text{(B2)}
$$

The amplitudes of the two polarization components are

$$
\begin{aligned}
A_c &= \mathcal{A} \frac{1}{2}(1 + \cos^2 \iota), \quad \text{and} \\
A_s &= \mathcal{A} \cos \iota.
\end{aligned}
\quad \text{(B3)}
$$

Now that we are including the pulsar terms, which have the same amplitude and frequency as the Earth terms (but a different phase) one can see that $c_3$ and $c_4$ are unequal and do not vanish.

One can work through the calculation of the previous Appendix, and everything carries through up to the point where one averages over pulsar positions on the sky. Then the calculation breaks down, because the $c_i$ depend upon the pulsar positions through the $\Delta_1$ and $\Delta_2$ terms, whereas our calculation assumed that the $c_i$ have constant values that are independent of the pulsar positions.

However, it is easy to work around this. We are interested in the angular correlation function for pulsars at different distances as well as at different sky locations. So, before averaging over the sky positions of the pulsars, we first average over their distances. This is easy, because the distances $L_1$ and $L_2$ only enter via the terms proportional to the sine and cosine of $\Delta_1$ and $\Delta_2$ and their difference.

We average by letting $L_1$ and $L_2$ vary independently over intervals $[L_{\min}, L_{\max}]$ where $0 \ll \omega L_{\min} \ll \omega L_{\max}$. This means that the lower limits and averaging intervals are much larger than the GW wavelength $2\pi/\omega$. This is sensible, because PTAs detect GW with wavelengths measured in years, whereas typical distances to pulsars are thousands of years. In the limit where the ranges are large compared to the GW wavelength, the result does not depend upon the starting values. In practice, this corresponds to treating $\Delta_1 - \Delta_2$ and $\Delta_1 + \Delta_2$ as independent variables, averaging them over any ranges which are large compared to $2\pi$.

This averaging removes the pulsar-position dependence of the $c_i$. Since

$$
\begin{aligned}
&2\sin\Delta_1 \sin\Delta_2 \cos(\Delta_2 - \Delta_1) \\
&= [\cos(\Delta_2 - \Delta_1) - \cos(\Delta_1 + \Delta_2)]\cos(\Delta_2 - \Delta_1),
\end{aligned}
\quad \text{(B4)}
$$

we can immediately see that after averaging over pulsar distances in the expressions for $\rho$ and $\rho^2$ one obtains:

$$
\begin{aligned}
\langle c_1 + c_2 \rangle &= \frac{1}{2}(A_c^2 + A_s^2), \\
\langle c_1 - c_2 \rangle &= \frac{1}{2}(A_c^2 - A_s^2), \\
\langle (c_1 + c_2)^2 \rangle &= \frac{5}{8}(A_c^2 + A_s^2)^2, \quad \text{and} \\
\langle (c_1 - c_2)^2 \rangle &= \frac{5}{8}(A_c^2 - A_s^2)^2.
\end{aligned}
\quad \text{(B5)}
$$

Here and in the next few equations, the angle brackets mean "average over $L_1$ and $L_2$." (To reproduce these: the average of $\cos^2$ is $1/2$ and the average of $\cos^4$ is $3/8$.)

To do the averaging for $c_3$ and $c_4$, first note that since $c_3 + c_4 = 0$, one has

$$
\langle c_3 + c_4 \rangle = \langle (c_3 + c_4)^2 \rangle = 0. \quad \text{(B6)}
$$

For the other two terms, use

$$
\begin{aligned}
&2\sin\Delta_1 \sin\Delta_2 \sin(\Delta_2 - \Delta_1) \\
&= [\cos(\Delta_2 - \Delta_1) - \cos(\Delta_1 + \Delta_2)]\sin(\Delta_2 - \Delta_1)
\end{aligned}
\quad \text{(B7)}
$$

to obtain

$$
\begin{aligned}
\langle c_3 - c_4 \rangle &= 0, \quad \text{and} \\
\langle (c_3 - c_4)^2 \rangle &= \frac{3}{2} A_c^2 A_s^2.
\end{aligned}
\quad \text{(B8)}
$$

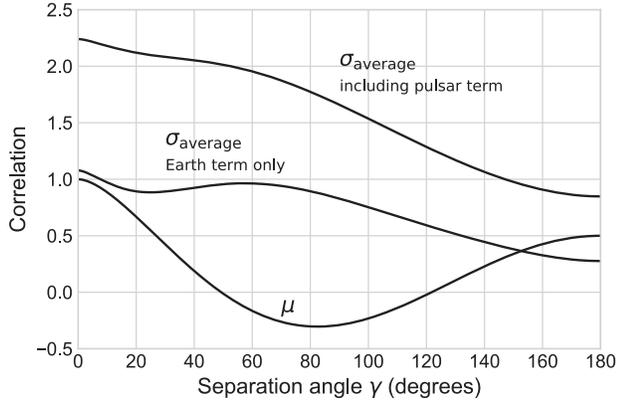

FIG. 7. The effect of the pulsar terms on the mean $\mu$ and standard deviation $\sigma$ of the Hellings-Downs correlation, for a population-averaged binary inspiral GW source. The Earth-term only case is Eq. (A32), and the case including the pulsar terms is Eq. (B11). They have identical means, but including the pulsar terms makes the variance about four times as large. (For convenience these plots are normalized by setting the squared GW amplitude $\mathcal{A}^2 = 15$ in the equations.)

The mean of the Hellings-Downs correlation follows immediately from Eq. (A24). Using $\langle c_1 + c_2 \rangle$ from Eq. (B5) gives

$$\mu = \frac{1}{4}(A_c^2 + A_s^2)\mu_\mathrm{u}, \qquad (B9)$$

which agrees with the result Eq. (A31), where only the Earth terms were included. This makes sense, since after averaging over pulsar distances, the pulsar terms have averaged away. (However, their squares will not average away, so the variance will change.)

To compute the variance, it is tempting to substitute the averages from Eqs. (B5) and (B8) into Eq. (A30). However, those expressions were obtained by subtracting the square of the mean from the second moment, and the pulsar distance averaging has $\langle (c_1 + c_2)^2 \rangle \neq \langle (c_1 + c_2) \rangle^2$. So we have to return to the original expression for the second moment Eq. (A27), evaluate that using Eqs. (B5) and (B8) and then form the variance using Eq. (B9).

For this binary-inspiral source model, which includes pulsar terms that have the same magnitude and frequency as the Earth terms, the variance of the Hellings-Downs correlation is

$$\sigma^2(\gamma) = \frac{1}{16}\left(A_c^2 + A_s^2\right)^2 \sigma_\mathrm{u}^2(\gamma) - \frac{3}{32}\left(A_c^2 - A_s^2\right)^2 \sigma_p^2(\gamma)$$
$$+ \left[\frac{3}{32}\left(A_c^2 + A_s^2\right)^2 + \frac{5}{64}\left(A_c^2 - A_s^2\right)^2\right]\sigma_c^2(\gamma). \quad (B10)$$

This should be compared with the Earth-term only equivalent Eq. (A31). The presence of pulsar terms (which here have the same magnitude as the Earth term) have left the mean unchanged but have modified the variance.

Averaging Eqs. (B9) and (B10) over the inclination angles $\iota$ gives the population averages

$$\mu_\mathrm{average} = \frac{1}{5}\mathcal{A}^2\mu_\mathrm{u}(\gamma), \quad \text{and}$$

$$\sigma_\mathrm{average}^2 = \frac{71}{1260}\mathcal{A}^4\sigma_\mathrm{u}^2(\gamma) + \frac{109}{1260}\mathcal{A}^4\sigma_c^2(\gamma) - \frac{1}{420}\mathcal{A}^4\sigma_p^2(\gamma).$$
(B11)

These are plotted in Fig. 7 and should be compared with Eq. (A32), which only include the Earth term. While the means are identical, the variance is larger by a factor of about four, because of the presence of the pulsar terms. While these terms average to zero (hence not affecting the mean) their squares do not average to zero, so they increase the variance ($\sigma_\mathrm{average}^2$ as defined here does not include all sources of variance [43]).

## APPENDIX C: THE CONVENTIONAL ENSEMBLE-AVERAGE APPROACH

In this Appendix, we discuss and employ the conventional frequency-domain Gaussian ensemble introduced in [16,17] and used for example in [21]. The Gaussian ensemble is introduced in Appendix C 1. Note that for PTAs these frequency-domain methods may be less appropriate than time-domain ones, see [22,44–46].

Appendix B demonstrated that including the pulsar terms in the Hellings-Downs correlation does not affect the mean $\mu(\gamma)$ but tends to increase the variance $\sigma^2(\gamma)$. In Appendix C 2 we investigate the conditions (on the GW spectrum) which justify dropping the pulsar terms, and compute the mean of the Hellings-Downs correlation.

In Appendix C 3 we compute the variance of the Hellings-Downs correlation, and show that the Gaussian ensemble results correspond exactly to the discrete confusion-noise source model of Sec. III A, in the limit of large numbers of weak sources. Hence, the Gaussian ensemble corresponds to a confusion-noise limit in which the closest (idealized) sources are arbitrarily close to Earth, and not at a fixed distance as in Sec. III. (Note that our calculations are based on a plane-wave radiation-zone description of GWs [47,48]. For this to apply, a real GW source would have to be at least a few times as distant as the PTA pulsars, say tens of kpc. This is still about three orders of magnitude closer to Earth than the closest relevant supermassive black-hole binaries [30,31].)

In Appendix C 4 we use identical techniques to compute the variance of the time-averaged squared strain $s = \overline{h_{ab}h^{ab}}$ and the variance the GW energy density $\rho_\mathrm{GW}$, which is proportional to the time average of $\dot{h}_{ab}\dot{h}^{ab}$. For GW backgrounds dominated by a single low frequency, these (cosmic) variances can be a large fraction of the (squared) mean.

In Appendix C 5, we use the pulsar-position-averaging technique introduced in Sec. IV to compute the cosmic variance. As for the mean and total variance, the cosmic variance corresponds exactly to the discrete confusion-noise source model of Sec. III A in the limit of large numbers of sources.

Finally, in Appendix C 6, we provide an alternative derivation of the cosmic variance for the Gaussian ensemble, using harmonic analysis methods originally developed for characterizing cosmic background radiation temperature fluctuations.

### 1. The Gaussian ensemble

We start by writing the general weak gravitational-wave solution as a plane wave expansion [16,17]. In this, we include two arbitrary functions associated with any propagation direction as a Fourier integral, obtaining

$$h_{ab}(t,\mathbf{x}) = \sum_A \int df \int d\mathbf{\Omega}\, e^{2\pi i f(t-\mathbf{\Omega}\cdot\mathbf{x})} h_A(f,\mathbf{\Omega}) e^A_{ab}(\mathbf{\Omega}). \quad (C1)$$

Here, $A = +, \times$ denotes polarization, and $h_A(f, \mathbf{\Omega})$ are two arbitrary complex functions of three variables, obeying $h_A(f, \mathbf{\Omega}) = (h_A(-f, \mathbf{\Omega}))^*$, where $*$ denotes complex conjugate. The polarization tensors $e^A_{ab}(\mathbf{\Omega})$ are defined in Eq. (D6), integration over frequency is $f \in (-\infty, \infty)$ unless specified otherwise, and the integral over the two-sphere is defined in Eq. (2.3).

Now, we imagine that the GW background arises as a central-limit-theorem process, so the functions $h_A(f, \mathbf{\Omega})$ become random variables. We have an ensemble $\mathcal{E}$ of many such functions, drawn from some distribution. From the central-limit theorem, these create a Gaussian process. Quantities of interest, which we denote $Q$, are characterized by their expectation values, which are the average value over the ensemble of functions. If we use $h$ as shorthand for $h_A(f, \mathbf{\Omega})$, and if $Q[h]$ is some quantity that depends upon $h$, then the expectation value is

$$\langle Q \rangle = \frac{1}{|\mathcal{E}|} \sum_{h \in \mathcal{E}} Q[h], \quad (C2)$$

where $|\mathcal{E}|$ denotes the number of functions in the ensemble.

We make two assumptions about our Gaussian ensemble. First, we assume that the mean of $h$ vanishes

$$\langle h_A(f, \mathbf{\Omega}) \rangle = 0. \quad (C3)$$

(A nonzero mean would be appropriate to describe fluctuations around a specific GW source.) Second, we assume that

$$\langle h_A^*(f, \mathbf{\Omega}) h_{A'}(f', \mathbf{\Omega}') \rangle = \delta_{AA'} \delta^2(\mathbf{\Omega}, \mathbf{\Omega}') \delta(f - f') H(f), \quad (C4)$$

where $H(f)$ is a positive real symmetric function $H(f) = H(-f)$ and $\delta^2(\mathbf{\Omega}, \mathbf{\Omega}')$ denote a two-dimensional delta function on the unit sphere [[49], first edition Eq. (3.56)].

The delta function in polarization indices means that the statistical properties of the two polarization degrees of freedom are identical but uncorrelated: an unpolarized ensemble. The delta function in frequency implies an ensemble that is second-order stationary in time, and the delta function on the sphere implies an ensemble that is second-order stationary in space. Some further discussion of these points may be found in [17,22].

The real function $H$ is a measure of the squared amplitude of perturbations at GW frequency $f$. It is related to the conventional stochastic background spectral function $\Omega_{\rm gw}(f)$ defined in [17] by

$$H(f) = \frac{3H_o^2}{32\pi^3} \frac{1}{|f|^3} \Omega_{\rm gw}(|f|), \quad (C5)$$

where $H_0$ is the present-day Hubble expansion rate. Much of the PTA literature uses a different measure of the spectrum known as the characteristic strain $h_c(f)$. Estimates for PTAs may be found, for example, in [30,31]. These quantities are related via

$$h_c^2(f) = \frac{3H_o^2}{2\pi^2} \frac{1}{|f|^2} \Omega_{\rm gw}(|f|) \quad (C6)$$

or equivalently via

$$H(f) = \frac{1}{16\pi} \frac{1}{|f|} h_c^2(|f|). \quad (C7)$$

An estimate of the characteristic strain $h_c(f)$ over the frequency band relevant to PTAs is shown in Fig. 8. This is reproduced from a semianalytic estimate of the GW stochastic background [[50], red dashed curve of Fig. 3]. This estimate includes both the effects of attenuation from environmental hardening effects as well as the emission of GW energy in $n > 2$ harmonics of the orbital period arising from eccentric orbits.

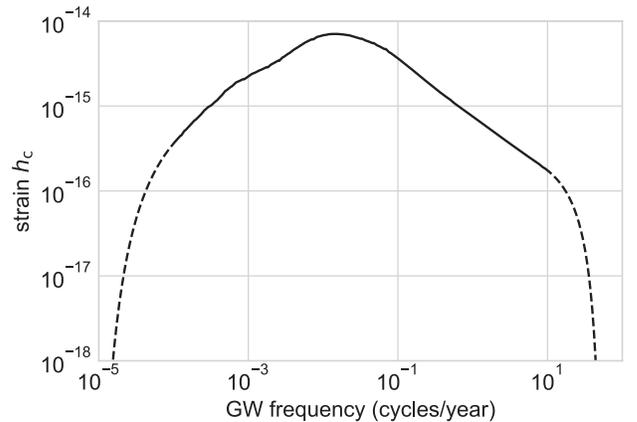

FIG. 8. Semi-analytic prediction for the spectrum of gravitational waves in the band relevant for PTAs, taken from [[50], red dashed curve of Fig. 3]. The dashed lines show how we cut off the spectrum to compute the autocorrelation function.

### 2. Neglecting pulsar terms and the GW autocorrelation function

Now we examine the Hellings-Downs correlation for a Gaussian ensemble. The pulse redshift at Earth time $t$ produced by the gravitational wave is obtained by summing Eq. (A2) via Eq. (C1) and making use of Eq. (A3), giving

$$Z(t) = \sum_A \int df \int d\mathbf{\Omega}\, F^A(\mathbf{\Omega}) h_A(f, \mathbf{\Omega}) \\ \times e^{2\pi i f t} \left[1 - e^{-2\pi i f L(1+\mathbf{\Omega}\cdot\mathbf{p})}\right], \quad (C8)$$

where the antenna pattern functions $F^A(\mathbf{\Omega})$ for $A = +, \times$ are defined in Eq. (A6), with the polarization tensors and $\mathbf{\Omega}$ coming from Eqs. (D6) and (D7). Here, the pulsar is at distance $L$ and spatial position $L\mathbf{p}$. If needed, we add a subscript to $F$ to label the pulsar, which enters via $\mathbf{p}$.

The corresponding formula for the timing residual is obtained by integrating once with respect to time, giving

$$\Delta\tau(t) = \sum_A \int df \int d\mathbf{\Omega}\, F^A(\mathbf{\Omega}) \frac{h_A(f, \mathbf{\Omega})}{2\pi i f} \\ \times e^{2\pi i f t} \left[1 - e^{-2\pi i f L(1+\mathbf{\Omega}\cdot\mathbf{p})}\right]. \quad (C9)$$

Both of these formulas give us exact values for any representative $h_A(f, \mathbf{\Omega})$ in the ensemble.

The expected correlation between the arrival time residuals for pulsars 1 and 2 is the expectation value of the quantity

$$\rho = \Delta\tau_1(t)\Delta\tau_2(t). \quad (C10)$$

We can evaluate $\langle\rho\rangle$ by substituting Eq. (C9) into Eq. (C10) and using Eq. (C4) to evaluate the expectation value, allowing us to carry out two of the integrals. We arrive at

$$\langle\rho\rangle = \int df \frac{H(|f|)}{4\pi^2 f^2} \int d\mathbf{\Omega} \left[1 - e^{-2\pi i f L_1(1+\mathbf{\Omega}\cdot\mathbf{p}_1)}\right] \\ \times \left[1 - e^{2\pi i f L_2(1+\mathbf{\Omega}\cdot\mathbf{p}_2)}\right] \sum_A F_1^A(\mathbf{\Omega}) F_2^A(\mathbf{\Omega}). \quad (C11)$$

A similar formula can be written for the redshift correlation, which removes the $4\pi^2 f^2$ factor from the denominator. No time averaging is needed to define $\rho$ because our GW statistical ensemble is stationary, and we have assumed ideal noise-free pulsars. In practice, time-averaging is needed to fit to a pulsar timing model [36,51,52]. It is also needed to compute the second moment of $\rho$.

Note that the same formula is obtained if we define (as we have done earlier in this paper) the correlation as the time average of Eq. (C10) over the interval $[-T/2, T/2]$. In this case, during the calculation, before the expectation value is taken, a factor of $\text{sinc}(\pi(f - f')T)$ is introduced, where $\text{sinc}(x) = \sin(x)/x$ [53]. However, this disappears after taking the expectation value. [Many authors define the correlation via a time integral rather than a time average, so a factor of the integration time $T$ appears on the rhs of Eq. (C11).]

It is instructive to compare Eq. (C11) with the mean of the Hellings-Downs correlation calculated in Appendix D and given by Eq. (D3). This is obtained from Eq. (C11) if the product of the two quantities in square brackets is set to unity. This corresponds to discarding the non-Earth terms.

We now return to the question of whether this is justified. The product of the quantities in square brackets on the final line of Eq. (C11) yields four terms. The product of "1" with "1" is the "Earth" term, the products of "1" with the exponentials are the Earth-pulsar interference terms, and the product of the two exponentials is the pulsar-pulsar interference term. If one averages over pulsars at different distances, then it is clear that all but the Earth term vanish. However, for a given set of pulsars in a particular experiment, such averaging is not justified: the pulsars have some particular positions and these do not change.

To understand when it might make sense to neglect the final three terms, it is helpful to introduce the autocorrelation function

$$C(T) = \frac{1}{4\pi^2} \int df\, e^{2\pi i f T} \frac{H(|f|)}{f^2} \\ = \frac{3H_0^2}{128\pi^5} \int df\, e^{2\pi i f T} \frac{\Omega_{\text{gw}}(|f|)}{f^5} \\ = \frac{1}{64\pi^3} \int df\, e^{2\pi i f T} \frac{h_c^2(|f|)}{f^3}, \quad (C12)$$

which we have written in three equivalent forms. The autocorrelation function is well known in digital and analog signal processing, where the Wiener-Khinchin theorem states that the Fourier transform of a signal's power spectrum is its autocorrelation function, and vice versa.

The autocorrelation function $C(T)$ provides useful information about the time/length scales on which a random process is correlated. Several properties of $C(T)$ follow immediately from its definition. First, it is symmetric, $C(T) = C(-T)$. Second, since $H(|f|)/f^2$ is non-negative, it follows immediately from the Schwarz inequality that $C(T)$ has its maximum at $T = 0$. Lastly, it follows from the definition (and can be proved via steepest descents or stationary phase) that for smooth functions of the types encountered in physics, the autocorrelation $C(T)$ vanishes for large $T$. The correlation time/length of a random process is typically defined as the smallest value of $T$ for which $C(T)/C(0)$ drops below $1/2$ or $1/e$. Figure 9 shows the autocorrelation $C(T)$ for the GW stochastic background spectrum shown in Fig. 8.

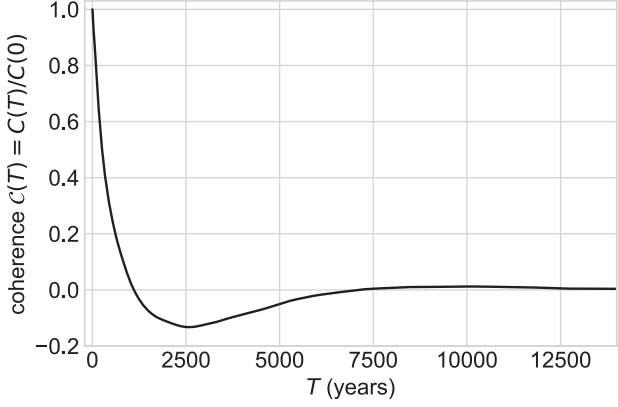

FIG. 9. Normalized autocorrelation function $\mathcal{C}(T)$ for the GW spectra shown in the previous figure. This falls off significantly for times $T$ greater than a few hundred years.

The useful information in the autocorrelation function is the value at the origin $C(0)$ and the rate of falloff. It is helpful to separate these two properties, defining a "total power" $C(0)$ and a normalized ratio $\mathcal{C}(T) = C(T)/C(0)$.

Using the autocorrelation function, we can carry out the integration over frequency in Eq. (C11), and the timing residual correlations may be written

$$\langle \rho \rangle = C(0) \int d\Omega \sum_A F_1^A(\mathbf{\Omega}) F_2^A(\mathbf{\Omega})$$
$$\times \left[ 1 + \mathcal{C}\Big(L_1(1+\mathbf{\Omega}\cdot\mathbf{p}_1)\Big) + \mathcal{C}\Big(L_2(1+\mathbf{\Omega}\cdot\mathbf{p}_2)\Big) \right.$$
$$\left. + \mathcal{C}\Big(L_1(1+\mathbf{\Omega}\cdot\mathbf{p}_1) - L_2(1+\mathbf{\Omega}\cdot\mathbf{p}_2)\Big) \right]. \quad \text{(C13)}$$

The three terms where the autocorrelation function appears are the two pulsar-Earth terms, and the pulsar-pulsar term.

If pulsars 1 and 2 are different, then we are justified in neglecting the pulsar terms in Eq. (C13) if $\mathcal{C}(T)$ is much smaller than unity when $T$ is comparable to the pulsar separations from Earth and each other. From Fig. 9, this is justified if the typical pulsar distances are thousands of years. In this case, the final three terms in Eq. (C13) will be small compared to unity, and it is valid to drop them. Only the Earth term in Eq. (C13) survives, and the timing residual correlations become

$$\langle \rho \rangle = C(0) \int d\Omega \sum_A F_1^A(\mathbf{\Omega}) F_2^A(\mathbf{\Omega})$$
$$= 4\pi C(0) \mu_u(\gamma), \quad \text{(C14)}$$

where the spherical average Eq. (2.3) of the antenna pattern functions Eq. (2.2) is given by Eq. (2.4) as $\mu_u(\gamma)$. The amplitude of the correlation is set by the GW power, as defined by $C(0)$ in Eq. (C12). This is the Gaussian ensemble equivalent of Eq. (3.36) in the discrete source calculation of Sec. III.

An interesting special case of Eq. (C13) is when pulsars 1 and 2 are the same (identical). Since they have the same direction, $\mathbf{p}_1 = \mathbf{p}_2$, which implies that $\gamma = 0$. Since they are at the same distance, $L_1 = L_2$. In that case, the last normalized autocorrelation function that appears in Eq. (C13) becomes $\mathcal{C}(0) = 1$, since its argument vanishes. Hence, the expected correlation between a pulsar and itself is twice as large as the expected correlation between that pulsar and another pulsar which lies along the same line of sight, but at a significantly different distance. Thus, for general choices of pulsars $\alpha$ and $\beta$, which might be the same or different, one has

$$\rho_{\alpha\beta} = \langle \rho \rangle = 4\pi C(0) \Big( \mu_u(\gamma_{\alpha\beta}) + \delta_{\alpha\beta} \mu_u(0) \Big), \quad \text{(C15)}$$

where $\gamma_{\alpha\beta}$ is the angle between the lines of sight to the two pulsars. Equation (C15) for $\langle \rho \rangle$ reduces to Eq. (C14) in the case where $\alpha \neq \beta$, but adds an overall factor of two when $\alpha = \beta$. Equation (C15) holds if the GW background is (statistically) stationary in time and has a normalized autocorrelation function that falls off at typical Earth-pulsar separations.

### 3. Total variance for the Gaussian ensemble

Here, we derive the total variance of the Hellings-Downs correlation for a Gaussian ensemble. A similar result is also given in the text before Eq. (9) of [21], and comes from a calculation corresponding to the last line of Eq. (5.4) in [17]. The result corresponds to a limit of the confusion-noise model for discrete point sources Eq. (3.48) found in Sec. III A.

For this derivation, and for the cosmic variance calculation in Sec. C 5, we exploit an important feature of a zero-mean Gaussian process first proved by Isserlis [54]: the expectation value of products of more than two $h$ can be written as sums of all possible products of the $h$ with at most two terms. This means that the expectation value of any even power of $h$ can be evaluated using Eq. (C4), and also implies that the expected values of odd powers of $h$ will vanish.

Begin with Eq. (C8) for the arrival time redshift of pulses at Earth time $t$, multiply its complex conjugate by the identical expression for the second pulsar, and form the time-averaged product over the range $t \in [-T/2, T/2]$. This gives

$$\rho = \overline{Z_1 Z_2}$$
$$= \sum_A \sum_{A'} \int df \int df' \int d\mathbf{\Omega} \int d\mathbf{\Omega}' R_1^{A*}(f, \mathbf{\Omega}) R_2^{A'}(f', \mathbf{\Omega}')$$
$$\times h_A^*(f, \mathbf{\Omega}) h_{A'}(f', \mathbf{\Omega}') \text{sinc}\Big(\pi(f-f')T\Big), \quad \text{(C16)}$$

where the subscripts "1" and "2" label the pulsars, and $\text{sinc}(x) = \sin(x)/x$ [53]. The pulsar term is incorporated via a modified antenna pattern function

$$R_n^A(f, \mathbf{\Omega}) = \left[1 - e^{-2\pi i f L_n(1+\mathbf{\Omega}\cdot\mathbf{p}_n)}\right] F_n^A(\mathbf{\Omega}), \quad (C17)$$

where $n = 1$ or $n = 2$ denotes the pulsar, and (as before) $L_n$ is the distance from the Earth to the pulsar. For any particular realization drawn from the ensemble, Eq. (C16) gives the correlation between the two pulsars. To compute the variance, we need the mean and second moment of $\overline{Z_1 Z_2}$.

The mean is easily computed by taking the ensemble average of Eq. (C16) by using Eq. (C4). One obtains

$$\langle \rho \rangle = \langle \overline{Z_1 Z_2} \rangle = h^2 \mu_\text{u}(\gamma), \quad (C18)$$

where we have assumed that the normalized autocorrelation function $\mathcal{C}$ is small for typical Earth-pulsar and pulsar-pulsar separations, which means that the three remaining exponential terms may be neglected. This is the same expression obtained in Eq. (C14), provided that a factor of $(2\pi f)^2$ is put into the numerator of Eq. (C12), since redshifts rather than timing residuals are being correlated.

For convenience, the overall scale in Eq. (C18) is expressed as

$$h^2 = 4\pi \int H(f) df = 8\pi \int_0^\infty H(f) df. \quad (C19)$$

This is a measure of the (squared) strain, expressed as an integral of the spectral function $H(f)$. In Eq. (C19) we have also expressed this in terms of one-sided integrals, making use of the symmetry $H(f) = H(-f)$ intrinsic in the definition Eq. (C4) of the two-point function for the Gaussian ensemble.

For the second moment, we multiply the correlation Eq. (C16) for any representative of the ensemble by its complex conjugate [55], obtaining

$$\rho^2 = \sum_A \sum_{A'} \sum_{A''} \sum_{A'''} \int df \int df' \int df'' \int df''' \int d\mathbf{\Omega} \int d\mathbf{\Omega}' \int d\mathbf{\Omega}'' \int d\mathbf{\Omega}''' \text{sinc}\left(\pi(f-f')T\right) \text{sinc}\left(\pi(f''-f''')T\right)$$
$$\times R_1^{A*}(f, \mathbf{\Omega}) R_2^{A'}(f', \mathbf{\Omega}') R_1^{A''}(f'', \mathbf{\Omega}'') R_2^{A'''*}(f''', \mathbf{\Omega}''') h_A^*(f, \mathbf{\Omega}) h_{A'}(f', \mathbf{\Omega}') h_{A''}(f'', \mathbf{\Omega}'') h_{A'''}^*(f''', \mathbf{\Omega}'''). \quad (C20)$$

To compute the ensemble average of $\rho^2$, we need the ensemble average of the four-point function. Isserlis' theorem [54] implies that for a Gaussian ensemble the four-point function is the sum of three two-point functions:

$$\langle h_A^*(f, \mathbf{\Omega}) \, h_{A'}(f', \mathbf{\Omega}') h_{A''}(f'', \mathbf{\Omega}'') \, h_{A'''}^*(f''', \mathbf{\Omega}''') \rangle =$$
$$\langle h_A^*(f, \mathbf{\Omega}) \, h_{A'}(f', \mathbf{\Omega}') \rangle \quad \langle h_{A''}^*(-f'', \mathbf{\Omega}'') \, h_{A'''}(-f''', \mathbf{\Omega}''') \rangle +$$
$$\langle h_A^*(f, \mathbf{\Omega}) \quad h_{A''}(f'', \mathbf{\Omega}'') \rangle \quad \langle h_{A'}^*(-f', \mathbf{\Omega}') \, h_{A'''}(-f''', \mathbf{\Omega}''') \rangle +$$
$$\langle h_A^*(f, \mathbf{\Omega}) \quad h_{A'''}(-f''', \mathbf{\Omega}''') \rangle \langle h_{A'}^*(-f', \mathbf{\Omega}') \, h_{A''}(f'', \mathbf{\Omega}'') \rangle, \quad (C21)$$

where we have used $h_A(f, \mathbf{\Omega}) = h_A^*(-f, \mathbf{\Omega})$ to write each of the three terms in a form directly computable from Eq. (C4). Using Eq. (C4) to evaluate Eq. (C21) gives

$$\left\langle h_A^*(f, \mathbf{\Omega}) h_{A'}(f', \mathbf{\Omega}') h_{A''}(f'', \mathbf{\Omega}'') h_{A'''}^*(f''', \mathbf{\Omega}''') \right\rangle =$$
$$\delta_{AA'} \, \delta_{A''A'''} \, \delta^2(\mathbf{\Omega}, \mathbf{\Omega}') \, \delta^2(\mathbf{\Omega}'', \mathbf{\Omega}''') \, \delta(f-f') \, \delta(f''-f''') H(f) H(f'') +$$
$$\delta_{AA''} \, \delta_{A'A'''} \, \delta^2(\mathbf{\Omega}, \mathbf{\Omega}'') \, \delta^2(\mathbf{\Omega}', \mathbf{\Omega}''') \, \delta(f-f'') \, \delta(f'-f''') \, H(f) H(f') +$$
$$\delta_{AA'''} \delta_{A'A''} \, \delta^2(\mathbf{\Omega}, \mathbf{\Omega}''') \delta^2(\mathbf{\Omega}', \mathbf{\Omega}'') \, \delta(f+f''') \delta(f'+f'') \, H(f) H(f'), \quad (C22)$$

where we have used $\delta(f) = \delta(-f)$ in several terms. To evaluate the second moment of $\rho$ we take the ensemble average of Eq. (C20) and use Eq. (C22) to obtain three terms. In the same order as Eq. (C22), the three terms are

$$\langle \rho^2 \rangle = \int df H(f) \int df'' H(f'') \sum_A \int d\mathbf{\Omega} R_1^{A*}(f, \mathbf{\Omega}) R_2^A(f, \mathbf{\Omega}) \sum_{A''} \int d\mathbf{\Omega}'' R_1^{A''}(f'', \mathbf{\Omega}'') R_2^{A''*}(f'', \mathbf{\Omega}'') +$$
$$\int df H(f) \int df' H(f') \text{sinc}^2\left(\pi(f-f')T\right) \sum_A \int d\mathbf{\Omega} R_1^{A*}(f, \mathbf{\Omega}) R_1^A(f, \mathbf{\Omega}) \sum_{A'} \int d\mathbf{\Omega}' R_2^{A'}(f', \mathbf{\Omega}') R_2^{A'*}(f', \mathbf{\Omega}') +$$
$$\int df H(f) \int df' H(f') \text{sinc}^2\left(\pi(f-f')T\right) \sum_A \int d\mathbf{\Omega} R_1^{A*}(f, \mathbf{\Omega}) R_2^A(f, \mathbf{\Omega}) \sum_{A'} \int d\mathbf{\Omega}' R_1^{A'}(f', \mathbf{\Omega}') R_2^{A'*}(f', \mathbf{\Omega}'). \quad (C23)$$

This second moment $\langle \rho^2 \rangle$ takes a simpler form if the normalized autocorrelation function $\mathcal{C}$ is small for typical Earth-pulsar and pulsar-pulsar separations.

In this case, replace the modified antenna pattern functions $R_n^A$ which appear in Eq. (C23) by their definitions from Eq. (C17). In the first and third line of Eq. (C23), all of the terms which contain exponentials may be neglected, and only the product of the unity terms remains (see Fig. 16 of Ref. [35]). In the second line of Eq. (C23), one has

$$R_1^{A*}(f,\mathbf{\Omega})R_1^A(f,\mathbf{\Omega}) = \left|1-e^{-2\pi i f L_1(1+\mathbf{\Omega}\cdot\mathbf{p}_1)}\right|^2 F_1^A(\mathbf{\Omega})F_1^A(\mathbf{\Omega})$$
$$= \left[2 - 2\cos\left(2\pi f L_1(1+\mathbf{\Omega}\cdot\mathbf{p}_1)\right)\right]$$
$$\times F_1^A(\mathbf{\Omega})F_1^A(\mathbf{\Omega})$$
$$\approx 2F_1^A(\mathbf{\Omega})F_1^A(\mathbf{\Omega}), \quad (\text{C24})$$

and a similar expression for the second pulsar term, since the rapidly varying cosine term, which arises from the two remaining exponential terms, may be neglected. Thus we obtain

$$\langle\rho^2\rangle = \quad (\text{C25})$$

$$h^4 \sum_A \int \frac{d\mathbf{\Omega}}{4\pi} F_1^A(\mathbf{\Omega})F_2^A(\mathbf{\Omega}) \sum_{A'} \int \frac{d\mathbf{\Omega}'}{4\pi} F_1^{A'}(\mathbf{\Omega}')F_2^{A'}(\mathbf{\Omega}')$$
$$+ 4\hbar^4 \sum_A \int \frac{d\mathbf{\Omega}}{4\pi} F_1^A(\mathbf{\Omega})F_1^A(\mathbf{\Omega}) \sum_{A'} \int \frac{d\mathbf{\Omega}'}{4\pi} F_2^{A'}(\mathbf{\Omega}')F_2^{A'}(\mathbf{\Omega}')$$
$$+ \hbar^4 \sum_A \int \frac{d\mathbf{\Omega}}{4\pi} F_1^A(\mathbf{\Omega})F_2^A(\mathbf{\Omega}) \sum_{A'} \int \frac{d\mathbf{\Omega}'}{4\pi} F_1^{A'}(\mathbf{\Omega}')F_2^{A'}(\mathbf{\Omega}').$$

The factor of four on the third line arises from (C24) because the pulsar terms contribute the same amount as the Earth terms. Fundamentally, this is because the Gaussian ensemble is stationary in time. Hence, when a pulsar is correlated with itself, the time-averaged square of the pulsar term is equal to the time-averaged square of the Earth term.

Two different scale factors appear in the second moment. The first is the squared strain $h^2$ defined by Eq. (C19). The second is

$$\hbar^4 = (4\pi)^2 \int df \int df' \text{sinc}^2\left(\pi(f-f')T\right) H(f)H(f'), \quad (\text{C26})$$

which is (another) measure of strain, determined by the spectral function $H(f)$ and the observation interval (averaging time) $T$. (In the Appendix of [38], this integral is evaluated for a simple cosmological model with $H(f) \propto |f|^{-7/3}$.)

The second moment Eq. (C25) can now be expressed in terms of the Hellings-Downs curve $\mu_u(\gamma)$. This is because each of the integrals over solid angle in Eq. (C25) has the "unpolarized" Hellings-Downs term in Eq. (2.2) as an integrand, whose average on the sphere is $\mu_u(\gamma)$ as given in Eq. (2.4). Thus we find

$$\langle\rho^2\rangle = h^4\mu_u^2(\gamma) + 4\hbar^4\mu_u^2(0) + \hbar^4\mu_u^2(\gamma). \quad (\text{C27})$$

Note that each of the integrals over solid angle on the second line of Eq. (C25) has the same form as the integrals on the following line, but with the directions to pulsars being identical, thus giving $\mu_u(0)$.

The variance $\sigma^2$ is formed by subtracting the square of the mean given in Eq. (C18) from the second moment of Eq. (C27). The variance $\sigma^2 = \langle\rho^2\rangle - \langle\rho\rangle^2$ of the Gaussian ensemble is then

$$\sigma^2(\gamma) = \hbar^4\left(\mu_u^2(\gamma) + 4\mu_u^2(0)\right). \quad (\text{C28})$$

These results can be compared to the confusion-noise model of Sec. III A in the limit of large numbers of weak sources. The Gaussian ensemble variance Eq. (C28) equals the confusion-noise model variance of Eq. (3.50) if $\hbar^4 = \mathcal{H}_2^2/8$, and the mean Eq. (C18) of the Gaussian model equals the mean of the confusion-noise model Eq. (3.51) if $h^2 = \mathcal{H}_2/2$. Thus, the confusion noise model has $\hbar^4 = h^4/2$.

The Gaussian ensemble variance has two interesting limits. To obtain these two limits, it is helpful to rewrite the definition of the strain measure $\hbar^4$ given by Eq. (C26) in terms of one-sided integrals, analogous to the second form given in Eq. (C19):

$$\hbar^4 = \frac{1}{2}(8\pi)^2 \int_0^\infty df H(f) \int_0^\infty df' H(f')$$
$$\times \left[\text{sinc}^2\left(\pi(f-f')T\right) + \text{sinc}^2\left(\pi(f+f')T\right)\right]. \quad (\text{C29})$$

Now consider the "single-sided" function $H(f)$ defined on the interval $f \geq 0$.

In the "narrowband" case, we assume that the support of $H(f)$ on the interval $f \geq 0$ has a narrow bandwidth $\Delta f \ll 1/T$ and that the central frequency $f_0$ of this support satisfies $f_0 T = z$ for some integer $z$. In this case, the first sinc function that occurs in Eq. (C29) gives $\text{sinc}(0) = 1$. The second sinc function vanishes since $\text{sinc}(2\pi z) = 0$ [53]. Comparison of Eqs. (C19) and (C29) then implies that in this narrowband case $\hbar^4 = h^4/2$, so the mean of Eq. (C18) and the variance of Eq. (C28) become

$$\mu(\gamma) = h^2\mu_u(\gamma), \quad \text{and}$$

$$\sigma^2(\gamma) = \frac{1}{2}h^4\left(\mu_u^2(\gamma) + 4\mu_u^2(0)\right). \quad (\text{C30})$$

These apply for a narrowband signal, such as that considered in Sec. III A. In fact, Eq. (C30) for the Gaussian ensemble mean and variance can be obtained from Eqs. (3.36) and (3.48) of Sec. III A, in the limit where $N_s$ becomes large with $N_s \mathcal{A}^2$ held constant, which identifies $h^2 = \mathcal{H}_2/2$ and sends $\mathcal{H}_4 \to 0$.

The assumption that the narrowband frequency $f_0$ satisfies $f_0 T =$ integer appears quite restrictive. But it is easy to see that dropping this assumption does not have much effect: it can increase the variance by up to a factor of two. This is because the narrow bandwidth assumption implies that the second sinc² in Eq. (C29) can be treated as a constant that lies in the range [0, 1]. Thus, allowing $f_0 T$ to be noninteger means that the factor $1/2$ on the second line of Eq. (C30) could instead lie anywhere in the interval $[1/2, 1]$.

The second case of interest is the "broadband signal" case, for which the observation time is long compared with the inverse bandwidth: $T \gg 1/\Delta f$. Here, one may replace $\text{sinc}^2(\pi f T)$ with $\delta(f)/T$, to obtain

$$\hbar^4 = \frac{(4\pi)^2}{T} \int H^2(f) df = \frac{32\pi^2}{T} \int_0^\infty H^2(f) df. \quad (C31)$$

In this broadband case the (single-pair or total) variance of the Hellings and Downs correlation is

$$\sigma^2(\gamma) = \frac{32\pi^2}{T} \int_0^\infty H^2(f) df \left( \mu_u^2(\gamma) + 4\mu_u^2(0) \right). \quad (C32)$$

The relative size of the variance $\sigma$ compared with the mean $\mu$ then depends upon the form of $H(f)$ and the observation time $T$. The scale of $\sigma^2$ is determined by the integral of $H^2(f)$ divided by $T$, whereas the scale of $\mu^2$ is determined by the square of the integral of $H$.

In contrast to the narrowband case of Eq. (C30), the broadband variance Eq. (C32) decreases with time. This is because in the Gaussian ensemble, the distinct frequency bands are independent. So, as the observation time $T$ increases, more frequency bands become "distinct." The variances then add in quadrature, resulting in a total variance that drops as $1/T$ with increasing observation time, as the effective number of distinct frequency bands grows. A related discussion may be found in [45].

Frequency-domain formulas for $\sigma^2$ that exhibit the same angular dependence as Eqs. (C30) and (C32) are implicitly given in [20,21]. Those formulas are also obtained from Gaussian ensemble calculations corresponding to the last line of Eq. (5.4) in [17]. For example, the text before Eq. (9) of [21] gives the variance as $\sigma_{ab}^2(f) = S_{aa}(f)S_{bb}(f) + S_{ab}^2(f)$ where $a$ and $b$ denote two pulsars. In our notation the autocorrelation is $S_{aa} = S_{bb} = 2(4\pi H(f)\mu_u(0))$ where the factor of two arises from the pulsar term, and the cross correlation is $S_{ab}(f) = 4\pi H(f)\mu_u(\gamma_{ab})$. One thus obtains $\sigma_{ab}^2(f) = (4\pi H(f))^2 (\mu_u^2(\gamma_{ab}) + 4\mu_u^2(0))$ where $\gamma_{ab} = \gamma$ is the angle between the lines of sight to pulsars $a$ and $b$. This variance has the same dependence on angle $\gamma$ as Eqs. (C30) and (C32), and integrating it over frequency gives the broadband result Eq. (C32).

### 4. Variance of the squared GW strain and GW energy-density for the Gaussian ensemble

Here, we find the variance of the time-averaged squared strain $s = \overline{h_{ab} h^{ab}}$ for the Gaussian ensemble. This is straightforward to compute, and has the same origins as the cosmic variance. This computation also gives the variance of the time-averaged GW energy density $\rho_{\text{GW}} = c^2 \overline{\dot{h}_{ab} \dot{h}^{ab}}/32\pi G$ [[17], Eq. (2.13)], provided that two factors of $\pi c^2 f^2/8G$ are inserted into the correct equations.

To carry out the calculation, we will need the spherical average Eq. (2.3) of the product of two polarization tensors, given by Eq. (D6). This spherical average is defined by

$$\eta_{abcd} = \frac{1}{4\pi} \int d\Omega\, e_{ab}(\mathbf{\Omega}) e^*_{cd}(\mathbf{\Omega})$$
$$= \frac{1}{4\pi} \int d\Omega \left( e^+_{ab}(\mathbf{\Omega}) e^+_{cd}(\mathbf{\Omega}) + e^\times_{ab}(\mathbf{\Omega}) e^\times_{cd}(\mathbf{\Omega}) \right). \quad (C33)$$

For completeness we have expressed this average both in terms of the linear polarization basis $A = +, \times$ and in terms of the (complex) circular polarization basis $e_{ab} = e^+_{ab} - i e^\times_{ab}$, which was introduced after Eq. (3.11). While $\eta_{abcd}$ could be evaluated by carrying out the integrals explicitly, symmetry arguments are simpler.

From rotational symmetry and index symmetries it follows that

$$\eta_{abcd} = \alpha \delta_{ab} \delta_{cd} + \beta(\delta_{ac}\delta_{bd} + \delta_{ad}\delta_{bc}), \quad (C34)$$

where $\delta_{ab}$ is the three-dimensional Kronecker delta and $\alpha$ and $\beta$ are dimensionless numerical constants. (The symmetry arguments are those of Ref. [[56] Appendix B].) The constants $\alpha$ and $\beta$ in Eq. (C34) are easily found. Since the polarization tensors are traceless, Eq. (C33) implies that $\eta^a{}_{acd} = 0$, which in Eq. (C34) implies that $3\alpha + 2\beta = 0$. Since we have normalized the polarization tensors so that $e^{ab} e^*_{ab} = 4$, Eq. (C33) implies that $\eta^{ab}{}_{ab} = 4$, which from Eq. (C34) implies that $3\alpha + 12\beta = 4$. The solution to this pair of linear equations is $\alpha = -4/15$ and $\beta = 2/5$. This simple formula for $\eta_{abcd}$ may also be obtained by taking the $\alpha \to 0$ limit of the more general overlap reduction function given in Ref. [[17], Eq. (3.34)].)

The first and second moments of $s = \overline{h_{ab} h^{ab}}$ are computed by replacing $\rho$ by $s$ in the calculations of Appendix C 3. The value of $s$ for any realization from the ensemble, which is analogous to Eq. (C16), is given by

$$s = \sum_A \sum_{A'} \int df \int df' \int d\Omega \int d\Omega'\, \text{sinc}\left(\pi(f - f')T\right)$$
$$\times e^A_{ab}(\mathbf{\Omega}) e^{ab}_{A'}(\mathbf{\Omega}') h^*_A(f, \mathbf{\Omega}) h_{A'}(f', \mathbf{\Omega}'). \quad (C35)$$

The ensemble expectation value of $s$ is evaluated using Eq. (C4). This gives

$$\langle s \rangle = \sum_A \int df H(f) \int d\mathbf{\Omega}\, e_{ab}^A(\mathbf{\Omega}) e_A^{ab}(\mathbf{\Omega}) = 4\pi \eta_{ab}{}^{ab} \int H(f) df = 4h^2, \quad (C36)$$

where $h^2$ is defined by Eq. (C19). The second moment of $s$ is obtained from the analog of Eq. (C20), which is

$$s^2 = \sum_A \sum_{A'} \sum_{A''} \sum_{A'''} \int df \int df' \int df'' \int df''' \int d\mathbf{\Omega} \int d\mathbf{\Omega}' \int d\mathbf{\Omega}'' \int d\mathbf{\Omega}''' \mathrm{sinc}\Big(\pi(f-f')T\Big)\mathrm{sinc}\Big(\pi(f''-f''')T\Big)$$
$$\times e_{ab}^A(\mathbf{\Omega}) e_{A'}^{ab}(\mathbf{\Omega}') e_{cd}^{A''}(\mathbf{\Omega}'') e_{A'''}^{cd}(\mathbf{\Omega}''') h_A^*(f,\mathbf{\Omega}) h_{A'}(f',\mathbf{\Omega}') h_{A''}(f'',\mathbf{\Omega}'') h_{A'''}^*(f''',\mathbf{\Omega}'''). \quad (C37)$$

Taking its expectation value for the Gaussian ensemble of universes gives the analog of Eq. (C25), where the four antenna-pattern functions are replaced by contractions of polarization tensors. The three terms arising from Isserlis' theorem are

$$\langle s^2 \rangle = h^4 \eta_{ab}{}^{ab}\eta_{cd}{}^{cd} + \hbar^4\Big(\eta_{abcd}\eta^{abcd} + \eta^{abcd}\eta_{abcd}\Big) = 16 h^4 + \frac{32}{5}\hbar^4, \quad (C38)$$

where $\hbar^4$ is defined by Eq. (C26). The final equality follows by using Eq. (C34) to compute $\eta^{ab}{}_{ab} = 3\alpha + 12\beta = 4$ and $\eta_{abcd}\eta^{abcd} = 6(\alpha + 4\beta)\beta = 16/5$.

Note that the $\hbar^4$ terms on the rhs of Eq. (C38) are equal. In contrast, when computing the variance of the Hellings and Downs correlation in Eq. (C25), the effect of the "pulsar term" is to make the second term four times as large as the third. Here, the squared strain is a local quantity: there are no pulsar terms.

The (cosmic) variance in $s$ is easily evaluated from Eqs. (C36) and (C38), giving

$$\sigma_s^2 = \langle s^2 \rangle - \langle s \rangle^2 = \frac{32}{5}\hbar^4. \quad (C39)$$

This implies that a local measurement of the squared GW strain or GW energy density has a fractional uncertainty

$$\frac{\sigma_s}{\langle s \rangle} = \sqrt{\frac{2}{5}}\frac{\hbar^2}{h^2}, \quad (C40)$$

where we have used Eqs. (C36) and (C39).

Corresponding formulas may be obtained for the mean GW energy density and its variance by inserting a factor of $\pi c^2 f^2/8G$ into the integral Eq. (C19) which defines $h^2$ and two such factors (one for $f$ and one for $f'$) into the integral Eq. (C26) which defines $\hbar^4$.

For the commensurate-frequency narrowband case with $\hbar^4/h^4 = 1/2$ this ratio corresponds to typical fluctuations of $\pm 1/\sqrt{5} \approx \pm 45\%$. These large fluctuations arise because in the narrowband case, the Universe contains a combination of standing and traveling waves. Depending upon the relative phases of the GW sources, Earth may be located near a node (small average squared GW strain) or an antinode (large average squared GW strain) of that pattern.

In contrast, for a broadband signal Eq. (C31) shows that the ratio of the fluctuations to the mean decreases proportional to $1/\sqrt{T}$ where $T$ is the observation or averaging time.

### 5. Cosmic variance of the Hellings and Downs correlation for the Gaussian ensemble

Here, we derive the cosmic variance of the Gaussian ensemble, using the same technique that we used for the discrete-source confusion-noise model in Sec. IV.

As explained at the start of Sec. IV, to compute the cosmic variance, we average the pulsar correlation Eq. (C16) over all pulsar-pair directions separated by angle $\gamma$ to obtain $\Gamma(\gamma)$ *before* computing the first and second moments. This removes the pulsar variance. If the pulsar distances are much larger than the wavelength $1/f$, then the exponential terms that appear from Eq. (C17) are rapidly oscillating functions of the pulsar directions. In contrast, the functions $F_n^A$ only vary slowly with the pulsar directions (like a quadrupole) so the product averages to zero. Thus, we can replace the $R_n^A$ with $F_n^A$. Using Eqs. (G6) and (G10) from Appendix G, the pulsar average converts the antenna patterns in Eq. (C16) into the Hellings-Downs two-point functions $\mu_{AA'}(\gamma,\beta)$. We obtain

$$\Gamma(\gamma) = \overline{\langle Z_1 Z_2 \rangle}_p = \sum_A \sum_{A'} \int df \int df' \mathrm{sinc}\Big(\pi(f-f')T\Big)$$
$$\times \int d\mathbf{\Omega} \int d\mathbf{\Omega}' \mu_{AA'}\Big(\gamma,\beta(\mathbf{\Omega},\mathbf{\Omega}')\Big) h_A^*(f,\mathbf{\Omega})$$
$$\times h_{A'}(f',\mathbf{\Omega}'), \quad (C41)$$

where $\beta(\mathbf{\Omega},\mathbf{\Omega}') = \cos^{-1}(\mathbf{\Omega}\cdot\mathbf{\Omega}')$ is the angle between the source vectors $\mathbf{\Omega}$ and $\mathbf{\Omega}'$. For any realization in the ensemble, $\Gamma(\gamma)$ is the pulsar-averaged Hellings-Downs correlation that would be obtained via measurement of and averaging over many pulsars, hence $\Gamma(\gamma)$ only contains cosmic variance. To estimate the magnitude of that variance, we compute $\langle \Gamma^2 \rangle - \langle \Gamma \rangle^2$, where the angle brackets now mean "ensemble average."

The ensemble average of $\Gamma(\gamma)$ is easily computed from Eq. (C41) by using Eq. (C4). It is

$$\langle \Gamma(\gamma) \rangle = \sum_A \int df \int d\Omega \, \mu_{AA}(\gamma, 0) H(f)$$

$$= 4\pi \int df \, H(f) \mu(\gamma, 0)$$

$$= h^2 \mu(\gamma, 0) = h^2 \mu_{\rm u}(\gamma), \qquad (C42)$$

where we have used Eq. (G7) to evaluate the polarization sum in the first line. Here, $\mu(\gamma, \beta)$ is the two-point function from Appendix G, $\mu(\gamma, 0) = \mu_{\rm u}(\gamma)$ is the conventional Hellings-Downs curve, and $h^2$ is defined by Eq. (C19). Thus, $\langle \Gamma(\gamma) \rangle$ is identical to the mean of $\rho$ before the pulsar average, given in Eq. (C18).

For the second moment, we start with the square of the pulsar-averaged correlation Eq. (C41) for any representative of the ensemble

$$\Gamma^2(\gamma) = \sum_A \sum_{A'} \sum_{A''} \sum_{A'''} \int df \int df' \int df'' \int df''' \int d\Omega \int d\Omega' \int d\Omega'' \int d\Omega''' \mu_{AA'}\left(\gamma, \beta(\Omega, \Omega')\right) \mu_{A''A'''}\left(\gamma, \beta(\Omega'', \Omega''')\right)$$

$$\times h_A^*(f, \Omega) h_{A'}(f', \Omega') h_{A''}(f'', \Omega'') h_{A'''}^*(f''', \Omega''') \operatorname{sinc}\left(\pi(f - f')T\right) \operatorname{sinc}\left(\pi(f'' - f''')T\right). \qquad (C43)$$

We evaluate its ensemble average by using the four-point ensemble average Eq. (C22). This gives [compare Eq. (C44) with Eqs. (C25) and (C27)]

$$\langle \Gamma(\gamma)^2 \rangle = \sum_A \sum_{A''} \int df H(f) \int df'' H(f'') \int d\Omega \int d\Omega'' \mu_{AA}(\gamma, 0) \mu_{A''A''}(\gamma, 0) +$$

$$\sum_A \sum_{A'} \int df H(f) \int df' H(f') \operatorname{sinc}^2\left(\pi(f-f')T\right) \int d\Omega \int d\Omega' \mu_{AA'}\left(\gamma, \beta(\Omega, \Omega')\right) \mu_{AA'}\left(\gamma, \beta(\Omega, \Omega')\right) +$$

$$\sum_A \sum_{A'} \int df H(f) \int df' H(f') \operatorname{sinc}^2\left(\pi(f-f')T\right) \int d\Omega \int d\Omega' \mu_{AA'}\left(\gamma, \beta(\Omega, \Omega')\right) \mu_{A'A}\left(\gamma, \beta(\Omega', \Omega)\right)$$

$$= h^4 \mu^2(\gamma, 0) + 2\hbar^4 \int \frac{d\Omega}{4\pi} \int \frac{d\Omega'}{4\pi} \left[ \mu_{++}^2\left(\gamma, \beta(\Omega, \Omega')\right) + \mu_{\times\times}^2\left(\gamma, \beta(\Omega, \Omega')\right) \right]$$

$$= h^4 \mu^2(\gamma, 0) + \hbar^4 \int_0^\pi \sin\beta d\beta \left[ \mu_{++}^2(\gamma, \beta) + \mu_{\times\times}^2(\gamma, \beta) \right]$$

$$= h^4 \mu_{\rm u}^2(\gamma) + 2\hbar^4 \tilde{\mu}^2(\gamma), \qquad (C44)$$

where $\hbar^4$ is defined by Eq. (C26), and we have used Eq. (G10) to set $\mu_{+\times}$ and $\mu_{\times+}$ to zero. Note that the first term on the rhs of Eq. (C44) is the square of the mean from Eq. (C42), and that the last two integrals over solid angle are equal. These are evaluated in Eq. (G12), where $\tilde{\mu}^2$ is the spherical average of the Hellings-Downs two-point function computed in Appendix G and given in Eq. (G11).

We obtain the variance $\sigma^2$ from the second moment $\langle \Gamma^2 \rangle$ by subtracting the square of the mean, which eliminates the first term in Eq. (C44), giving a cosmic variance

$$\sigma_{\rm cosmic}^2(\gamma) = \langle \Gamma(\gamma)^2 \rangle - \langle \Gamma(\gamma) \rangle^2 = 2\hbar^4 \tilde{\mu}^2(\gamma). \qquad (C45)$$

This should be compared with the total variance (pulsar plus cosmic) given in Eq. (C28). The cosmic variance of Eq. (C45) corresponds exactly to the cosmic variance Eq. (4.8) for the confusion-noise model of Sec. III A in the limit of large numbers of weak sources, where $\hbar^4 = \mathcal{H}_2^2/8$. That is the same identification needed to make the total variances match in this limit. As discussed following Eq. (C29), the overall amplitude in Eq. (C45) takes different forms for narrowband and broadband signals, as in Eqs. (C30) and (C32).

### 6. Cosmic variance from harmonic analysis

An alternative approach to obtaining the cosmic variance for the Gaussian ensemble uses techniques which were originally developed for the analysis of cosmic background radiation temperature fluctuation maps [57–60]. In [61,62], pulsar timing redshift maps are analyzed using these harmonic analysis methods. The existence of such a map is equivalent to assuming that one has data from a large number of low-noise pulsars, uniformly distributed on the sky.

The pulsar redshift $Z$ induced by a GW background as a function of the pulsar sky direction $\Omega$ is decomposed into spherical harmonics

$$Z(\Omega) = \sum_{lm} a_{lm} Y_{lm}(\Omega), \qquad (C46)$$

where $\mathbf{\Omega}$ denotes coordinates on $S^2$ or equivalently the coordinate pair $\theta, \phi$. The complex expansion coefficients $a_{lm}$ have specific values in any given realization of the universe. An ensemble of universes corresponds to an ensemble of $a_{lm}$; here angle brackets denote averages over that ensemble. While the physical redshift is real, the $Z$ defined by Eq. (C46) is complex because it corresponds to the Fourier transform of data from a single frequency bin [61].

In [61] the complex coefficients $a_{lm}$ are first found for a point source of gravitational waves located along the $z$-axis. From these, one can infer the values of the $a_{lm}$ for a point source with an arbitrary sky position. These coefficients are characterized by the "rotation invariant" quantity

$$C_l = \frac{1}{2l+1} \sum_{m=-l}^{l} |a_{lm}|^2. \tag{C47}$$

The redshift $Z$ is a linear functional of the GW amplitude $h_{ab}$, as shown in Eqs. (A2) and (A3). Thus, if the GW background is described by a Gaussian ensemble, it follows from Eq. (C46) that the $a_{lm}$ are also Gaussian random variables. We can then employ Eq. (C47) to (statistically) characterize the $a_{lm}$.

For a given value of $l$, there are $2l+1$ coefficients $a_{lm}$, for $m = -l, ..., l$. Each coefficient is a complex number, whose real and imaginary parts are independent random variables. These $4l+2$ independent variables are all drawn from the same zero-mean Gaussian distribution. The second moment of that distribution (up to a factor of $4l+2$) is determined by Eq. (21) of [61] as

$$\langle C_l \rangle = \begin{cases} 0 & \text{if } l < 2 \\ \frac{4\pi}{(l+2)(l+1)l(l-1)} h^2 & \text{if } l \geq 2 \end{cases}. \tag{C48}$$

The statistical independence of the real and imaginary parts of $a_{lm}$ for different $l$ and $m$ imply that for the Gaussian ensemble, each of the $C_l$ is an independent random variable. These are drawn from a (central) chi-squared distribution $\chi_k^2$ with $k = 4l+2$ degrees of freedom, rescaled so that the mean is $\langle C_l \rangle$ rather than $k = 4l+2$.

Note that since we use the normalization $\mu_u(0) = 1/3$ whereas [61] have $1/2$, in Eq. (C48) we have changed their $6\pi$ to $4\pi$. We have also inserted a factor of $h^2$ to permit easy comparison with the earlier parts of this Appendix.

The Hellings and Downs correlation for any member of the ensemble is given by Eq. (19) of [61] as

$$\mu(\gamma) = \sum_{l=0}^{\infty} C_l \frac{2l+1}{4\pi} P_l(\cos\gamma), \tag{C49}$$

where $P_l(x)$ denotes the Legendre polynomial of order $l$ in $x$. In [63] it is shown that the expected value of this sum is precisely the Hellings and Downs curve:

$$\langle \mu(\gamma) \rangle = h^2 \mu_u(\gamma)$$
$$= h^2 \sum_{l=2}^{\infty} \frac{2l+1}{(l+2)(l+1)l(l-1)} P_l(\cos\gamma). \tag{C50}$$

That is to say, the Hellings-Downs curve may be expressed as this particular sum of Legendre polynomials of $\cos\gamma$.

It is also straightforward to compute the variance of the Hellings and Downs correlation from the results of [61]. Start with Eq. (C49), take the ensemble average of its square and subtract the square of its ensemble average to obtain

$$\sigma^2(\gamma) = \langle \mu^2 \rangle - \langle \mu \rangle^2$$
$$= \sum_{l,l'=0}^{\infty} \left( \langle C_l C_{l'} \rangle - \langle C_l \rangle \langle C_{l'} \rangle \right) \frac{(2l+1)(2l'+1)}{(4\pi)^2}$$
$$\times P_l(\cos\gamma) P_{l'}(\cos\gamma)$$
$$= \sum_{l=0}^{\infty} \left( \langle C_l^2 \rangle - \langle C_l \rangle^2 \right) \left( \frac{2l+1}{4\pi} \right)^2 P_l^2(\cos\gamma). \tag{C51}$$

To obtain the final line, we have used $\langle C_l C_{l'} \rangle = \langle C_l \rangle \langle C_{l'} \rangle$ for $l \neq l'$, which follows from the statistical independence of the $C_l$ for different values of $l$. Now consider the other case, for which $l = l'$. Since a $\chi_k^2$-distributed random variable has variance $2k$, after rescaling we find

$$\langle C_l^2 \rangle - \langle C_l \rangle^2 = 2k \left[ \frac{\langle C_l \rangle}{k} \right]^2 = \frac{\langle C_l \rangle^2}{2l+1}. \tag{C52}$$

The rescaling factor $\langle C_l \rangle / k$ is discussed immediately after Eq. (C48); our Eq. (C52) is consistent with Eq. (22) from [61].

The cosmic variance is obtained by substituting Eq. (C52) into Eq. (C51) and using Eq. (C48), giving

$$\sigma_{\text{cosmic}}^2(\gamma) = h^4 \sum_{l=2}^{\infty} \frac{2l+1}{(l+2)^2(l+1)^2 l^2 (l-1)^2}$$
$$\times P_l^2(\cos\gamma) = h^4 \tilde{\mu}^2(\gamma). \tag{C53}$$

While we have not given a rigorous proof, a simple numerical check (it is sufficient to include $l = 2, 3, ..., 10$ in the sum) convincingly demonstrates the final equality in Eq. (C53). Comparison of Eq. (C50) with Eq. (C18) and Eq. (C53) with Eq. (C45) shows that this is exactly the cosmic variance, with $\hbar^4 = h^4/2$ as discussed before Eq. (C30). This agreement makes sense, since as stressed earlier, the existence of the sky map is predicated on the existence of a large number of low-noise pulsars. The cosmic covariance may be found with a similar calculation, see Ref. [38].

(Note: several weeks after Eq. (C53) appeared on the arXiv the same result was derived independently in [64].)

## APPENDIX D: UNPOLARIZED POINT SOURCE MEAN $\mu_u(\gamma)$ AND VARIANCE $\sigma_u^2(\gamma)$

Here, we calculate the mean and variance of the standard "unpolarized" term in the pulsar redshift or timing residual correlation between two pulsars. The mean is the pulsar correlation for a single GW source of unit amplitude at a fixed point in the sky, averaged over all possible sky positions of the pulsar pair where their separation angle is fixed at $\gamma$. The variance characterizes the scale of the fluctuations in that correlation.

As shown in [25], and as we have also demonstrated in Appendix A, this is equivalent to fixing the pulsar pair at definite positions, and calculating the mean and variance as the location of the GW source is shifted uniformly around the sky. So this is how our calculation is structured, which is also how it was conceived by Hellings and Downs. Our calculation closely follows that given in Appendix C of Ref. [18], where the mean (but not the variance) is found. Note the sign typo one line before Eq. (C9) of Ref. [18].

Our starting point is the correlation of pulsar timing residuals Eq. (C14) for an unpolarized unit amplitude source

$$\rho(\mathbf{\Omega}) = F_1^+(\mathbf{\Omega})F_2^+(\mathbf{\Omega}) + F_1^\times(\mathbf{\Omega})F_2^\times(\mathbf{\Omega}), \qquad (D1)$$

where 1 and 2 label the pulsars at sky directions $\mathbf{p}_1$ and $\mathbf{p}_2$. Note that different authors use different choices of normalization for this expression. Ours is consistent with Hellings and Downs and corresponds to the normalization condition $\langle \rho \rangle (\gamma = 0) = \mu_u(\gamma = 0) = 1/3$.

Here, we use $\langle Q \rangle$ to denote the direction average of some functional of $\mathbf{\Omega}$, so that

$$\langle Q \rangle = \frac{1}{4\pi} \int d\mathbf{\Omega}\, Q(\mathbf{\Omega}), \qquad (D2)$$

where the normalization ensures that $\langle 1 \rangle = 1$. Hence, the Hellings-Downs function is the mean

$$\mu_u(\gamma) = \langle \rho \rangle = \frac{1}{4\pi} \int d\mathbf{\Omega} \sum_A F_1^A(\mathbf{\Omega}) F_2^A(\mathbf{\Omega}), \qquad (D3)$$

and the variance in the Hellings-Downs function is

$$\sigma_u^2(\gamma) = \langle (\Delta\rho)^2 \rangle = \langle (\rho - \langle\rho\rangle)^2 \rangle = \langle \rho^2 \rangle - \langle\rho\rangle^2, \qquad (D4)$$

where the subscript "u" means "unpolarized" and

$$\langle \rho^2 \rangle = \frac{1}{4\pi} \int d\mathbf{\Omega} \left( \sum_A F_1^A(\mathbf{\Omega}) F_2^A(\mathbf{\Omega}) \right)^2 \qquad (D5)$$

is the second moment.

To evaluate these quantities, we follow the approach of [[18], Appendix C]. We write the transverse traceless symmetric polarization tensors as

$$\begin{aligned} e_{ab}^+(\mathbf{\Omega}) &= m_a m_b - n_a n_b, \quad \text{and} \\ e_{ab}^\times(\mathbf{\Omega}) &= m_a n_b + n_a m_b, \end{aligned} \qquad (D6)$$

where $\mathbf{n}$ and $\mathbf{m}$ are a pair of orthogonal unit vectors which are both orthogonal to the propagating direction $\mathbf{\Omega}$. Note that while $e_{ab}^+$ and $e_{ab}^\times$ depend upon $\mathbf{\Omega}$, we only indicate this functional dependence when needed for clarity.

Closed forms for $\mathbf{\Omega}$, $\mathbf{n}$ and $\mathbf{m}$ are given by [17] as:

$$\begin{aligned} \mathbf{\Omega} &= \hat{\mathbf{x}} \cos\phi \sin\theta + \hat{\mathbf{y}} \sin\phi \sin\theta + \hat{\mathbf{z}} \cos\theta, \\ \mathbf{m} &= \hat{\mathbf{x}} \sin\phi \qquad\qquad - \hat{\mathbf{y}} \cos\phi, \\ \mathbf{n} &= \hat{\mathbf{x}} \cos\phi \cos\theta + \hat{\mathbf{y}} \sin\phi \cos\theta - \hat{\mathbf{z}} \sin\theta. \end{aligned} \qquad (D7)$$

The reader unfamiliar with these forms should quickly verify that the three vectors are unit length and mutually orthogonal. We note in passing that $\mathbf{m}$ and $\mathbf{n}$ could be replaced by any orthogonal linear combination

$$\begin{aligned} \mathbf{m}' &= \phantom{-}\mathbf{m} \cos\epsilon + \mathbf{n} \sin\epsilon, \quad \text{and} \\ \mathbf{n}' &= -\mathbf{m} \sin\epsilon + \mathbf{n} \cos\epsilon, \end{aligned} \qquad (D8)$$

where $\epsilon$ is an arbitrary function of $\mathbf{\Omega}$. This has the effect of "rotating" the polarization vectors into linear combinations of each other, defining a different basis for "+" and "×", but not making any difference to the treatment. The angle $\epsilon$ may vary *arbitrarily* as a function of $\mathbf{\Omega}$, because we are free to pick *any* GW polarization basis, provided that the two basis vectors are orthogonal to one another, and orthogonal to the wave propagation direction $\mathbf{\Omega}$.

Without loss of generality, let pulsars 1 and 2, separated by angle $\gamma$, have unit direction vectors

$$\mathbf{p}_1 = \hat{\mathbf{z}} \quad \text{and} \quad \mathbf{p}_2 = \hat{\mathbf{x}} \sin\gamma + \hat{\mathbf{z}} \cos\gamma. \qquad (D9)$$

Employ the definitions Eq. (A6) for $F$, Eq. (D6) for the polarization tensors, and Eq. (D7) for the vectors $\mathbf{\Omega}$, $\mathbf{m}$ and $\mathbf{n}$. For the first pulsar, one has

$$\begin{aligned} F_1^\times(\mathbf{\Omega}) &= 0, \quad \text{and} \\ F_1^+(\mathbf{\Omega}) &= \frac{1}{2} \frac{p_1^a p_1^b}{1 + \mathbf{\Omega} \cdot \mathbf{p}_1} e_{ab}^+(\mathbf{\Omega}) \\ &= \frac{1}{2} \frac{(\mathbf{p}_1 \cdot \mathbf{m})^2 - (\mathbf{p}_1 \cdot \mathbf{n})^2}{1 + \mathbf{\Omega} \cdot \mathbf{p}_1} \\ &= -\frac{1}{2} \frac{\sin^2\theta}{1 + \cos\theta} \\ &= \frac{1}{2}(\cos\theta - 1). \end{aligned} \qquad (D10)$$

For the second pulsar, the expression for $F_2^\times(\mathbf{\Omega})$ is not needed, since according to Eq. (D10) it would be multiplied by zero. The plus polarization term is

$$\begin{aligned}F_2^+(\mathbf{\Omega}) &= \frac{1}{2}\frac{(\mathbf{p}_2\cdot\mathbf{m})^2 - (\mathbf{p}_2\cdot\mathbf{n})^2}{1+\mathbf{\Omega}\cdot\mathbf{p}_2}\\ &= \frac{(\sin\gamma\sin\phi)^2 - (\sin\gamma\cos\phi\,\cos\theta - \cos\gamma\sin\theta)^2}{2(1+\sin\gamma\cos\phi\sin\theta + \cos\gamma\cos\theta)}\\ &= \frac{\sin^2\gamma\,\sin^2\phi - \sin^2\gamma\,\cos^2\phi\,\cos^2\theta - \cos^2\gamma\sin^2\theta + 2\sin\gamma\,\cos\gamma\,\sin\theta\cos\theta\cos\phi}{2(1+\sin\gamma\,\cos\phi\,\sin\theta + \cos\gamma\,\cos\theta)}.\end{aligned} \quad (D11)$$

Since the product of the $\times$ terms vanishes, the product of Eqs. (D10) and (D11) gives $\rho(\mathbf{\Omega}) = F_1^+ F_2^+$.

We want to compute the mean and variance of $\rho(\mathbf{\Omega})$. To put it into computationally tractable form, a bit of algebra is helpful. Let $q$ denote part of the expression in the denominator of $F_2^+$ in the final line of Eq. (D11), namely

$$q = \sin\gamma\cos\phi\sin\theta + \cos\gamma\cos\theta. \quad (D12)$$

The product $(1+q)(q-1)$ can be simplified to

$$\begin{aligned}(1+q)(q-1) = &-\sin^2\gamma\sin^2\phi - \sin^2\gamma\cos^2\phi\cos^2\theta\\ &- \cos^2\gamma\sin^2\theta\\ &+ 2\sin\gamma\,\cos\gamma\,\sin\theta\cos\theta\cos\phi,\end{aligned} \quad (D13)$$

which is identical to the numerator that appears in the final line of Eq. (D11), apart from the sign of the first term, $\sin^2\gamma\sin^2\phi$. Hence, by adding twice this term, we can write

$$\begin{aligned}\rho(\mathbf{\Omega}) &= F_1^+ F_2^+\\ &= \frac{1}{2}(\cos\theta - 1)\frac{(1+q)(q-1) + 2\sin^2\gamma\,\sin^2\phi}{2(1+q)}\\ &= \frac{1}{4}(\cos\theta - 1)\left[q - 1 + \frac{2\sin^2\gamma\,\sin^2\phi}{1+q}\right]. \quad (D14)\end{aligned}$$

We now proceed to the integrals.

Our goal is to average $\rho(\mathbf{\Omega})$ and its square over the GW directions. In that averaging, we'll do the integral with respect to $\phi$ first, so it is helpful to show the $\phi$ dependence explicitly. For this, we write

$$\rho(\mathbf{\Omega}) = \frac{1}{4}\left[u + v\cos\phi + w\frac{\sin^2\phi}{r + s\cos\phi}\right], \quad (D15)$$

where the quantities $u$, $v$, $w$, $r$ and $s$ are independent of $\phi$. From Eq. (D14) and the definition of $q$, we read off:

$$\begin{aligned}u &= (\cos\theta - 1)(\cos\gamma\cos\theta - 1),\\ v &= (\cos\theta - 1)\sin\gamma\sin\theta,\\ w &= (\cos\theta - 1)2\sin^2\gamma,\\ r &= 1 + \cos\gamma\cos\theta, \quad \text{and}\\ s &= \sin\gamma\sin\theta. \quad (D16)\end{aligned}$$

In integrating $\rho(\mathbf{\Omega})$ and $\rho^2(\mathbf{\Omega})$ with respect to $\phi$, we can now see that there are six different types of integrals which appear.

Three are trivial: the integral of a constant, of $\cos\phi$ and of $\cos^2\phi$. The three nontrivial integrals are

$$\begin{aligned}I_1(\theta) &= \int_0^{2\pi}\frac{\sin^2\phi}{r+s\cos\phi}d\phi,\\ I_2(\theta) &= \int_0^{2\pi}\frac{\sin^2\phi\cos\phi}{r+s\cos\phi}d\phi, \quad \text{and}\\ I_3(\theta) &= \int_0^{2\pi}\frac{\sin^4\phi}{(r+s\cos\phi)^2}d\phi. \quad (D17)\end{aligned}$$

For reasons that will be clear later, we indicate their dependence on $\theta$, but not the dependence on $\gamma$. Fortunately, all three of these integrals can be easily evaluated using the same contour integration technique. (The reader who is unfamiliar with these methods is advised to first study the textbook Example 3-38 from [65].) We first carry out this evaluation and then return to the main calculation.

To evaluate the integrals $I_{1,2,3}$, we make the change of variable $z = e^{i\phi}$ and write them as contour integrals along the closed path shown in Fig. 10, which loops once around the origin in the counterclockwise direction, at unit distance

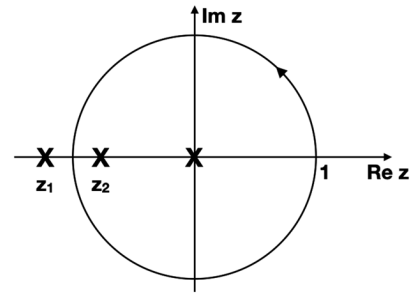

FIG. 10. The contour of integration for the integrals in Eq. (D20) goes around the unit circle once in the counterclockwise direction. The integrands have poles along the negative real axis at $z = z_1$ and $z = z_2$, and a third pole at the origin. If $\theta < \pi - \gamma$, then $z_1$ lies within the circle, and $z_2$ is outside. We have depicted the other case: if $\theta > \pi - \gamma$, then $z_2$ is within and $z_1$ is outside. The residues of the poles are listed in Table I.

TABLE I. Residues of integrals in Eq. (D20), *without* the factors that appear before the integral symbol. Note that the final row can be obtained from the previous row by swapping $z_1$ and $z_2$.

| Location | Name | Value for $I_1$ | Value for $I_2$ | Value for $I_3$ |
|---|---|---|---|---|
| $z=0$ | $R_0$ | $\frac{z_1+z_2}{z_1^2 z_2^2}$ | $-\frac{z_1^2 z_2^2 - z_1^2 - z_1 z_2 - z_2^2}{z_1^3 z_2^3}$ | $-\frac{4z_1^2 z_2^2 - 3z_1^2 - 4z_1 z_2 - 3z_2^2}{z_1^4 z_2^4}$ |
| $z=z_1$ | $R_1$ | $\frac{(z_1-1)^2(z_1+1)^2}{z_1^2(z_1-z_2)}$ | $\frac{(z_1-1)^2(z_1+1)^2(z_1^2+1)}{z_1^3(z_1-z_2)}$ | $\frac{(z_1-1)^3(z_1+1)^3(3z_1^3-5z_1^2 z_2+5z_1-3z_2)}{z_1^4(z_1-z_2)^3}$ |
| $z=z_2$ | $R_2$ | $-\frac{(z_2-1)^2(z_2+1)^2}{z_2^2(z_1-z_2)}$ | $-\frac{(z_2-1)^2(z_2+1)^2(z_2^2+1)}{z_2^3(z_1-z_2)}$ | $\frac{(z_2-1)^3(z_2+1)^3(-5z_1 z_2^2-3z_1+3z_2^2+5z_2)}{z_2^4(z_2-z_1)^3}$ |

from the origin. Since $\cos\phi = (z+1/z)/2$, one can write the denominators as

$$r + s\cos\phi = \frac{\sin\theta\sin\gamma}{2z}(z-z_1)(z-z_2), \quad (D18)$$

or as the square of this quantity. The two roots are found by substituting $r$ and $s$ into the lhs, rewriting $\cos\phi$ in terms of $z$, and factoring the quadratic. Both roots lie on the negative real axis, at locations

$$z_1 = -\frac{(1-\cos\gamma)(1-\cos\theta)}{\sin\gamma\sin\theta}, \quad \text{and}$$

$$z_2 = -\frac{(1+\cos\gamma)(1+\cos\theta)}{\sin\gamma\sin\theta}. \quad (D19)$$

Making use of $\sin\phi = (z-1/z)/2i$ and $d\phi = -idz/z$, one has:

$$I_1 = \frac{i}{2\sin\theta\sin\gamma}\oint dz\frac{(z-1/z)^2}{(z-z_1)(z-z_2)},$$

$$I_2 = \frac{i}{4\sin\theta\sin\gamma}\oint dz\frac{(z-1/z)^2(z+1/z)}{(z-z_1)(z-z_2)}, \quad \text{and}$$

$$I_3 = \frac{-i}{4\sin^2\theta\sin^2\gamma}\oint dz\frac{z(z-1/z)^4}{(z-z_1)^2(z-z_2)^2}. \quad (D20)$$

The path of integration is shown in Fig. 10. Note that the locations of the poles depend upon $\theta$ and $\gamma$. If $\theta < \pi - \gamma$, then $z_1$ lies within the unit circle, and $z_2$ is outside, otherwise $z_2$ is within the circle and $z_1$ is outside.

All three integrands are ratios of polynomials, so are analytic on the entire plane apart from the poles at $z=0$, $z=z_1$ and $z=z_2$. There are no branch points, branch cuts, essential singularities, or other complications. By the residue theorem, the integrals are just $2\pi i$ times the sum of the residues within the unit circle. These residues are computed by expanding the integrands in a Laurent series at the three poles; we have listed them in Table I. In effect, these integrals are evaluated by taking derivatives to find the residues.

The first integral is evaluated as follows. For $\theta < \pi - \gamma$, there are contributions from the poles at $z=0$ and $z=z_1$, and one obtains

$$I_{1_<} = \frac{i(2\pi i)(R_0 + R_1)}{2\sin\theta\sin\gamma} = \frac{2\pi}{(1+\cos\gamma)(1+\cos\theta)}, \quad (D21)$$

whereas for $\theta > \pi - \gamma$, there are contributions from the poles at $z=0$ and $z=z_2$, and one obtains

$$I_{1_>} = \frac{i(2\pi i)(R_0 + R_2)}{2\sin\theta\sin\gamma} = \frac{2\pi}{(1-\cos\gamma)(1-\cos\theta)}. \quad (D22)$$

Here, $R_0$, $R_1$, and $R_2$ are the appropriate residues from Table I, and we have added "<" and ">" as subscripts to $I$ to indicate the range of $\theta$.

The behavior is similar for the remaining two integrals. For $\theta < \pi - \gamma$, the second integral is

$$I_{2_<} = \frac{i(2\pi i)(R_0 + R_1)}{4\sin\theta\sin\gamma}$$

$$= \frac{-\pi(1-\cos\gamma)(1-\cos\theta)}{\sin\gamma\sin\theta(1+\cos\gamma)(1+\cos\theta)}, \quad (D23)$$

and for $\theta > \pi - \gamma$, we obtain

$$I_{2_>} = \frac{i(2\pi i)(R_0 + R_2)}{4\sin\theta\sin\gamma}$$

$$= \frac{-\pi(1+\cos\gamma)(1+\cos\theta)}{\sin\gamma\sin\theta(1-\cos\gamma)(1-\cos\theta)}. \quad (D24)$$

For $\theta < \pi - \gamma$, the third integral evaluates to

$$I_{3_<} = \frac{-i(2\pi i)(R_0 + R_1)}{4\sin^2\theta\sin^2\gamma}$$

$$= \frac{3\pi}{(1+\cos\gamma)^2(1+\cos\theta)^2}, \quad (D25)$$

and for $\theta > \pi - \gamma$, it is

$$I_{3_>} = \frac{-i(2\pi i)(R_0 + R_2)}{4\sin^2\theta\sin^2\gamma}$$

$$= \frac{3\pi}{(1-\cos\gamma)^2(1-\cos\theta)^2}. \quad (D26)$$

With these three integrals in hand, only integration over $\theta$ remains.

We begin by computing the Hellings-Downs function $\langle\rho\rangle$, starting from Eqs. (D2) and (D15), which give

$$\langle\rho\rangle = \frac{1}{16\pi}\int d\Omega\left[u + v\cos\phi + w\frac{\sin^2\phi}{r+s\cos\phi}\right]. \quad (D27)$$

The integral over $\phi$ gives

$$\langle\rho\rangle = \frac{1}{16\pi}\int_0^\pi d\theta\,\sin\theta\left[2\pi u + wI_1(\theta)\right], \quad (D28)$$

where the $v\cos\phi$ term has integrated to zero, $I_1$ is defined by Eq. (D17) and $u$ and $w$ are given by Eq. (D16). The integral of $u\sin\theta$ is easily evaluated, but because of the different functional dependencies in Eqs. (D21) and (D22), the remaining integral must be split into two ranges, giving

$$\begin{aligned}\mu_u(\gamma) = \langle \rho \rangle &= \frac{1}{4}\left[1 + \frac{\cos\gamma}{3} + \frac{1}{4\pi}\left(\int_0^{\pi-\gamma} d\theta \sin\theta \, w I_{1_<}(\theta) + \int_{\pi-\gamma}^{\pi} d\theta \sin\theta \, w I_{1_>}(\theta)\right)\right] \\ &= \frac{1}{4}\left[1 + \frac{1}{3}\cos\gamma + (1-\cos\gamma)\left(1+\cos\gamma + 2\log\left(\frac{1-\cos\gamma}{2}\right)\right) - \sin^2\gamma\right] \\ &= \frac{1}{4} + \frac{1}{12}\cos\gamma + \frac{1}{2}(1-\cos\gamma)\log\left(\frac{1-\cos\gamma}{2}\right). \end{aligned} \quad (D29)$$

This is the standard Hellings-Downs curve. As $\gamma \to 0$, this expression approaches $1/3$, which agrees with the Hellings-Downs normalization (Eq. 5 in Ref. [11]) but note that many authors normalize it to $1/2$ at zero angle.

Now we proceed to the second moment and variance. To simplify the notation in what follows, we define the function

$$\begin{aligned}\Psi(\gamma) &= (1-\cos\gamma)\log\left(\frac{1-\cos\gamma}{2}\right) \\ &= 4\sin^2(\gamma/2)\log\left(\sin(\gamma/2)\right). \end{aligned} \quad (D30)$$

To calculate $\langle \rho^2 \rangle$, we begin with Eqs. (D2) and (D15), squaring the latter to obtain

$$\begin{aligned}\langle \rho^2 \rangle &= \frac{1}{64\pi}\int d\Omega\left[u^2 + 2uv\cos\phi + v^2\cos^2\phi + 2uw\frac{\sin^2\phi}{r+s\cos\phi} + 2vw\frac{\cos\phi\sin^2\phi}{r+s\cos\phi} + w^2\frac{\sin^4\phi}{(r+s\cos\phi)^2}\right] \\ &= \frac{1}{64\pi}\int_0^\pi d\theta \sin\theta\left[2\pi u^2 + \pi v^2 + 2uw I_1(\theta) + 2vw I_2(\theta) + w^2 I_3(\theta)\right], \end{aligned} \quad (D31)$$

where the functions $u$, $v$ and $w$ are defined in Eq. (D16). On the second line, we integrated over $\phi$, making use of the functions defined in Eq. (D17).

The $\theta$ integration of Eq. (D31) is straightforward. We can integrate the $u^2$ and $v^2$ terms immediately, obtaining

$$\frac{13}{120} + \frac{1}{12}\cos\gamma + \frac{1}{120}\cos^2\gamma. \quad (D32)$$

The remaining three integrals in Eq. (D31), containing $I_1$, $I_2$, and $I_3$ respectively, are computed in the same way as $\langle \rho \rangle$ was computed in Eq. (D29). Each $\theta$ integral is the sum of an integral from 0 to $\pi-\gamma$ and an integral from $\pi-\gamma$ to $\pi$, and may be evaluated using the explicit forms in Eqs. (D21)–(D26). The logarithmic divergences around $\pi-\gamma$ cancel, and the three integrals simplify to

$$I_1: \frac{1}{24}(1-\cos^2\gamma)(3+6\cos\gamma+\cos^2\gamma)$$
$$+\frac{1}{2}(1+\cos\gamma)\Psi(\gamma), \quad (D33)$$

$$I_2: \frac{1}{24}(1-\cos\gamma)^2(7+8\cos\gamma+\cos^2\gamma)$$
$$+\frac{1}{2}(1-\cos\gamma)\Psi(\gamma), \quad \text{and} \quad (D34)$$

$$I_3: \frac{3}{4}(1-\cos\gamma)\Big(1+\cos\gamma+\Psi(\gamma)\Big). \quad (D35)$$

Summing together Eqs. (D32)–(D35) yields

$$\langle \rho^2 \rangle = \frac{51}{40} + \frac{1}{12}\cos\gamma - \frac{139}{120}\cos^2\gamma + \frac{1}{4}(7-3\cos\gamma)\Psi(\gamma). \quad (D36)$$

This completes the $\theta$ integration in Eq. (D31).

From the first and second moments of $\rho$ in Eqs. (D29) and (D36), we can compute the variance in the unpolarized term, which is

$$\begin{aligned}\sigma_u^2(\gamma) = \langle \Delta\rho^2 \rangle &= \langle \rho^2 \rangle - \langle \rho \rangle^2 \\ &= \frac{97}{80} + \frac{1}{24}\cos\gamma - \frac{839}{720}\cos^2\gamma \\ &+ \frac{1}{12}\Big(18 - 10\cos\gamma - 3\Psi(\gamma)\Big)\Psi(\gamma). \quad (D37)\end{aligned}$$

We note that cubic and higher moments of $\rho$ could be computed with the same methods.

## APPENDIX E: POLARIZED POINT SOURCE MEAN AND VARIANCE $\sigma_p^2(\gamma)$

Here, we calculate the mean and variance of the polarized correlation term in the Hellings-Downs correlation

$$\rho(\mathbf{\Omega}) = F_1^+(\mathbf{\Omega})F_2^\times(\mathbf{\Omega}) - F_1^\times(\mathbf{\Omega})F_2^+(\mathbf{\Omega}). \quad (E1)$$

The formalism and notation are the same as those of Appendix D, apart from using the subscript "p" for "polarized." We will see that the mean value $\mu_p(\gamma) = \langle \rho \rangle$ vanishes so that the variance is equal to the second moment $\sigma_p^2(\gamma) = \langle \rho^2 \rangle$.

Following the same steps as in Appendix D, we arrive at

$$\rho(\mathbf{\Omega}) = \frac{1}{4}\left[\frac{\bar{u}\sin\phi + \bar{v}\sin\phi\cos\phi}{r + s\cos\phi}\right], \quad (E2)$$

which should be compared with Eq. (D15). Here, we have defined:

$$\bar{u} = -2(\cos\theta - 1)\sin\gamma\cos\gamma\sin\theta,$$
$$\bar{v} = 2(\cos\theta - 1)\sin^2\gamma\cos\theta,$$
$$r = 1 + \cos\gamma\cos\theta, \quad \text{and}$$
$$s = \sin\gamma\sin\theta. \quad (E3)$$

Note that the quantity in square brackets in Eq. (E2) is precisely the quantity $UW = (1 - \cos\theta)^2 AB/(1+q)$ from Eq. (A18), whose square appears as the polarization term $\langle U^2 W^2 \rangle$ in the variance Eq. (A28).

The integral of Eq. (E2) over $\phi \in [0, 2\pi]$ vanishes, because the integrand changes sign under reflection about $\phi = \pi$. This shows that the mean $\langle \rho \rangle$ vanishes, so this term does not contribute to the mean correlation.

To calculate the variance $\langle \rho^2 \rangle$ of this polarization term, three integrals are needed:

$$I_4(\theta) = \int_0^{2\pi} \frac{\sin^2\phi}{(r + s\cos\phi)^2} d\phi,$$
$$I_5(\theta) = \int_0^{2\pi} \frac{\sin^2\phi \cos\phi}{(r + s\cos\phi)^2} d\phi, \quad \text{and}$$
$$I_6(\theta) = \int_0^{2\pi} \frac{\sin^2\phi \cos^2\phi}{(r + s\cos\phi)^2} d\phi. \quad (E4)$$

Following exactly the same methods as used in Appendix D starting from Eq. (D17), these evaluate to:

$$I_4(\theta) = \frac{\mp 2\pi}{(\cos\gamma + \cos\theta)(1 \mp \cos\gamma)(1 \mp \cos\theta)},$$
$$I_5(\theta) = \frac{\pm 2\pi(1 \pm \cos\gamma)(1 \pm \cos\theta)}{(\cos\gamma + \cos\theta)\sin\gamma\sin\theta(1 \mp \cos\gamma)(1 \mp \cos\theta)},$$
$$I_6(\theta) = \frac{\mp 2\pi}{(\cos\gamma + \cos\theta)(1 \mp \cos\gamma)(1 \mp \cos\theta)}$$
$$- \frac{\mp 3\pi}{(1 \mp \cos\gamma)^2(1 \mp \cos\theta)^2}. \quad (E5)$$

Here, the upper sign is for $\theta > \pi - \gamma$ and the lower sign is for $\theta < \pi - \gamma$.

The integrals needed to evaluate the variance are

$$\frac{1}{4\pi}\int_0^{\pi-\gamma} d\theta \sin\theta\left[\bar{u}^2 I_4(\theta) + 2\bar{u}\,\bar{v}\, I_5(\theta) + \bar{v}^2 I_6(\theta)\right]$$
$$= \frac{1}{3}\left(\cos^2\gamma - 1\right)\left(\cos^3\gamma + 3\cos^2\gamma + 58\right)$$
$$+ 4\left(3\cos\gamma - 7\right)\left(1 - \cos\gamma\right)\log\frac{1 - \cos\gamma}{2} \quad (E6)$$

and

$$\frac{1}{4\pi}\int_{\pi-\gamma}^{\pi} d\theta \sin\theta\left[\bar{u}^2 I_4(\theta) + 2\bar{u}\,\bar{v}\, I_5(\theta) + \bar{v}^2 I_6(\theta)\right]$$
$$= \frac{1}{3}\left(1 - \cos^2\gamma\right)\left(\cos^3\gamma + 3\cos^2\gamma + 2\right). \quad (E7)$$

Note that in Appendix D, the integrals for $I_1$, $I_2$ and $I_3$ were carried out separately. That is not possible here, because the integrands formed from $I_4$, $I_5$ and $I_6$ have a pole at $\theta = \pi - \gamma$, which leads to a logarithmic divergence if they are integrated individually. However, this divergence cancels when the three terms are summed, so we have evaluated them together.

Adding together Eqs. (E6) and (E7) and reinserting the factor of $(1/4)^2$ from Eq. (E2), we obtain the second moment $\langle \rho^2 \rangle$. Since the mean $\langle \rho \rangle$ is zero, the second moment is also the variance. Thus, we obtain

$$\sigma_p^2(\gamma) = \langle \rho^2 \rangle$$
$$= \frac{7}{6}\left(\cos^2\gamma - 1\right) + \frac{1}{4}\left(3\cos\gamma - 7\right)\left(1 - \cos\gamma\right)$$
$$\times \log\left(\frac{1 - \cos\gamma}{2}\right) \quad (E8)$$

for the variance of the polarized term in the Hellings-Downs correlation.

The square root of this quantity (the standard deviation) is shown in Fig. 11, where it is compared with that of the unpolarized Hellings-Downs expressions in Eqs. (2.4) and (2.6).

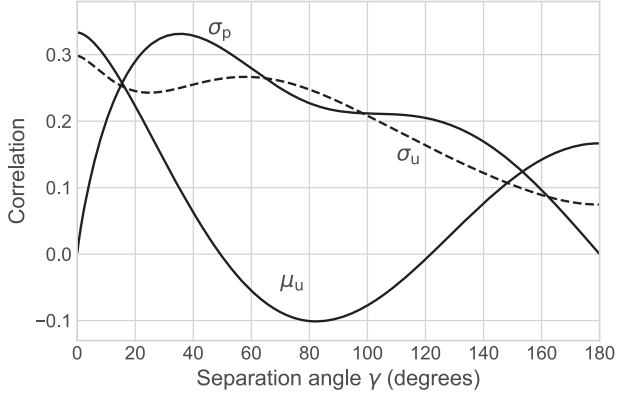

FIG. 11. The standard deviation $\sigma_p(\gamma)$ of the polarization term in the correlation. The polarization term has zero mean, so does not affect the expectation for the correlation $\mu(\gamma)$. For comparison the dashed curve is the standard deviation for the unpolarized Hellings-Downs term.

## APPENDIX F: CROSS-POLARIZED VARIANCE $\sigma_c^2(\gamma)$

The final quantity needed to describe the variance of the Hellings-Downs correlation for a general polarized GW source is the average value of the cross term

$$\left(F_1^+ F_2^+ + F_1^\times F_2^\times\right)^2 + \left(F_1^+ F_2^\times - F_1^\times F_2^+\right)^2$$
$$= \left(F_1^+ F_1^+ + F_1^\times F_1^\times\right)\left(F_2^+ F_2^+ + F_2^\times F_2^\times\right). \quad (F1)$$

In the final line, the terms associated with the first pulsar are collected on the left, and those with the second pulsar on the right, which is particularly convenient for computation.

Starting from the final line of Eq. (F1), the cross term in the variance is defined as the sky average

$$\sigma_c^2(\gamma) = \frac{1}{4\pi} \int d\Omega \left(F_1^+(\Omega)F_1^+(\Omega) + F_1^\times(\Omega)F_1^\times(\Omega)\right)\left(F_2^+(\Omega)F_2^+(\Omega) + F_2^\times(\Omega)F_2^\times(\Omega)\right)$$
$$= \frac{1}{4\pi} \int d\Omega \left(\frac{1-\cos\theta}{2}\right)^2 \left(\frac{1-\sin\gamma\cos\phi\sin\theta-\cos\gamma\cos\theta}{2}\right)^2$$
$$= \frac{1}{32} \int_0^\pi d\theta \sin\theta(1-\cos\theta)^2\left((1-\cos\gamma\cos\theta)^2 + \frac{1}{2}\sin^2\gamma\sin^2\theta\right)$$
$$= \frac{13}{120} + \frac{1}{12}\cos\gamma + \frac{1}{120}\cos^2\gamma. \quad (F2)$$

Note that the second line follows directly from the equality

$$(\mathbf{p}\cdot\mathbf{m})^2 + (\mathbf{p}\cdot\mathbf{n})^2 = 1 - (\mathbf{p}\cdot\boldsymbol{\Omega})^2 = (1+\mathbf{p}\cdot\boldsymbol{\Omega})(1-\mathbf{p}\cdot\boldsymbol{\Omega}), \quad (F3)$$

which holds for the unit vectors pointing to either pulsar, since $\mathbf{m}$, $\mathbf{n}$ and $\boldsymbol{\Omega}$ form an orthonormal basis. The third line in Eq. (F2) is obtained by integrating $\phi \in [0, 2\pi]$ and the last line from evaluating the integral over $\theta$.

A helpful consistency check follows from the first line of Eq. (F1), which implies that

$$\sigma_c^2(\gamma) = \mu_u^2(\gamma) + \sigma_u^2(\gamma) + \sigma_p^2(\gamma). \quad (F4)$$

The first two terms on the rhs of Eq. (F4) are given by Eq. (D36). The third term on the rhs of Eq. (F4) is given by Eq. (E8). Adding them together correctly reproduces the final line of Eq. (F2).

## APPENDIX G: THE TWO-POINT HELLINGS-DOWNS FUNCTION $\mu(\gamma,\beta)$ AND ITS AVERAGES $\tilde{\mu}(\gamma)$ AND $\tilde{\mu^2}(\gamma)$

Here, we calculate the two-point Hellings-Downs function $\mu(\gamma,\beta)$ and its second moment with respect to the angular separation $\beta$ between GW sources. The function is a generalization of the Hellings-Downs correlation, giving the average cross correlation produced by two unpolarized coherent GW sources [66] at different locations on the sky. It is defined by

$$\mu(\gamma,\beta) = \langle F_1^+(\boldsymbol{\Omega}_A)F_2^+(\boldsymbol{\Omega}_B) + F_1^\times(\boldsymbol{\Omega}_A)F_2^\times(\boldsymbol{\Omega}_B)\rangle, \quad (G1)$$

where the average is over all pulsar directions $\mathbf{p}_1$ and $\mathbf{p}_2$ separated by angle $\gamma$. After averaging over pulsar directions, the rhs is only a function of the two variables $\cos\gamma = \mathbf{p}_1 \cdot \mathbf{p}_2$ and $\cos\beta = \boldsymbol{\Omega}_A \cdot \boldsymbol{\Omega}_B$, where $\beta$ is the angle between the directions to the two sources $A$ and $B$.

A potentially related function of two variables, denoted by $\gamma(x, y)$, is defined in Eq. (24) and plotted in Fig. 6 of Ref. [25]. (Note that in Ref. [25], the symbol $\gamma$ is the name given to the function, and is unrelated to the angle $\gamma$ that we use in this work.) However, no closed form is given for $\gamma(x, y)$ in [25], and we have been unable to establish, based on its definition, if $\gamma(x, y)$ is related to our two-point function $\mu(x, y)$. One conjecture is that $\gamma(x, y) = c(\mu(x, y))^2$, where $c$ is a positive constant.

Our computation of $\mu(\gamma,\beta)$ employs the same techniques as previous calculations in Appendixes D and E of this paper. We locate the two pulsars following Eqs. (A12) and (A13), and place the GW sources at $\mathbf{\Omega}_A = \hat{x}\sin\beta + \hat{z}\cos\beta$ and $\mathbf{\Omega}_B = \hat{z}$ with corresponding choices for the polarization vectors $\mathbf{m}$ and $\mathbf{n}$ from Eq. (D7) and polarization tensors from Eq. (D6). The antenna pattern functions are then defined by Eq. (A6) and the combination $\rho = F_1^+(\mathbf{\Omega}_A)F_2^+(\mathbf{\Omega}_B) + F_1^\times(\mathbf{\Omega}_A)F_2^\times(\mathbf{\Omega}_B)$ is formed. Note that we have assigned the more "complicated" source (in the sense of computational complexity) to the "simpler" pulsar and vice versa.

We now need to average $\rho$ over the three variables $\theta$, $\phi$ and $\lambda$ that define the possible pulsar pairs at separation angle $\gamma$, where Eq. (A14) followed by Eq. (A15) defines the average. The first integral to perform is over $\lambda$ and is evaluated as a contour integral, defining $z = \exp(i\lambda)$ and writing $\rho$ as a function of $z$. The denominator of $\rho$ can be written as a quadratic function of $z$ using the transformations in Eqs. (D18) and (D19) with $\gamma$ replaced by $\beta$. The integrand has a fourth-order pole at the origin and first-order poles along the negative real axis, exactly as depicted in Fig. 10, provided that in the caption of that figure, $\gamma$ is replaced by $\beta$. As shown there, the sums of residues for $\theta < \pi - \beta$ and for $\theta > \pi - \beta$ are different. The sums of the residues are nonzero for $\theta < \pi - \beta$ but vanish for $\theta > \pi - \beta$. Multiplying the sum of the residues by $2\pi i$ completes the first integral.

The integral over $\phi$ is also done as a contour integral, this time with $z = \exp(i\phi)$. One can use the same transformations as in Eqs. (D18) and (D19), and now the description of Fig. 10 applies without modification. The integrand has a pole of second order at $z = 0$ and poles of first order at $z_1$ and $z_2$. After some simplification, one obtains the following quantities, which must still be integrated with respect to $\theta$:

$$I(\theta,\gamma,\beta) = \begin{cases} 0 & \text{for } \theta > \pi - \beta \\ I_<(\theta,\gamma,\beta) = \dfrac{(\cos\theta - 1)^2(\cos\theta + \cos\beta)(\cos\theta\cos\gamma + 2\cos\gamma - 1)}{(1+\cos\theta)^2(1+\cos\beta)^2} & \text{for } \theta \leq \pi - \beta \text{ and } \theta < \pi - \gamma \\ I_>(\theta,\gamma,\beta) = \dfrac{(\cos\theta + \cos\beta)(\cos\theta\cos\gamma - 2\cos\gamma - 1)}{(1+\cos\beta)^2} & \text{for } \theta \leq \pi - \beta \text{ and } \theta > \pi - \gamma. \end{cases} \quad (G2)$$

Note that (because of the integral over $\lambda$) the integrand vanishes if $\theta > \pi - \beta$.

Integration of these expressions then yields $\mu(\gamma,\beta)$. For $\beta < \gamma$, one has

$$\mu(\gamma,\beta) = \frac{1}{4}\int_0^{\pi-\gamma} d\theta \sin\theta \, I_<(\theta,\gamma,\beta) + \frac{1}{4}\int_{\pi-\gamma}^{\pi-\beta} d\theta \sin\theta \, I_>(\theta,\gamma,\beta) \tag{G3}$$

and for $\beta > \gamma$, one has

$$\mu(\gamma,\beta) = \frac{1}{4}\int_0^{\pi-\beta} d\theta \sin\theta \, I_<(\theta,\gamma,\beta). \tag{G4}$$

These integrals evaluate to

$$\mu(\gamma,\beta) = \begin{cases} \frac{1}{48}\left(33 - 18\cos\beta - 3\cos^2\beta + (32 - 21\cos\beta - 6\cos^2\beta - \cos^3\beta)\cos\gamma\right)\sec^4\left(\frac{\beta}{2}\right) \\ + \left(1 - \frac{1}{2}\cos\beta - \frac{1}{2}\cos\gamma\right)\sec^4\left(\frac{\beta}{2}\right)\log\left(\sin^2\left(\frac{\gamma}{2}\right)\right) & \text{for } \beta < \gamma, \text{ and} \\ \frac{1}{24}\left(33 - 3\cos\beta - (16 + 5\cos\beta + \cos^2\beta)\cos\gamma\right)\sec^2\left(\frac{\beta}{2}\right) \\ + \left(1 - \frac{1}{2}\cos\beta - \frac{1}{2}\cos\gamma\right)\sec^4\left(\frac{\beta}{2}\right)\log\left(\sin^2\left(\frac{\beta}{2}\right)\right) & \text{for } \beta > \gamma, \end{cases} \quad (G5)$$

which is shown in Fig. 12. The function $\mu(\gamma,\beta)$ reduces to the normal Hellings-Downs function in the limit $\beta \to 0$, where the two GW sources are at the same sky location.

In addition to the two-point function defined by the sum in Eq. (G1) we also require the two-point functions for the individual polarizations. These are defined by

$$\mu_{++}(\gamma,\beta) = \langle F_1^+(\mathbf{\Omega}_A)F_2^+(\mathbf{\Omega}_B)\rangle, \quad \text{and}$$
$$\mu_{\times\times}(\gamma,\beta) = \langle F_1^\times(\mathbf{\Omega}_A)F_2^\times(\mathbf{\Omega}_B)\rangle, \tag{G6}$$

where the averages are over all pulsar pairs whose directions are separated by angle $\gamma$, and $\beta$ is the angle between the source directions $\mathbf{\Omega}_A$ and $\mathbf{\Omega}_B$. It follows immediately from Eqs. (G1) and (G6) that

$$\mu(\gamma,\beta) = \mu_{++}(\gamma,\beta) + \mu_{\times\times}(\gamma,\beta). \tag{G7}$$

With some simple symmetry arguments, we can obtain both $\mu_{++}$ and $\mu_{\times\times}$ from $\mu$.

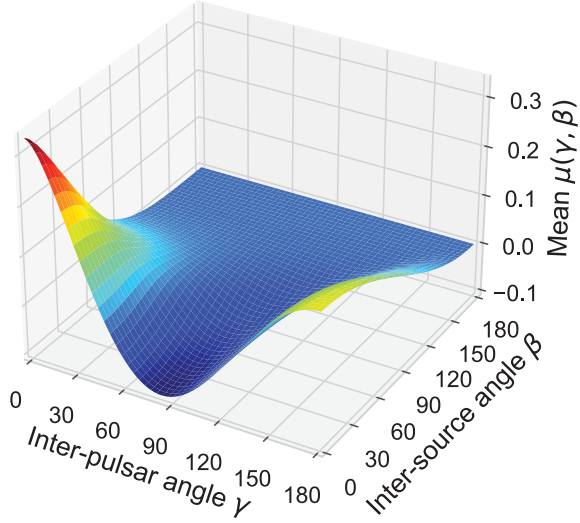

FIG. 12. The two-point function $\mu(\gamma, \beta)$ is given by Eq. (G5). It is the mean correlation between two pulsars separated by angle $\gamma$ arising from two coherent GW point sources separated by angle $\beta$. When $\beta = 0$, it reduces to the normal Hellings-Downs correlation $\mu_u(\gamma)$.

Consider the effect on these three functions of changing the sign of the direction to pulsar 1, and simultaneously changing the sign of the direction to the source point $\mathbf{\Omega}_A$. This involution transformation moves one pulsar and one source to their antipodal points on the celestial sphere. It corresponds to shifting $\gamma \to \pi - \gamma$ and $\beta \to \pi - \beta$, which reflects both angles about $\pi/2$. How does this transformation affect the two-point functions?

Under this transformation, the polarization basis vectors corresponding to the source direction $\mathbf{\Omega}_A$ transform as $\mathbf{m} \to -\mathbf{m}$ and $\mathbf{n} \to \mathbf{n}$. To see this, change $\mathbf{\Omega}$ to $-\mathbf{\Omega}$ via $\phi \to \phi + \pi$ and $\theta \to \pi - \theta$ in Eq. (D7). This means that the polarization basis tensors defined in Eq. (D6) transform as $e_{ab}^+(\mathbf{\Omega}_A) \to e_{ab}^+(\mathbf{\Omega}_A)$ and $e_{ab}^\times(\mathbf{\Omega}_A) \to -e_{ab}^\times(\mathbf{\Omega}_A)$. Since both $\mathbf{p}_1$ and $\mathbf{\Omega}_A$ have changed sign, the quantity $1 + \mathbf{p}_1 \cdot \mathbf{\Omega}_A$ is invariant. Hence, the antenna pattern functions defined by Eq. (A6) transform as $F_1^+(\mathbf{\Omega}_A) \to F_1^+(\mathbf{\Omega}_A)$ and $F_1^\times(\mathbf{\Omega}_A) \to -F_1^\times(\mathbf{\Omega}_A)$. Referring back to the two-point function definitions Eq. (G6) this implies that

$$\mu_{++}(\gamma,\beta) = \mu_{++}(\pi-\gamma,\pi-\beta), \quad \text{and}$$
$$\mu_{\times\times}(\gamma,\beta) = -\mu_{\times\times}(\pi-\gamma,\pi-\beta), \quad (G8)$$

since the antenna pattern functions at the second pulsar are unaffected by the transformation.

These simple transformation rules make it straightforward to obtain both $\mu_{\times\times}$ and $\mu_{++}$ from $\mu$. It follows immediately from Eqs. (G7) and (G8) that the two-point functions for the individual polarizations may be obtained from Eq. (G5) as

$$\mu_{++}(\gamma,\beta) = \frac{1}{2}\Big(\mu(\gamma,\beta) + \mu(\pi-\gamma,\pi-\beta)\Big), \quad \text{and}$$
$$\mu_{\times\times}(\gamma,\beta) = \frac{1}{2}\Big(\mu(\gamma,\beta) - \mu(\pi-\gamma,\pi-\beta)\Big). \quad (G9)$$

If desired, these relations may be used to obtain explicit closed forms analogous to Eq. (G5).

Similar symmetry arguments show that if we average uniformly over all pulsar locations at fixed angular separation $\gamma$, the products that involve one plus and one cross polarization vanish:

$$\mu_{+\times}(\gamma,\beta) = \langle F_1^+(\mathbf{\Omega}_A)F_2^\times(\mathbf{\Omega}_B)\rangle = 0, \quad \text{and}$$
$$\mu_{\times+}(\gamma,\beta) = \langle F_1^\times(\mathbf{\Omega}_A)F_2^+(\mathbf{\Omega}_B)\rangle = 0. \quad (G10)$$

The argument is straightforward: shift the pulsar locations $\mathbf{p}_1$ and $\mathbf{p}_2$ to $-\mathbf{p}_1$ and $-\mathbf{p}_2$, and shift the source locations from $\mathbf{\Omega}_A$ and $\mathbf{\Omega}_B$ to $-\mathbf{\Omega}_A$ and $-\mathbf{\Omega}_B$. This cannot affect the pulsar average, since $\beta$ and $\gamma$ are unchanged. But under this transformation $F^+$ is invariant and $F^\times$ changes sign. Thus, the average value of the product must vanish, since zero is the only value which is invariant under a change of sign.

We stress that the relationships expressed in Eqs. (G9) and (G10) have nothing to do with the GW sources, which might all be highly polarized and not symmetrically placed or distributed. They are consequences of the pulsar averaging, which erases these terms for *any* configuration of sources.

In Sec. IV, the second moment of $\mu(\gamma,\beta)$ with respect to $\beta$ is needed to compute the cosmic variance in the confusion-noise limit. This moment is an average over all separation angles $\beta_{jk}$ between the source directions, with $\cos\beta_{jk}$ uniformly distributed on $[-1, 1]$. We indicate this average with an over-tilde, so the second moment is [67]

$$\tilde{\mu^2}(\gamma) = \frac{1}{2}\int_0^\pi d\beta\, \sin\beta\, \mu^2(\gamma,\beta)$$
$$= -\frac{5}{48} + \frac{49}{432}\cos^2\gamma - \frac{1}{6}\left(\cos^2\gamma + 3\right)\log\left(\frac{1-\cos\gamma}{2}\right)\log\left(\frac{1+\cos\gamma}{2}\right)$$
$$+ \frac{1}{12}(\cos\gamma - 1)(\cos\gamma + 3)\log\left(\frac{1-\cos\gamma}{2}\right) + \frac{1}{12}(\cos\gamma + 1)(\cos\gamma - 3)\log\left(\frac{1+\cos\gamma}{2}\right), \quad (G11)$$

which is illustrated in Fig. 4. The second moment $\tilde{\mu}^2(\gamma)$ is symmetric under reflection about $\gamma = \pi/2$, which is equivalent to $\gamma \to \pi - \gamma$ or $\cos\gamma \to -\cos\gamma$.

For computing the cosmic variance, we also need the spherical average of $\mu_{++}^2 + \mu_{\times\times}^2$. This is easily obtained from the previous integral, since

$$\frac{1}{2}\int_0^\pi d\beta \, \sin\beta \left[\mu_{++}^2(\gamma,\beta) + \mu_{\times\times}^2(\gamma,\beta)\right]$$
$$= \frac{1}{4}\int_0^\pi d\beta \, \sin\beta \left[\mu^2(\gamma,\beta) + \mu^2(\pi-\gamma,\pi-\beta)\right]$$
$$= \frac{1}{4}\int_0^\pi d\beta \, \sin\beta \left[\mu^2(\gamma,\beta) + \mu^2(\pi-\gamma,\beta)\right]$$
$$= \frac{1}{2}\left[\tilde{\mu}^2(\gamma) + \tilde{\mu}^2(\pi-\gamma)\right] = \tilde{\mu}^2(\gamma). \quad \text{(G12)}$$

On the second line of Eq. (G12) we used Eq. (G9) to write $\mu_{++}$ and $\mu_{\times\times}$ in terms of $\mu$, since the cross terms cancel. The third line follows by changing variables in the second term to $u = \pi - \beta$ and using $\sin\beta = \sin u$ and $d\beta = -du$. The fourth line comes from the definition Eq. (G11) of $\tilde{\mu}^2$, and the final equality follows because $\tilde{\mu}^2(\gamma)$ is symmetric under reflection about $\gamma = \pi/2$.

We conclude this part of the Appendix with some comments about the behavior of the two-point functions under a change of GW polarization basis as given by Eq. (D8). A useful discussion of the physical interpretation (in terms of the spin) may be found in Sec. 3.4 of Ref. [68]. This change of polarization basis corresponds to a local gauge transformation of the gravitational field: any physical observables such as the cosmic variance must be invariant under such a basis change. The calculations in this paper correspond to a particular choice of GW gauge: in Eqs. (D6) and (D7) we have defined a specific basis for the polarization vectors and tensors, as a function of the GW propagation direction $\mathbf{\Omega}$. However, one is free to choose any other basis, defined by Eq. (D8) where the angle $\epsilon(\mathbf{\Omega})$ is an *arbitrary function* of the propagation direction.

Under this gauge transformation of the gravitational field, the quantities $F_1^+(\mathbf{\Omega})F_2^+(\mathbf{\Omega})$ and $F_1^\times(\mathbf{\Omega})F_2^\times(\mathbf{\Omega})$ are not gauge-invariant, but their sum is. Since this sum is the integrand for the normal Hellings-Downs curve, that quantity is gauge-invariant. In contrast, the quantity $F_1^+(\mathbf{\Omega})F_2^+(\mathbf{\Omega}') + F_1^\times(\mathbf{\Omega})F_2^\times(\mathbf{\Omega}')$ which appears in the two-point function $\mu(\gamma,\beta)$ is not gauge-invariant, because the gauge transformations $\epsilon(\mathbf{\Omega})$ and $\epsilon(\mathbf{\Omega}')$ may differ. However, combinations such as the one that appears in Eq. (G12) are gauge-invariant, as can be easily seen from working in a circular polarization basis.

---